\documentclass[manuscript, screen]{acmart}
\setcopyright{none}
\renewcommand\footnotetextcopyrightpermission[1]{} 

\usepackage{amsmath,amsfonts}
\usepackage{algorithmic}
\usepackage{algorithm}
\usepackage{array}
\usepackage{textcomp}
\usepackage{stfloats}
\usepackage{url}
\usepackage{verbatim}
\usepackage{graphicx}

\usepackage[htt]{hyphenat}
\usepackage{multirow}
\usepackage{fontawesome}
\usepackage[font=small]{caption}
\usepackage{subcaption}
\usepackage{hhline}
\usepackage{xurl}
\usepackage{enumitem}
\usepackage{amsfonts} 

\definecolor{tgray2}{HTML}{C0C0C0}
\definecolor{twhite}{HTML}{FFFFFF}
\definecolor{tgray}{HTML}{EFEFEF}
\definecolor{tblack}{HTML}{000000}

\begin{document}

\title{Crypto-Ransomware and Their Defenses: In-depth Behavioral Characterization, Discussion of Deployability, and New Insights}

\author{Wenjia Song}
\email{wenjia7@vt.edu}
\affiliation{%
  \institution{Virginia Tech}
  \city{Blacksburg}
  \state{VA}
  \country{USA}
  \postcode{24061}
}

\author{Sanjula Karanam}
\email{sanjula@vt.edu}
\affiliation{%
  \institution{Virginia Tech}
  \city{Blacksburg}
  \state{VA}
  \country{USA}
  \postcode{24061}
}

\author{Ya Xiao}
\email{yax99@vt.edu}
\affiliation{%
  \institution{Virginia Tech}
  \city{Blacksburg}
  \state{VA}
  \country{USA}
  \postcode{24061}
}

\author{Jingyuan Qi}
\email{jingyq1@vt.edu}
\affiliation{%
  \institution{Virginia Tech}
  \city{Blacksburg}
  \state{VA}
  \country{USA}
  \postcode{24061}
}

\author{Nathan Dautenhahn}
\email{ndd@rice.edu}
\affiliation{%
  \institution{Rice University}
  \city{Houston}
  \state{TX}
  \country{USA}
  \postcode{77005}
}

\author{Na Meng}
\email{nm8247@vt.edu}
\affiliation{%
  \institution{Virginia Tech}
  \city{Blacksburg}
  \state{VA}
  \country{USA}
  \postcode{24061}
}

\author{Elena Ferrari}
\email{elena.ferrari@uninsubria.it}
\affiliation{%
  \institution{University of Insubria}
  \city{Varese}
  \country{Italy}}

\author{Danfeng (Daphne) Yao}
\email{danfeng@vt.edu}
\affiliation{%
  \institution{Virginia Tech}
  \city{Blacksburg}
  \state{VA}
  \country{USA}
  \postcode{24061}
}

\renewcommand{\shortauthors}{Song et al.}

\begin{abstract}
Crypto-ransomware has caused an unprecedented scope of impact in recent years with an evolving level of sophistication. An extensive range of studies have been on defending against ransomware and reviewing the efficacy of various protections. However, for practical defenses, deployability holds equal significance as detection accuracy. Therefore, in this study, we review 117 published ransomware defense works, categorize them by the level they are implemented at, and discuss the deployability. API-based solutions are easy to deploy and most existing works focus on machine learning-based classification. To provide more insights, we quantitively characterize the runtime behaviors of real-world ransomware samples. Based on our experimental findings, we present a possible future detection direction with our consistency analysis and API-contrast-based refinement. Moreover, we experimentally evaluate various commercial defenses and identify the security gaps. Our findings help the field understand the deployability of ransomware defenses and create more effective, practical solutions.

\end{abstract}




\maketitle

\section{Introduction}
\label{sec:intro}

Crypto-ransomware extorts money from victims by encrypting their files. It first appeared in 1989 and has had a resurgence recently. In 2017, WannaCry hit around 230,000 computers across 150 countries, causing a loss of \$4 billion~\cite{wannacry230}.  The notorious Colonial Pipeline hack affected nearly half of the U.S. east coast gas supply and roughly \$5 million was paid for recovery in May 2021~\cite{pipeline}. 37\% organizations reported being attacked by ransomware in 2021 and the average cost of recovering was \$1.85 million \cite{sophos2021}. Even worse, on average, only 65\% of data was recovered after the ransom payment~\cite{sophos2021}. One possible reason for this low recovery rate is file corruption.

Many efforts have been made to defend against ransomware attacks. However, in successful ransomware attacks, 77\% of victims are running up-to-date endpoint protection, implying inadequacy in current solutions in practice~\cite{77protection}. There have also been academic solutions proposed to detect ransomware threats. Monitoring the file system is a widely used approach~\cite{unveil,ransomspector,cryptolock,ShieldFS,r-locker,Mehnaz2018rwguard}. Hardware performance~\cite{posse,hardware-profile}, API call occurrence~\cite{api-2023,Api-based,Hwang2020two-stage}, and network activities~\cite{Alhawi2018NetCoverse,Almashhadani2019multi-classifier,Cabaj2016cryptowall} could also reveal the malicious purpose of a program. Survey works have also been published to summarize the knowledge and effectiveness of defenses in recent years~\cite{oz2022survey,mcintosh2021survey,moussaileb2021survey}.

Despite these research advances, there seems to be limited literature focusing on the deployability of existing ransomware defenses. While effectiveness is crucial for a defense system, the significant amount of deployment effort could be a reason that stops the state-of-the-art detection system from turning into prevalent endpoint protection solutions. Existing file system-based solutions (e.g., UNVEIL~\cite{unveil} and ShieldFS~\cite{ShieldFS}) focus on low-level file system activities, requiring kernel-level modifications. API-based approaches could be a deployable alternative, taking advantage of fine-grained program behavior information. API-based solutions monitor the target’s behaviors at the program level and require fewer system modifications. Current API-based detection works~\cite{api-2023,Api-based} mostly rely on machine learning classification. In-depth analyses of ransomware-specific execution patterns could complement and strengthen the detection. To close this gap, we aim to answer the following important research questions:

\noindent\textbf{RQ1:} What are the existing ransomware defenses? What are the more deployable approaches? (Section~\ref{sec:existing})

\noindent\textbf{RQ2:} For highly deployable commercial defenses, what are the ransomware behaviors that trigger detection? What are the security gaps? (Section~\ref{sec:commercial})

\noindent\textbf{RQ3:} What are the quantitative ransomware API invocation behaviors? How do they systematically compare with benign software? How to quantify the unique ransomware API usage patterns for detection? (Sections~\ref{sec:character} and \ref{sec:classification})

The objective of this work is to comprehensively review various existing ransomware defenses and discuss their deployability, provide experimental insights on the unique ransomware API invocation patterns, and demonstrate a deployable detection direction by quantifying program API usage. Specifically, we review 117 published ransomware defense works, up to 2023, and categorize them by the level they are implemented at. We discuss the deployability of each category. Furthermore, we experimentally evaluate 3 types of commercial tools, including 6 crypto decryptors, around 70 malware scanners, and 8 antivirus software. Additionally, we execute and analyze 2 sets of real-world ransomware, with over 400 samples in total.

We summarize our experimental findings below.
\begin{itemize}

\begin{sloppypar}
\item \textbf{API-usage profiling.} {\sloppy We quantitatively analyze the behaviors of ransomware samples through two sets of experiments. We first manually inspect 54 real-world ransomware samples from 35 families, including the notorious WannaCry, Sodinokibi, Babuk, and the most active ransomware families in 2021, LockBit, Mespinoza, and Hive, with a focus on encryption activities. We find that ransomware has a distinct file access behavior pattern during execution. We further collect the occurrence frequency of 288 Windows APIs from 348 ransomware samples. We discover differences in API occurrences and invocation frequencies between benign and ransomware executions, which is beneficial for improving detection accuracy. We leverage these for computing the API contrast score.}
\end{sloppypar}

\item \textbf{API-based classification.} Based on our observation of API usage, we discuss a possible classification mechanism, consisting of consistency analysis and API usage contrast refinement operations. In consistency analysis, we use multiple mathematical methods to capture the unique ransomware execution features, focusing on the fundamental encryption nature. In refinement, we further examine the positive cases by an API contrast score to filter wrongly classified cases. We further provide in-depth case analysis of API usage contrast with attack context. We carry out a feasibility assessment of our classification approach against 29 sets of execution traces. The results show that our consistency analysis effectively identifies all malicious executions. Our findings could help inspire future detection works in this field.

\item \textbf{Evaluation of commercial tools.} We extensively evaluate 3 types of commercial defenses. Through our experiments, we find that the success rate of commercial decryptors is low (1 success out of 6), suffering from low generality across variants of ransomware. Anti-virus software detects only generic malicious behaviors, being insensitive to core ransomware encryption. Malware scanners use signature-based detection and miss unknown or obfuscated samples. The observations reveal that behavioral detection is necessary and our new API findings could help strengthen ransomware-specific protection.

\end{itemize}

The rest of the paper is organized as follows. \textbf{Section~\ref{sec:related}} reviews the related survey works. \textbf{Section~\ref{sec:existing}} categorizes existing ransomware defesnes and discusses their deployability. \textbf{Section~\ref{sec:commercial}} evaluates various commercial tools. \textbf{Section~\ref{sec:character}} reports our experimental findings on ransomware API-based behavioral characteristics, with systematic comparison to benign software. \textbf{Section~\ref{sec:classification}} shows the results of the feasibility test of API-based classification. \textbf{Section~\ref{sec:discussion}} further discusses our insights and limitations of our work. \textbf{Section~\ref{sec:conclusion}} concludes the work.

\section{Related Work}
\label{sec:related}

Many survey works have been done to summarize the knowledge of ransomware and its defenses from different aspects. We review existing surveys in this section and discuss how the focus of our work differs.

\textbf{Surveys of ransomware evolution.} Ransomware first appeared in 1989 as PC Cyborg and has been evolving for several decades. Techniques utilized in each attack stage have exhibited new features and increased complexity. Research has been done to study the evolution~\cite{oz2022survey,berrueta2019survey,garg2018past,zimba2019understanding,rehman2019security,desai2019survey,maniath2019survey,Sultan2018,survey23}. Specifically, Oz et al.~\cite{oz2022survey} summarized the evolution of ransomware families from 1989 to 2020. Zimba et al.~\cite{zimba2019understanding} reviewed the transformation of ransomware attack structure in the Internet-of-Things (IoT) system. Maniath et al.~\cite{maniath2019survey} discussed the change of encryption key management strategies used by ransomware. Alwashali et al. \cite{ali2021} conducted a survey on Ransomware-as-a-Service (RaaS), a newly emerged business selling malware to anyone who wants to launch an attack.

\textbf{Surveys of ransomware on different platforms.} Ransomware targets various systems and platforms. Comprehensive studies has been done on ransomware on different platforms, such as personal computers (PCs)~\cite{gonzalez2017detection, popoola2017ransomware,dargahi2019cyber,maigida2019systematic,keshavarzi2020i2ce3,kok2019ransomware,alzahrani2019review,popoola2017ransomware,berrueta2019survey,silva2019survey,bijitha2020survey,gonzalez2017detection,garg2018past,kiru2019age,maniath2019survey}, mobile devices~\cite{abraham2019survey,dargahi2019cyber,maigida2019systematic,keshavarzi2020i2ce3,silva2019survey,bijitha2020survey}, and cyber-physical systems (CPS)~\cite{dargahi2019cyber, keshavarzi2020i2ce3,popoola2017ransomware,humayun2021internet,benmalek2024ransomware}. Some less common platform targeted by ransomware includes cloud~\cite{ibarra2019ransomware} and wearable devices~\cite{bander2018}. Besides, works have also been done analyzing ransomware infecting specific operating systems, such as Windows~\cite{aurangzeb2017ransomware,naseer2020windows,moussaileb2021survey} and Android~\cite{mohan2017efficacy,desai2019survey,alzahrani2019review,Olaimat2021}.

\textbf{Surveys of ransomware taxonomy.} Ransomware could be categorized by various features. By attack methods, two main types are identified, namely locker-ransomware and crypto-ransomware~\cite{bander2018,oz2022survey}. Besides these two types, Razaulla et al.~\cite{survey23} further included scareware and leakware as categories of ransomware. By targeted victims, they divide ransomware into consumer ransomware and organization ransomware~\cite{bander2018,oz2022survey}. As mentioned above, by platform, ransomware could be divided into PC ransomware, mobile ransomware, cloud ransomware, IoT ransomware, and ransomwear. Bajpai et al.~\cite{bajpai2018key} also introduced a taxonomy based on the key management strategies used by ransomware.

\textbf{Surveys of ransomware defenses.} The detection and mitigation is also an important field of ransomware study. Extensive research has been done reviewing a wide range of defensive strategies~\cite{mohan2017efficacy,maigida2019systematic,kok2019ransomware, alzahrani2019review,berrueta2019survey,bijitha2020survey,shinde2016ransomware,gonzalez2017detection,humayun2021internet,bander2018,kiru2019age,desai2019survey,maniath2019survey,mcintosh2021survey,moussaileb2021survey,cen2024ransomware,alqahtani2022survey,aldauiji2022utilizing,begovic2023cryptographic,Larsen2021,Beaman2021}. While most of the works discuss all types of defensive strategies, some of them have a specific focus. Particularly, Cen et al.~\cite{cen2024ransomware} reviewed detections targeting the early attack stage. Begovic et al.~\cite{begovic2023cryptographic} and Larsen et al.~\cite{Larsen2021} analyzed encryption stage detection methods. Machine learning and deep learning solutions have also been studied~\cite{abraham2019survey,alzahrani2019review,Larsen2021}. McIntosh et al.~\cite{mcintosh2021survey} analyzed various existing ransomware mitigation techniques and emphasized the need for unified evaluation metrics.

\textbf{Our work.} Our work’s focus is on the deployability of ransomware defense. While effectiveness is one of the most important metrics in evaluating a defense strategy and the focus of the majority works in this field, the deployability gap hinders many solutions from being prevalent in the real world. Therefore, to understand the deployability of different methods, we analyze and categorize existing solutions by the layer where they are deployed in the system and discuss the challenges of deploying a solution at each level. Beyond reviewing existing work, we further provide experimental insights on the evaluation of commercial ransomware defenses and dynamic features of core crypto-ransomware behaviors, which could help the future development of deployable ransomware detection methods.

\section{Existing Ransomware Defenses and Their Deployability}
\label{sec:existing}



\begin{table}[]
\caption{Summary of ransomware defenses based on the system level they are implemented at. Hybrid methods may appear in multiple categories. Generally, the lower the level a solution is implemented at, the more difficult to deploy it to a new system or device.}

\small

\begin{tabular}{
>{\columncolor[HTML]{EFEFEF}}l l
>{\columncolor[HTML]{EFEFEF}}l l
>{\columncolor[HTML]{EFEFEF}}l 
>{\columncolor[HTML]{EFEFEF}}l }
\hline
\cellcolor[HTML]{C0C0C0}                              & \cellcolor[HTML]{C0C0C0}\textbf{Methods}                                                           & \cellcolor[HTML]{C0C0C0}\textbf{Features}                             & \cellcolor[HTML]{C0C0C0}\textbf{Existing Solutions}                                                                                                                                                                                                                                                                                                                                                                                                                                                                                                                                                                                                                                                                                                                                                                                                                           & \multicolumn{2}{l}{\cellcolor[HTML]{C0C0C0}\textbf{Advantage}}                                                                                                                                                                                                                                                                     \\ \hline
\cellcolor[HTML]{EFEFEF}                              & Antivirus Software                                                                                 &                                                                       &                                                                                                                                                                                                                                                                                                                                                                                                                                                                                                                                                                                                                                                                                                                                                                                                                                                                               &                                                                                                                                                 & \cellcolor[HTML]{EFEFEF}                                                                                                                                                         \\ \hhline{|>{\arrayrulecolor{tgray}}->{\arrayrulecolor{tblack}}|-|-|-|-|>{\arrayrulecolor{tgray}}-|}
\cellcolor[HTML]{EFEFEF}                              &                                                                                                    & Opcode                                                                & \cite{opcode2019}, \cite{opcode2020}, \cite{karimi2017android}, \cite{ferrante2018extinguishing}                                                                                                                                                                                                                                                                                                                                                                                                                                                                                                                                                                                                                                                                                                                                                                              & \cellcolor[HTML]{EFEFEF}                                                                                                                        & \cellcolor[HTML]{EFEFEF}                                                                                                                                                         \\ \hhline{|>{\arrayrulecolor{tgray}}->{\arrayrulecolor{twhite}}|->{\arrayrulecolor{tblack}}|-|->{\arrayrulecolor{tgray}}|-|-|}
\cellcolor[HTML]{EFEFEF}                              &                                                                                                    & Bytecode                                                              & \cite{cimitile2018talos}, \cite{khan2020digital}, \cite{reddy2021machine}, \cite{verma2018analysing}, \cite{hill2018cryptoknight}, \cite{yamany2022new}                                                                                                                                                                                                                                                                                                                                                                                                                                                                                                                                                                                                                                                                                                                       & \cellcolor[HTML]{EFEFEF}                                                                                                                        & \cellcolor[HTML]{EFEFEF}                                                                                                                                                         \\ \hhline{|>{\arrayrulecolor{tgray}}->{\arrayrulecolor{twhite}}|->{\arrayrulecolor{tblack}}|-|->{\arrayrulecolor{tgray}}|-|-|}
\cellcolor[HTML]{EFEFEF}                              &                                                                                                    & API                                                                   & \begin{tabular}[c]{@{}l@{}}\cite{electronics8080868}, \cite{randetector}, \cite{faris2020optimizing}, \cite{maiorca2017r}, \cite{medhat2018new}, \\ \cite{su2018detecting}, \cite{ferrante2018extinguishing}, \cite{gharib2017dna}, \cite{hasan2017ranshunt}, \cite{andronio2015heldroid}\end{tabular}                                                                                                                                                                                                                                                                                                                                                                                                                                                                                                                                                                        & \cellcolor[HTML]{EFEFEF}                                                                                                                        & \cellcolor[HTML]{EFEFEF}                                                                                                                                                         \\ \hhline{|>{\arrayrulecolor{tgray}}->{\arrayrulecolor{twhite}}|->{\arrayrulecolor{tblack}}|-|->{\arrayrulecolor{tgray}}|-|-|}
\cellcolor[HTML]{EFEFEF}                              &                                                                                                    & \begin{tabular}[c]{@{}l@{}}Text and \\ Other Files\end{tabular}       & \begin{tabular}[c]{@{}l@{}}\cite{randetector}, \cite{alzahrani2018randroid}, \cite{cuzzocrea2018novel}, \cite{medhat2018new}, \cite{su2018detecting}, \\ \cite{alsoghyer2020effectiveness}, \cite{andronio2015heldroid}, \cite{gharib2017dna}, \cite{hasan2017ranshunt}, \cite{ahmed2022mitigating},\\ \cite{keong2020voterchoice}, \cite{shaukat2018ransomwall}, \cite{verma2018analysing}, \cite{ahmadian20162entfox}\end{tabular}                                                                                                                                                                                                                                                                                                                                                                                                                                          & \cellcolor[HTML]{EFEFEF}                                                                                                                        & \cellcolor[HTML]{EFEFEF}                                                                                                                                                         \\ \hhline{|>{\arrayrulecolor{tgray}}->{\arrayrulecolor{twhite}}|->{\arrayrulecolor{tblack}}|-|->{\arrayrulecolor{tgray}}|-|-|}
\cellcolor[HTML]{EFEFEF}                              &                                                                                                    & Program Header                                                        & \cite{keong2020voterchoice}, \cite{shaukat2018ransomwall}, \cite{verma2018analysing}, \cite{zuhair2020multi}, \cite{aurangzeb2022bigrc}                                                                                                                                                                                                                                                                                                                                                                                                                                                                                                                                                                                                                                                                                                                                       & \cellcolor[HTML]{EFEFEF}                                                                                                                        & \cellcolor[HTML]{EFEFEF}                                                                                                                                                         \\ \hhline{|>{\arrayrulecolor{tgray}}->{\arrayrulecolor{twhite}}|->{\arrayrulecolor{tblack}}|-|->{\arrayrulecolor{tgray}}|-|-|}
\cellcolor[HTML]{EFEFEF}                              & \multirow{-6}{*}{Static Analysis}                                                                  & Domain Name                                                           & \cite{raptor}                                                                                                                                                                                                                                                                                                                                                                                                                                                                                                                                                                                                                                                                                                                                                                                                                                                                 & \multirow{-6}{*}{\cellcolor[HTML]{EFEFEF}\begin{tabular}[c]{@{}l@{}}Fast and \\ efficient. \\ No execution\\ is required.\end{tabular}}         & \cellcolor[HTML]{EFEFEF}                                                                                                                                                         \\ \hhline{|>{\arrayrulecolor{tgray}}->{\arrayrulecolor{tblack}}|-|-|-|-|>{\arrayrulecolor{tgray}}-|}

\cellcolor[HTML]{EFEFEF}                              &                                                                                                    & Honey Pot                                                             & \begin{tabular}[c]{@{}l@{}}\cite{keong2020voterchoice}, \cite{shaukat2018ransomwall}, \cite{wang2018automatically}, \cite{zuhair2020multi}, \cite{honeypot2016}, \\ \cite{r-locker}, \cite{moussaileb2018ransomware}, \cite{lee2017make}\end{tabular}                                                                                                                                                                                                                                                                                                                                                                                                                                                                                                                                                                                                                         & \cellcolor[HTML]{EFEFEF}                                                                                                                        & \cellcolor[HTML]{EFEFEF}                                                                                                                                                         \\ \hhline{|>{\arrayrulecolor{tgray}}->{\arrayrulecolor{twhite}}|->{\arrayrulecolor{tblack}}|-|->{\arrayrulecolor{tgray}}|-|-|}
\cellcolor[HTML]{EFEFEF}                              &                                                                                                    & File Entropy                                                          & \cite{jung2018ransomware}, \cite{wang2018automatically}, \cite{chen2017uncovering}, \cite{lee2019machine}                                                                                                                                                                                                                                                                                                                                                                                                                                                                                                                                                                                                                                                                                                                                                                     & \cellcolor[HTML]{EFEFEF}                                                                                                                        & \cellcolor[HTML]{EFEFEF}                                                                                                                                                         \\ \hhline{|>{\arrayrulecolor{tgray}}->{\arrayrulecolor{twhite}}|->{\arrayrulecolor{tblack}}|-|->{\arrayrulecolor{tgray}}|-|-|}
\cellcolor[HTML]{EFEFEF}                              & \multirow{-3}{*}{\begin{tabular}[c]{@{}l@{}}User File \\ Configuration\\ or Analysis\end{tabular}} & Data Decryption                                                       & \cite{Kim2022}, \cite{partyticket}, \cite{Lee2020Magniber}                                                                                                                                                                                                                                                                                                                                                                                                                                                                                                                                                                                                                                                                                                                                                                                                                    & \multirow{-3}{*}{\cellcolor[HTML]{EFEFEF}\begin{tabular}[c]{@{}l@{}}Only limited \\ monitoring is \\ required.\end{tabular}}                    & \cellcolor[HTML]{EFEFEF}                                                                                                                                                         \\ \hhline{|>{\arrayrulecolor{tgray}}->{\arrayrulecolor{tblack}}|-|-|-|>{\arrayrulecolor{tgray}}-|-|}

\cellcolor[HTML]{EFEFEF}                              &                                                                                                    & \begin{tabular}[c]{@{}l@{}}API Occurrence \\ or Sequence\end{tabular} & \begin{tabular}[c]{@{}l@{}}\cite{ALRIMY2019476},  \cite{ferrante2018extinguishing}, \cite{gharib2017dna}, \cite{hasan2017ranshunt}, \\ \cite{api-2023},\cite{Kok2022}, \cite{Api-based}, \cite{Hwang2020two-stage}, \cite{detection-svm}, \cite{call-graph}, \\ \cite{EldeRan}, \cite{ahmed2020automated}, \cite{ahmed2020system}, \cite{homayoun2019drthis}, \cite{lu2017ransomware}, \cite{fernando2024fesad}\\ \cite{zuhair2019rands}, \cite{al2018zero}, \cite{bae2020ransomware}, \cite{lu2020ransomware}, \cite{maniath2017deep}, \\ \cite{scalas2019effectiveness}, \cite{sharmeen2020avoiding}, \cite{al2021redundancy}, \cite{kok2020evaluation},  \cite{vinayakumar2017evaluating}, \\ \cite{oz2023rob}, \cite{zhou2020evaluation}, \cite{ahmed2022mitigating}, \cite{aurangzeb2022bigrc}, \cite{gazzan2023enhanced},\end{tabular} & \cellcolor[HTML]{EFEFEF}                                                                                                                        & \cellcolor[HTML]{EFEFEF}                                                                                                                                                         \\ \hhline{|>{\arrayrulecolor{tgray}}->{\arrayrulecolor{twhite}}|->{\arrayrulecolor{tblack}}|-|->{\arrayrulecolor{tgray}}|-|-|}
\cellcolor[HTML]{EFEFEF}                              &                                                                                                    & File Access API                                                       & \cite{verma2018analysing}, \cite{honda2018ransomware}, \cite{may2019combating}, \cite{ahmadian20162entfox}, \cite{chen2019automated}, \cite{homayoun2017know}                                                                                                                                                                                                                                                                                                                                                                                                                                                                                                                                                                                                                                                                                                                 & \cellcolor[HTML]{EFEFEF}                                                                                                                        & \cellcolor[HTML]{EFEFEF}                                                                                                                                                         \\ \hhline{|>{\arrayrulecolor{tgray}}->{\arrayrulecolor{twhite}}|->{\arrayrulecolor{tblack}}|-|->{\arrayrulecolor{tgray}}|-|-|}
\cellcolor[HTML]{EFEFEF}                              &                                                                                                    & Crypto API                                                            & \begin{tabular}[c]{@{}l@{}}\cite{shaukat2018ransomwall}, \cite{verma2018analysing}, \cite{al2020pseudo}, \cite{palisse2017ransomware}, \cite{ahmadian20162entfox}, \\ \cite{paybreak}, \cite{lee2018ransomware}, \cite{Yuste2021Avaddon}, \cite{Cheng2019dptCry}\end{tabular}                                                                                                                                                                                                                                                                                                                                                                                                                                                                                                                                                                                                 & \cellcolor[HTML]{EFEFEF}                                                                                                                        & \cellcolor[HTML]{EFEFEF}                                                                                                                                                         \\ \hhline{|>{\arrayrulecolor{tgray}}->{\arrayrulecolor{twhite}}|->{\arrayrulecolor{tblack}}|-|->{\arrayrulecolor{tgray}}|-|-|}
\cellcolor[HTML]{EFEFEF}                              &                                                                                                    & Registry-related API                                                  & \cite{hasan2017ranshunt}, \cite{keong2020voterchoice}, \cite{ahmadian20162entfox}, \cite{chen2017automated}, \cite{homayoun2017know}                                                                                                                                                                                                                                                                                                                                                                                                                                                                                                                                                                                                                                                                                                                                          & \cellcolor[HTML]{EFEFEF}                                                                                                                        & \cellcolor[HTML]{EFEFEF}                                                                                                                                                         \\ \hhline{|>{\arrayrulecolor{tgray}}->{\arrayrulecolor{twhite}}|->{\arrayrulecolor{tblack}}|-|->{\arrayrulecolor{tgray}}|-|-|}
\cellcolor[HTML]{EFEFEF}                              &                                                                                                    & Network API                                                           & \cite{wang2018automatically}, \cite{chen2019automated}                                                                                                                                                                                                                                                                                                                                                                                                                                                                                                                                                                                                                                                                                                                                                                                                                        & \cellcolor[HTML]{EFEFEF}                                                                                                                        & \cellcolor[HTML]{EFEFEF}                                                                                                                                                         \\ \hhline{|>{\arrayrulecolor{tgray}}->{\arrayrulecolor{twhite}}|->{\arrayrulecolor{tblack}}|-|->{\arrayrulecolor{tgray}}|-|-|}
\cellcolor[HTML]{EFEFEF}                              &                                                                                                    & Cilpboard API                                                         & \cite{wang2018automatically}                                                                                                                                                                                                                                                                                                                                                                                                                                                                                                                                                                                                                                                                                                                                                                                                                                                  & \cellcolor[HTML]{EFEFEF}                                                                                                                        & \cellcolor[HTML]{EFEFEF}                                                                                                                                                         \\ \hhline{|>{\arrayrulecolor{tgray}}->{\arrayrulecolor{twhite}}|->{\arrayrulecolor{tblack}}|-|->{\arrayrulecolor{tgray}}|-|-|}
\cellcolor[HTML]{EFEFEF}                              & \multirow{-7}{*}{\begin{tabular}[c]{@{}l@{}}Dynamic API \\ Monitoring\end{tabular}}                & Process-related API                                                   & \cite{wang2018automatically}, \cite{chen2019automated}                                                                                                                                                                                                                                                                                                                                                                                                                                                                                                                                                                                                                                                                                                                                                                                                                        & \cellcolor[HTML]{EFEFEF}                                                                                                                        & \cellcolor[HTML]{EFEFEF}                                                                                                                                                         \\ \hhline{|>{\arrayrulecolor{tgray}}->{\arrayrulecolor{tblack}}|-|-|-|-|>{\arrayrulecolor{tgray}}-|}

\cellcolor[HTML]{EFEFEF}                              &                                                                                                    & Query Sequences                                                       & \cite{database-ransomware}                                                                                                                                                                                                                                                                                                                                                                                                                                                                                                                                                                                                                                                                                                                                                                                                                                                    & \cellcolor[HTML]{EFEFEF}                                                                                                                        & \cellcolor[HTML]{EFEFEF}                                                                                                                                                         \\ \hhline{|>{\arrayrulecolor{tgray}}->{\arrayrulecolor{twhite}}|->{\arrayrulecolor{tblack}}|-|->{\arrayrulecolor{tgray}}|-|-|}
\cellcolor[HTML]{EFEFEF}                              &                                                                                                    & Foreground Analysis                                                   & \cite{chen2017uncovering}                                                                                                                                                                                                                                                                                                                                                                                                                                                                                                                                                                                                                                                                                                                                                                                                                                                     & \cellcolor[HTML]{EFEFEF}                                                                                                                        & \cellcolor[HTML]{EFEFEF}                                                                                                                                                         \\ \hhline{|>{\arrayrulecolor{tgray}}->{\arrayrulecolor{twhite}}|->{\arrayrulecolor{tblack}}|-|->{\arrayrulecolor{tgray}}|-|-|}
\cellcolor[HTML]{EFEFEF}                              &                                                                                                    & Memory Dumps                                                          & \cite{cohen2018}, \cite{ahmed2022mitigating}, \cite{lu2017ransomware}                                                                                                                                                                                                                                                                                                                                                                                                                                                                                                                                                                                                                                                                                                                                                                                                  & \cellcolor[HTML]{EFEFEF}                                                                                                                        & \cellcolor[HTML]{EFEFEF}                                                                                                                                                         \\ \hhline{|>{\arrayrulecolor{tgray}}->{\arrayrulecolor{twhite}}|->{\arrayrulecolor{tblack}}|-|->{\arrayrulecolor{tgray}}|-|-|}
\cellcolor[HTML]{EFEFEF}                              &                                                                                                    & System Logs                                                           & \cite{roy2021deepran}                                                                                                                                                                                                                                                                                                                                                                                                                                                                                                                                                                                                                                                                                                                                                                                                                                                         & \cellcolor[HTML]{EFEFEF}                                                                                                                        & \cellcolor[HTML]{EFEFEF}                                                                                                                                                         \\ \hhline{|>{\arrayrulecolor{tgray}}->{\arrayrulecolor{twhite}}|->{\arrayrulecolor{tblack}}|-|->{\arrayrulecolor{tgray}}|-|-|}
\cellcolor[HTML]{EFEFEF}                              & \multirow{-5}{*}{\begin{tabular}[c]{@{}l@{}}Other Dynamic \\ Behavirol \\ Analysis\end{tabular}}   & Other                                                                 & \cite{zuhair2019rands}, \cite{zuhair2020multi}, \cite{ferrante2018extinguishing}                                                                                                                                                                                                                                                                                                                                                                                                                                                                                                                                                                                                                                                                                                                                                                                              & \cellcolor[HTML]{EFEFEF}                                                                                                                        & \cellcolor[HTML]{EFEFEF}                                                                                                                                                         \\ \hhline{|>{\arrayrulecolor{tgray}}->{\arrayrulecolor{tblack}}|-|-|-|>{\arrayrulecolor{tgray}}-|-|}

\cellcolor[HTML]{EFEFEF}                              &                                                                                                    & Number of Packets                                                     & \cite{ferrante2018extinguishing}, \cite{verma2018analysing}, \cite{lu2017ransomware}, \cite{Almashhadani2019multi-classifier}, \cite{modi2020detecting}                                                                                                                                                                                                                                                                                                                                                                                                                                                                                                                                                                                                                                                                                                                       & \cellcolor[HTML]{EFEFEF}                                                                                                                        & \cellcolor[HTML]{EFEFEF}                                                                                                                                                         \\ \hhline{|>{\arrayrulecolor{tgray}}->{\arrayrulecolor{twhite}}|->{\arrayrulecolor{tblack}}|-|->{\arrayrulecolor{tgray}}|-|-|}
\cellcolor[HTML]{EFEFEF}                              &                                                                                                    & \begin{tabular}[c]{@{}l@{}}Text and Packet\\ Information\end{tabular} & \begin{tabular}[c]{@{}l@{}}\cite{andronio2015heldroid}, \cite{keong2020voterchoice}, \cite{wang2018automatically}, \cite{Alhawi2018NetCoverse}, \cite{Almashhadani2019multi-classifier}, \\ \cite{Cabaj2016cryptowall}, \cite{cusack2018machine}, \cite{fernandez2019intelligent}, \cite{modi2020detecting}, \cite{wani2020ransomware}, \\ \cite{ahmadian2015connection}, \cite{SINGH2023108601}\end{tabular}                                                                                                                                                                                                                                                                                                                                                                                                                                                                 & \cellcolor[HTML]{EFEFEF}                                                                                                                        & \cellcolor[HTML]{EFEFEF}                                                                                                                                                         \\ \hhline{|>{\arrayrulecolor{tgray}}->{\arrayrulecolor{twhite}}|->{\arrayrulecolor{tblack}}|-|->{\arrayrulecolor{tgray}}|-|-|}
\cellcolor[HTML]{EFEFEF}                              &                                                                                                    & HTTP Requests                                                         & \cite{bortolameotti2017decanter}, \cite{cabaj2018software}                                                                                                                                                                                                                                                                                                                                                                                                                                                                                                                                                                                                                                                                                                                                                                                                                    & \cellcolor[HTML]{EFEFEF}                                                                                                                        & \cellcolor[HTML]{EFEFEF}                                                                                                                                                         \\ \hhline{|>{\arrayrulecolor{tgray}}->{\arrayrulecolor{twhite}}|->{\arrayrulecolor{tblack}}|-|->{\arrayrulecolor{tgray}}|-|-|}
\multirow{-26}{*}{\cellcolor[HTML]{EFEFEF}User Mode}  & \multirow{-4}{*}{\begin{tabular}[c]{@{}l@{}}Network Packet \\ Monitoring\end{tabular}}             & request command                                                       & \cite{morato2018ransomware}                                                                                                                                                                                                                                                                                                                                                                                                                                                                                                                                                                                                                                                                                                                                                                                                                                                   & \multirow{-16}{*}{\cellcolor[HTML]{EFEFEF}\begin{tabular}[c]{@{}l@{}}Effectively \\ captures \\ program \\ execution \\ patterns.\end{tabular}} & \multirow{-26}{*}{\cellcolor[HTML]{EFEFEF}\begin{tabular}[c]{@{}l@{}}Easy to \\ deploy. \\ Less or \\ no system\\ modification \\ or \\ dependency \\ is required.\end{tabular}} \\ \hhline{|>{\arrayrulecolor{tblack}}-|-|-|-|-|-|}

\cellcolor[HTML]{EFEFEF}                              &                                                                                                    & IRP                                                                   & \begin{tabular}[c]{@{}l@{}}\cite{shaukat2018ransomwall}, \cite{cryptolock}, \cite{unveil}, \cite{ransomspector}, \cite{ShieldFS}, \\ \cite{Mehnaz2018rwguard}, \cite{peeler}, \cite{ramesh2020automated}, \cite{ayub2020request}, \cite{hirano2019machine}, \\ \cite{kim2018white}, \cite{palisse2017data}, \cite{moussaileb2018ransomware}, \cite{ayub2023rwarmor}, \cite{elkhail2023seamlessly}\end{tabular}                                                                                                                                                                                                                                                                                                                                                                                                                                                                & \cellcolor[HTML]{EFEFEF}                                                                                                                        &                                                                                                                                                                                  \\ \hhline{|>{\arrayrulecolor{tgray}}->{\arrayrulecolor{twhite}}|->{\arrayrulecolor{tblack}}|-|->{\arrayrulecolor{tgray}}|-|-|} \cline{6-6} 
\cellcolor[HTML]{EFEFEF}                              & \multirow{-2}{*}{\begin{tabular}[c]{@{}l@{}}File System I/O \\ Analysis\end{tabular}}              & Buffer Entropy                                                        & \cite{cryptolock}                                                                                                                                                                                                                                                                                                                                                                                                                                                                                                                                                                                                                                                                                                                                                                                                                                                             & \cellcolor[HTML]{EFEFEF}                                                                                                                        &                                                                                                                                                                                  \\ \hhline{|>{\arrayrulecolor{tgray}}->{\arrayrulecolor{tblack}}|-|-|-|>{\arrayrulecolor{tgray}}-|-|}
 \cline{6-6} 
\cellcolor[HTML]{EFEFEF}                              & Process Analysis                                                                                   & \begin{tabular}[c]{@{}l@{}}Process Access\\ Control\end{tabular}      & \cite{access}                                                                                                                                                                                                                                                                                                                                                                                                                                                                                                                                                                                                                                                                                                                                                                                                                                                                 & \cellcolor[HTML]{EFEFEF}                                                                                                                        &                                                                                                                                                                                  \\ \hhline{|>{\arrayrulecolor{tgray}}->{\arrayrulecolor{tblack}}|-|-|-|>{\arrayrulecolor{tgray}}-|-|}
 \cline{6-6} 
\multirow{-4}{*}{\cellcolor[HTML]{EFEFEF}Kernel Mode} & \begin{tabular}[c]{@{}l@{}}Hardware \\ Performance \\ Monitoring\end{tabular}                      & Counter                                                               & \begin{tabular}[c]{@{}l@{}}\cite{ferrante2018extinguishing}, \cite{posse}, \cite{hardware-profile}, \cite{hardware23}, \cite{ketzaki2020behaviour}, \\ \cite{ramesh2020automated},  \cite{palisse2017data}, \cite{alam2020rapper}, \cite{celdran2023behavioral}, \cite{mcintosh2023applying}\end{tabular}                                                                                                                                                                                                                                                                                                                                                                                                                                                                                                                                                                     & \multirow{-4}{*}{\cellcolor[HTML]{EFEFEF}Difficult to evade.}                                                                                   &                                                                                                                                                                                  \\ \hhline{|>{\arrayrulecolor{tblack}}-|-|-|-|-|-|}

Hardware                                              & \begin{tabular}[c]{@{}l@{}}Firmware \\ Modification\end{tabular}                                   &                                                                       & \cite{ssd-insider}, \cite{flashguard}, \cite{RSSD}, \cite{MimosaFTL}                                                                                                                                                                                                                                                                                                                                                                                                                                                                                                                                                                                                                                                                                                                                                                                                          & \multicolumn{2}{l}{\cellcolor[HTML]{EFEFEF}}                                                                                                                                                                                                                                                                                       \\ \hhline{|>{\arrayrulecolor{tblack}}-|-|-|-|-|-|}

External Devices                                      & External Backup                                                                                    &                                                                       & \cite{castiglione2019dynamic}                                                                                                                                                                                                                                                                                                                                                                                                                                                                                                                                                                                                                                                                                                                                                                                                                                                 & \multicolumn{2}{l}{\cellcolor[HTML]{EFEFEF}}                                                                                                                                                                                                                                                                                       \\ \hhline{|>{\arrayrulecolor{tblack}}-|-|-|-|-|-|}

\end{tabular}

\label{tab:existing}
\end{table}

In this section, we discuss existing ransomware defenses concentrating on their deployability. A summary is shown in Table~\ref{tab:existing}. Generally, the lower the level in the system a solution is implemented at, the more difficult it is to deploy it on a new system or device. One common reason is that lower-level solutions require more dependency and manual configurations.

\subsection{Defenses at user mode level}

Solutions implemented at the user mode level are easy to deploy as little system modifications are required. We discuss approaches that analyze or monitor various features in this section.

\textbf{Static analysis.} Static analysis extracts useful information from the target program without execution. This method is fast and efficient as no environment setup is required. A range of static information could be useful for making security decisions. Opcode sequences are found to be an effective feature for machine learning models to perform classification between ransomware and benign programs~\cite{opcode2019, opcode2020,karimi2017android,ferrante2018extinguishing}. Bytecode sequences~\cite{cimitile2018talos,khan2020digital,reddy2021machine,verma2018analysing,hill2018cryptoknight} and APIs~\cite{andronio2015heldroid,electronics8080868,randetector,faris2020optimizing,maiorca2017r,medhat2018new,su2018detecting,ferrante2018extinguishing,gharib2017dna,hasan2017ranshunt} are some other useful static features. Besides code, other text and files could also reveal the purpose of a program. Specifically, Android permission file is widely used in Android ransomware detection~\cite{andronio2015heldroid,randetector,alzahrani2018randroid, alsoghyer2020effectiveness,gharib2017dna,ahmed2022mitigating}. Detecting potential threatening and ransom notes in strings could also help identify malicious programs~\cite{su2018detecting,andronio2015heldroid,ahmed2022mitigating,shaukat2018ransomwall,gharib2017dna,hasan2017ranshunt}. Furthermore, Cuzzocrea et al.~\cite{cuzzocrea2018novel} and Yamany et al.~\cite{yamany2022new} utilize the structural entropy and code n-grams to determine if a sample is similar to a ransomware sample. Program header information, such as META-INF for Android applications and import table for executables, could also aid the detection~\cite{keong2020voterchoice, shaukat2018ransomwall, verma2018analysing, zuhair2020multi, aurangzeb2022bigrc}. YARA\footnote{https://yara.readthedocs.io/en/stable/index.html} is a widely used tool that could be utilized to define static rules for malware detection. Medhat et al.~\cite{medhat2018new} used the combination of matched YARA rules of a sample to detect ransomware.

\textbf{Dynamic analysis.} Dynamic analysis collects traces during program execution and could reveal program behaviors that are obfuscated in static files. Dynamic analysis usually happens at the program level, e.g., program instrumentation, requiring relatively light effort to deploy. Programs exhibit a variety of behaviors during execution and many of them have been proven to be valuable for ransomware detection. One of the most popular approaches is to use API call occurrences or sequences as features and classify samples using machine learning models~\cite{ALRIMY2019476,ferrante2018extinguishing,ferrante2018extinguishing,gharib2017dna,hasan2017ranshunt,api-2023,Kok2022,Api-based,Hwang2020two-stage,detection-svm,call-graph,EldeRan,ahmed2020automated,ahmed2020system,homayoun2019drthis,lu2017ransomware,zuhair2019rands,al2018zero,bae2020ransomware,lu2020ransomware,maniath2017deep,scalas2019effectiveness,sharmeen2020avoiding,al2021redundancy,kok2020evaluation,vinayakumar2017evaluating,oz2023rob,zhou2020evaluation,ahmed2022mitigating,aurangzeb2022bigrc,gazzan2023enhanced,fernando2024fesad}. Beyond leveraging recorded traces directly, Takeuchi et al.~\cite{detection-svm} used API call n-grams to detect ransomware execution. Chen et al.~{call-graph} constructed a call graph to capture the relationship between calls. Coglio et al.~\cite{api-2023} selected the most informative APIs as input features. While most works leverage API traces at all execution stages, classification based on only early-stage (i.e., pre-encryption) calls has also been studied~\cite{Kok2022, api-2023}. Approaches that capture program behaviors with a focus on a single or a few API categories, such as file~\cite{verma2018analysing,honda2018ransomware,may2019combating,ahmadian20162entfox,chen2019automated,homayoun2017know}, crypto~\cite{shaukat2018ransomwall,verma2018analysing,al2020pseudo,palisse2017ransomware,ahmadian20162entfox,lee2018ransomware}, network~\cite{wang2018automatically,chen2019automated}, and registry~\cite{hasan2017ranshunt,keong2020voterchoice,ahmadian20162entfox,chen2017automated,homayoun2017know}, are also effective. Recovery strategies that try to restore the decryption key by hooking crypto-related APIs have been proposed as well~\cite{paybreak,Yuste2021Avaddon,Cheng2019dptCry}. API instrumentation (a.k.a., API hooking) could be easily done by utilizing tools such as Intel PIN\footnote{https://www.intel.com/content/www/us/en/developer/articles/tool/pin-a-dynamic-binary-instrumentation-tool.html}. While instrumentation may result in notable runtime overhead during deployment, strategies like selective monitoring can serve to alleviate this issue. We discuss more in Section~\ref{sec:discussion}. The balance between deployment effort and the amount of valuable data attainable from API monitoring makes API-based detection a promising direction.

\textbf{Other behaviral features.} Moreover, besides API monitoring, Sendner et al.~\cite{database-ransomware} utilized query sequences to detect database ransomware. Chen et al.~\cite{chen2017uncovering} studied the foreground behaviors of programs. Cohen et al.~\cite{cohen2018} analyzed the program memory dump. Roy et al.~\cite{roy2021deepran} used system logs to identify malicious ransomware behaviors. Some research works also define customized behavioral rules to detect ransomware~\cite{zuhair2019rands,zuhair2020multi,verma2018analysing}.

\subsection{Defenses at kernel mode level} Kernel-level solutions monitor program behaviors at a lower level and could be more difficult for malware to evade, as obfuscation usually happens at a higher-level. However, system version dependency and modifications of the kernel drivers make the deployment of such solutions to new devices very challenging. Many detection researches have been done analyzing the I/O Request Packets (IRP) to detect malicious file access patterns~\cite{shaukat2018ransomwall,cryptolock,unveil,ransomspector,ShieldFS,Mehnaz2018rwguard,peeler,ramesh2020automated, ayub2020request, hirano2019machine,kim2018white,palisse2017data,moussaileb2018ransomware,ayub2023rwarmor,elkhail2023seamlessly}. Encrypted data usually has a higher entropy. Therefore, the entropy of the reading and writing buffer is also a useful feature~\cite{cryptolock}. McIntosh et al.~\cite{access} proposed Staged Event-Driven Access Control to limit processes’ access during execution. Examining hardware counters to identify intensive computation for encryption behaviors~\cite{ferrante2018extinguishing,posse,hardware-profile,hardware23,ketzaki2020behaviour,ramesh2020automated,palisse2017data,alam2020rapper,celdran2023behavioral,mcintosh2023applying}.

\subsection{Defenses at hardware level} Solid State Drives (SSDs) have the property of out-of-place writing, which could help recover ransomware encrypted data. Solutions~\cite{ssd-insider,flashguard,RSSD,MimosaFTL} built into SSD firmware have been developed to recover data when a ransomware attack is detected. While adding an extra layer of mitigating the risk of data loss, such built-in solutions require acquiring certain hardware.

\section{Evaluation of Commercial Defense}
\label{sec:commercial}

Commercial tools are in the format of download and run executables. Minimum effort is required to deploy such end-point protection. In this section, we report the findings from our experimental evaluation of decryptors, antivirus software, and malware scanners and answer \textbf{RQ2}. Our observations show that there exist security gaps.

\subsection{Experimental Setup}

We describe the experimental setup for evaluating various types of commercial defense in this section.

\noindent \textbf{Decryptor.} For the ransomware samples we successfully run, we found decryptors for 6 of them on the NO MORE RANSOM website\footnote{https://www.nomoreransom.org/en/decryption-tools.html}. The 6 decryptors are specifically designed for 6 distinct families (Table \ref{tab:decryptor}). Different decrypting tools require different information, such as uploading the ransom note or providing file pairs (i.e. both unencrypted and encrypted versions) for key cracking.

\noindent \textbf{Commercial Antivirus Software.}
We tested 8 different commercial antivirus software that provides protection against ransomware, namely Antivirus A (anonymized), Antivirus B (anonymized), Bitdefender, Malwarebytes, Kaspersky, McAfee, Norton, and 360 Total Security (Table~\ref{tab:av-version} in the appendix). We anonymized antivirus tools A and B following their user terms and conditions. All 8 antivirus software are generic for all types of malware with ransomware detection feature. We conducted three series of experiments. We first evaluated them with 20 real ransomware samples from 20 families active from 2015 to 2020 (marked in Table~\ref{tab:sha256} in the appendix). Each sample was compressed with the password ``infected’’ at the beginning. We uncompressed the sample and if the antivirus had no reaction, we then executed it. The machine was disconnected from the internet during all real ransomware execution to prevent spreading. Because 360 Total Security performed poorly in the offline setting, we added an online setting for it, in which we only uncompressed the samples without executing them. Second, we used two publicly available ransomware simulators to test more attack scenarios. RanSim\footnote{https://www.knowbe4.com/ransomware-simulator} simulates 23 ransomware attack scenarios. Quickbuck\footnote{https://github.com/NextronSystems/ransomware-simulator} simulates several typical ransomware behaviors. Lastly, we also used our own script to further test the behavioral detection because most antivirus tools block public simulators as malware, giving them no chance to execute. Our script simulates behaviors such as iterating and encrypting files, appending random or known ransomware file extensions, deleting volume shadow copies, and dropping ransom notes. Those ransomware-like behaviors can be run separately or together. The script is in Python 3.8. To mimic real ransomware, we statically link the crypto libraries used (PyAesCrypt\footnote{https://github.com/marcobellaccini/pyAesCrypt} and Cryptography\footnote{https://github.com/pyca/cryptography}) so that no external dependencies are needed. All tests were run on a Windows 7 VM with 2 CPUs and 4096 MB memory. The machine was reversed to the initial image after each test.

\noindent \textbf{Commercial Malware Scanner.} We test over 70 commercial malware scanners available on VirusTotal~\footnote{https://www.virustotal.com/gui/} using the 54 samples (Table~\ref{tab:sha256} in the appendix). All scanners are general purpose malware scanners in a black-box manner (i.e., the algorithm used is unknown). For each sample we scanned, the number of available scanners slightly varies (Table \ref{tab:scanner} in the appendix). We tested the scanners in three settings. First, we used them to scan the plain executable file (.exe). Second, we scanned the compressed file with a simple password “infected”. This password is used as a convention for sharing malware samples. Lastly, we made a compressed file with a random, complex password (i.e., DDfA3WFxcPMUsrsA) to test the scanners’ ability to detect obfuscated samples. We chose these settings because password obfuscation is a very low-cost way for the attacker to distribute malware samples.

\subsection{Crypto Decryptors}
\begin{table*}[]
\centering
\caption{List of decryptors we test. SHA-256 column shows the first 4 digits of the ransomware sample that generates the encrypted files. File extension is the extension appended by ransomware after encryption. Notes explain the reason if we are not able to get the decryption step. The Trend Micro tool is designed for decrypting files infected by multiple families. Different families can be selected before decryption. }

\scriptsize\begin{tabular}{
>{\columncolor[HTML]{EFEFEF}}l l
>{\columncolor[HTML]{EFEFEF}}l l
>{\columncolor[HTML]{EFEFEF}}l l
>{\columncolor[HTML]{EFEFEF}}l }
\hline
\cellcolor[HTML]{C0C0C0}\textbf{\begin{tabular}[c]{@{}l@{}}Ransomware\\ family\end{tabular}} & \cellcolor[HTML]{C0C0C0}\textbf{SHA-256} & \cellcolor[HTML]{C0C0C0}\textbf{\begin{tabular}[c]{@{}l@{}}File\\ extension\end{tabular}} & \cellcolor[HTML]{C0C0C0}\textbf{\begin{tabular}[c]{@{}l@{}}Decryptor \\ provider\end{tabular}} & \cellcolor[HTML]{C0C0C0}\textbf{Requirements} & \cellcolor[HTML]{C0C0C0}\textbf{Success} & \cellcolor[HTML]{C0C0C0}\textbf{Notes}         \\ \hline
Alcatraz                                                                                     & 9185                                     & .alcatraz                                                                                 & Avast                                                                                          & A pair of files                               & Yes                                      &                                                \\ \hline
Babuk                                                                                        & eb18                                     & .doydo                                                                                    & Avast                                                                                          &                                               & No                                       &                                                \\ \hline
Jigsaw                                                                                       & 9074                                     & .v316                                                                                     & Trend Micro                                                                                    &                                               & No                                       &                                                \\ \hline
Ragnarok                                                                                     & db8b                                     & .ragnarok\_cry                                                                            & Emsisoft                                                                                       & Ransom note                                   & No                                       & Ransom note file is not supported by decryptor \\ \hline
Sodinokibi                                                                                   & fd16                                     & .031j2adrq8                                                                               & BitDefender                                                                                    & Internet access                               & No                                       &                                                \\ \hline
Xorist                                                                                       & fb54                                     & .locks                                                                                    & Trend Micro                                                                                    & A pair of files                               & No                                       & Not able to proceed after adding file pair     \\ \hline
\end{tabular}

\label{tab:decryptor}
\end{table*}

Decryption is a ransomware-specific recovery strategy that aims to recover files without payment. Generally, it works by inspecting the encryption algorithm and inferring the key. Strategies include finding implementation flaws of encryption functions, brute forcing the key in a certain scope, and monitoring the key generation. 

 Among the 6 decryptors we test, only the decryptor for Alcatraz successfully recovers the files. The decryptor requires a pair of original and encrypted files for cracking the password and finds the password in seconds. To minimize the effect of randomness and confirm the effectiveness, we run the attack three times. For each attack, the decryptor crack the password in 12119, 3724, and 4879 tries, respectively. Alcatraz first appeared in 2016 and uses AES-256 with Base64 encoding for encryption. It is computationally infeasible to brute force the AES-256 key as the key space is $2^{256}$~\cite{AES-brute-force}. 
  Therefore, we suspect that there might be a design flaw in Alcatraz’s encryption function that the decryptor uses as a shortcut to search for the correct key. This low success rate of recovery further necessitates early detection.

\subsection{Commercial Antivirus Software} 
\label{sec:antivirus}

\begin{table*}[]
\centering
\caption{Evaluation results of 8 commercial antivirus software. \# detection before execution: number of samples that are blocked before the malware executes. \# detection during execution: number of samples that execute but get terminated during execution. \# completely rolled back: among the attacks that started, the number of those being completely rolled back to the state with no trace of infected files. \# failed: number of attacks the tool failed to generate alerts or make a reaction. Empty means zero. Ransomware simulators are software that mimics ransomware behaviors for security evaluation purposes. Blocked and detected are reactions taken by antivirus software. -- represents no reaction.}
\vspace{-2mm}

\scriptsize \begin{tabular}{|
>{\columncolor[HTML]{EFEFEF}}l 
>{\columncolor[HTML]{EFEFEF}}l |c
>{\columncolor[HTML]{EFEFEF}}c c
>{\columncolor[HTML]{EFEFEF}}c |c
>{\columncolor[HTML]{EFEFEF}}c |c
>{\columncolor[HTML]{EFEFEF}}c c
>{\columncolor[HTML]{EFEFEF}}c |}
\hline
\multicolumn{2}{|l|}{\cellcolor[HTML]{C0C0C0}}                                                              & \multicolumn{4}{c|}{\cellcolor[HTML]{C0C0C0}\textbf{Real ransomware samples}}                                                                                                                                                                                                                                                                                                                                        & \multicolumn{2}{c|}{\cellcolor[HTML]{C0C0C0}\textbf{Ransomware simulators}}                                                                                           & \multicolumn{4}{c|}{\cellcolor[HTML]{C0C0C0}\textbf{Our script}}                                                                                                                                                                                                                                                                                                                                                                            \\ \hline
\multicolumn{2}{|l|}{\cellcolor[HTML]{C0C0C0}\textbf{\begin{tabular}[c]{@{}l@{}}Antivirus \\ software\end{tabular}}} & \multicolumn{1}{c|}{\cellcolor[HTML]{C0C0C0}\textbf{\begin{tabular}[c]{@{}c@{}}\# detection\\  before\\  execution\end{tabular}}} & \multicolumn{1}{c|}{\cellcolor[HTML]{C0C0C0}\textbf{\begin{tabular}[c]{@{}c@{}}\# detection\\  during\\  execution\end{tabular}}} & \multicolumn{1}{c|}{\cellcolor[HTML]{C0C0C0}\textbf{\begin{tabular}[c]{@{}c@{}}\# completely\\rolled\\ back\end{tabular}}} & \cellcolor[HTML]{C0C0C0}\textbf{\# failed} & \multicolumn{1}{c|}{\cellcolor[HTML]{C0C0C0}\textbf{RanSim}}                                  & \cellcolor[HTML]{C0C0C0}\textbf{\begin{tabular}[c]{@{}c@{}}Quick-\\Buck\end{tabular}}           & \multicolumn{1}{c|}{\cellcolor[HTML]{C0C0C0}\textbf{\begin{tabular}[c]{@{}c@{}}File \\ traverse\end{tabular}}} & \multicolumn{1}{c|}{\cellcolor[HTML]{C0C0C0}\textbf{\begin{tabular}[c]{@{}c@{}}File \\encryption\end{tabular}}} & \multicolumn{1}{c|}{\cellcolor[HTML]{C0C0C0}\textbf{\begin{tabular}[c]{@{}c@{}}Ransom \\ note\\ dropping\end{tabular}}} & \cellcolor[HTML]{C0C0C0}\textbf{\begin{tabular}[c]{@{}c@{}}Volume \\ shadow\\ deletion\end{tabular}} \\ \hline
\multicolumn{2}{|l|}{\cellcolor[HTML]{EFEFEF}Antivirus A}                                                   & \multicolumn{1}{c|}{20}                                                                                                  & \multicolumn{1}{c|}{\cellcolor[HTML]{EFEFEF}}                                                                            & \multicolumn{1}{c|}{}                                                                                             &                                   & \multicolumn{1}{c|}{Blocked}                                                         & Blocked                                                               & \multicolumn{1}{c|}{--}                                                                               & \multicolumn{1}{c|}{\cellcolor[HTML]{EFEFEF}--}                                                             & \multicolumn{1}{c|}{--}                                                                                        & --                                                                                          \\ \hline
\multicolumn{2}{|l|}{\cellcolor[HTML]{EFEFEF}Antivirus B}                                                   & \multicolumn{1}{c|}{20}                                                                                                  & \multicolumn{1}{c|}{\cellcolor[HTML]{EFEFEF}}                                                                            & \multicolumn{1}{c|}{}                                                                                             &                                   & \multicolumn{1}{c|}{Blocked}                                                         & Blocked                                                               & \multicolumn{1}{c|}{--}                                                                               & \multicolumn{1}{c|}{\cellcolor[HTML]{EFEFEF}--}                                                             & \multicolumn{1}{c|}{--}                                                                                        & --                                                                                          \\ \hline
\multicolumn{2}{|l|}{\cellcolor[HTML]{EFEFEF}Bitdefender}                                                   & \multicolumn{1}{c|}{19}                                                                                                  & \multicolumn{1}{c|}{\cellcolor[HTML]{EFEFEF}1}                                                                           & \multicolumn{1}{c|}{1}                                                                                            &                                   & \multicolumn{1}{c|}{\begin{tabular}[c]{@{}c@{}}Passed all \\ scenarios\end{tabular}} & Blocked                                                               & \multicolumn{1}{c|}{--}                                                                               & \multicolumn{1}{c|}{\cellcolor[HTML]{EFEFEF}--}                                                             & \multicolumn{1}{c|}{--}                                                                                        & Detected                                                                                    \\ \hline
\multicolumn{2}{|l|}{\cellcolor[HTML]{EFEFEF}Malwarebytes}                                                  & \multicolumn{1}{c|}{17}                                                                                                  & \multicolumn{1}{c|}{\cellcolor[HTML]{EFEFEF}3}                                                                           & \multicolumn{1}{c|}{3}                                                                                            &                                   & \multicolumn{1}{c|}{Blocked}                                                         & Blocked                                                               & \multicolumn{1}{c|}{--}                                                                               & \multicolumn{1}{c|}{\cellcolor[HTML]{EFEFEF}--}                                                             & \multicolumn{1}{c|}{--}                                                                                        & --                                                                                          \\ \hline
\multicolumn{2}{|l|}{\cellcolor[HTML]{EFEFEF}Kaspersky}                                                     & \multicolumn{1}{c|}{12}                                                                                                  & \multicolumn{1}{c|}{\cellcolor[HTML]{EFEFEF}7}                                                                           & \multicolumn{1}{c|}{5}                                                                                            & 1                                 & \multicolumn{1}{c|}{Blocked}                                                         & Blocked                                                               & \multicolumn{1}{c|}{--}                                                                               & \multicolumn{1}{c|}{\cellcolor[HTML]{EFEFEF}--}                                                             & \multicolumn{1}{c|}{--}                                                                                        & --                                                                                          \\ \hline
\multicolumn{2}{|l|}{\cellcolor[HTML]{EFEFEF}McAfee}                                                        & \multicolumn{1}{c|}{16}                                                                                                  & \multicolumn{1}{c|}{\cellcolor[HTML]{EFEFEF}}                                                                            & \multicolumn{1}{c|}{}                                                                                             & 4                                 & \multicolumn{1}{c|}{Blocked}                                                         & Blocked                                                               & \multicolumn{1}{c|}{--}                                                                               & \multicolumn{1}{c|}{\cellcolor[HTML]{EFEFEF}--}                                                             & \multicolumn{1}{c|}{--}                                                                                        & --                                                                                          \\ \hline
\multicolumn{2}{|l|}{\cellcolor[HTML]{EFEFEF}Norton}                                                        & \multicolumn{1}{c|}{18}                                                                                                  & \multicolumn{1}{c|}{\cellcolor[HTML]{EFEFEF}}                                                                            & \multicolumn{1}{c|}{}                                                                                             & 2                                 & \multicolumn{1}{c|}{Blocked}                                                         & \begin{tabular}[c]{@{}c@{}}Detected\\ marco\\ simulation\end{tabular} & \multicolumn{1}{c|}{--}                                                                               & \multicolumn{1}{c|}{\cellcolor[HTML]{EFEFEF}--}                                                             & \multicolumn{1}{c|}{--}                                                                                        & --                                                                                          \\ \hline
\multicolumn{1}{|l|}{\cellcolor[HTML]{EFEFEF}}                                     & offline                 & \multicolumn{1}{c|}{3}                                                                                                   & \multicolumn{1}{c|}{\cellcolor[HTML]{EFEFEF}}                                                                            & \multicolumn{1}{c|}{}                                                                                             & 17                                & \multicolumn{1}{c|}{\begin{tabular}[c]{@{}c@{}}Failed all\\ scenarios\end{tabular}}  & --                                                                    & \multicolumn{1}{c|}{--}                                                                               & \multicolumn{1}{c|}{\cellcolor[HTML]{EFEFEF}--}                                                             & \multicolumn{1}{c|}{--}                                                                                        & --                                                                                          \\ \hhline{|>{\arrayrulecolor{tgray}}->{\arrayrulecolor{tblack}}|-|-|-|-|-|-|-|-|-|-|-|} 
\multicolumn{1}{|l|}{\multirow{-2}{*}{\cellcolor[HTML]{EFEFEF}\begin{tabular}[c]{@{}l@{}}360 Total\\Security\end{tabular}}}                & online                & \multicolumn{1}{c|}{20}                                                                                                  & \multicolumn{1}{c|}{\cellcolor[HTML]{EFEFEF}}                                                                            & \multicolumn{1}{c|}{}                                                                                             &                                   & \multicolumn{1}{c|}{Blocked}                                                         & Blocked                                                               & \multicolumn{1}{c|}{--}                                                                               & \multicolumn{1}{c|}{\cellcolor[HTML]{EFEFEF}--}                                                             & \multicolumn{1}{c|}{--}                                                                                        & Detected                                                                                    \\ \hline
\end{tabular}
\vspace{-1mm}

\label{tab:antivirus}
\end{table*}

We test the antivirus software with real ransomware samples and simulations of ransomware behaviors. First, we report their effectiveness and analyze the features (static or behavioral) used for detection based on their reactions. The results are summarized in Table~\ref{tab:antivirus}.

\noindent\textbf{Detection before execution.} All 8 commercial tools detect and block the majority of threats before they start execution and make any modifications to the system. They do so by moving the executable file to quarantine or prohibiting it from execution. It is likely that they run a signature matching similar to the malware scanners. Antivirus tools A, B, and 360 (online) detect all malware executables immediately upon decompression, suggesting that they have up-to-date malware databases. Bitdefender, Malwarebytes, Kaspersky, McAfee, and Norton block 19, 17, 12, 16, and 18 samples, respectively, before execution (Table~\ref{tab:antivirus}).

\noindent\textbf{Detection during execution.} We execute a sample to test behavioral detection if it is not blocked by scanning. Bitdefender and Malwarebytes catch all threats they missed before execution and completely roll back the malicious behaviors, with no signs of infection left in the system. Kaspersky successfully detects all 7 attacks missed earlier and revokes 5 of them. In the other two cases, a few infected files are left in the system but all original data is accessible. This shows that antivirus software has real-time behavioral detection to catch ongoing threats.

While all other attacks are terminated with no additional information, Bitdefender gives two warnings before the remediation of the Cerber ransomware attack starts. One of the firewall rules is triggered first, followed by the discovery of an infected file, which is the ransom note in this case.

\noindent\textbf{Failed cases.} Kaspersky, McAfee, and Norton miss 1, 4, and 2 attacks, respectively. One special case is the Hive sample, which Kaspersky and McAfee miss. It runs in the background but no attack behaviors are observed in 10 minutes. In all other cases, a large number of files are encrypted, but no warning or action is triggered. In an offline setting, 360 only catches 3 out of 20 threats, implying cloud computation for detection.

\noindent\textbf{Detection on ransomware-like behaviors.} We further investigate the ability of antivirus software to detect various ransomware-like behaviors by running simulations. Most of them block public simulators as real malware. A few exceptions include Bitdefender, which passes all RanSim scenarios, Norton, which detects the macro simulation, and 360 (offline), which has no reaction at all. Emulating various ransomware behaviors, we find that antivirus software only reacts to the deletion of volume shadow copy, which is a backup copy of computer volume. Intensive encryption and file access do not invoke any warning.

\subsection{Commercial Malware Scanners}

We also evaluate the efficacy of general-purpose malware scanners. We find that the scanners’ detection capability could be significantly weakened by simple obfuscation, i.e., compression with passwords. Therefore, dynamic behavioral detection is necessary to complement the protection.

Malware scanners provide pre-execution scanning, an early layer of protection against infection.
Our experiments show that the majority of the scanners can effectively identify the threat in plain executable files. On average, 56 out of over 70 scanners raise an alert. However, the detection capability significantly reduces on password-protected samples. When compressed with the simple password ``infected'', WannaCry triggers three alarms, which is the most among the 54 samples. In the complex password setting, 37 out of 54 samples completely evade detection, i.e., marked as safe by all scanners. This result suggests that the scanners are likely signature-based and have very limited detection capability against unknown and obfuscated samples. Full results are shown in Table \ref{tab:scanner} in the appendix. Obfuscating a malicious sample with encryption requires little effort but significantly increases the chance of escaping. However, forcing breaking advanced encryption algorithms is impossible and should not be the goal of a malware scanner. One possible approach to strengthen security is to warn the user that the file is encrypted and suggest further scanning.

\section{Characterization of Ransomware API Usage}
\label{sec:character}

To develop new insights into building deployable behavioral detection against ransomware, we experimentally analyze the API utilization of real-world ransomware samples. In this section, we present our findings on the ransomware API usage pattern with detailed analyses. With a comparison to benign software behaviors, we find that API usage and frequency are informative in terms of revealing malicious behaviors. We present a possible classification approach based on these findings in Section~\ref{sec:classification}. As API monitoring happens at the program level, the deployability of such approaches is also high. We discuss more about its deployability in Section~\ref{sec:discussion}.

\subsection{Encryption and File Access Behaviors}
\label{sec:file-access}

\begin{figure*}

     \centering
     \begin{subfigure}[b]{0.32\textwidth}
         \centering
         \includegraphics[width=\textwidth]{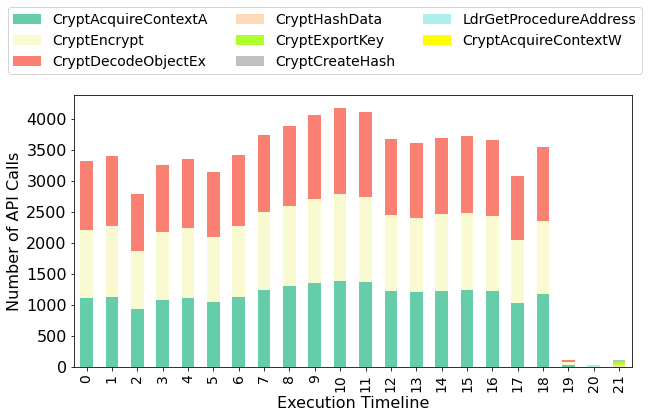}
         \vspace*{-6mm}
         \caption{ Avoslocker's crypto API frequency} 
         \label{fig:avoslocker-crypto-calls}
     \end{subfigure}
      \begin{subfigure}[b]{0.32\textwidth}
         \centering
         \includegraphics[width=\textwidth]{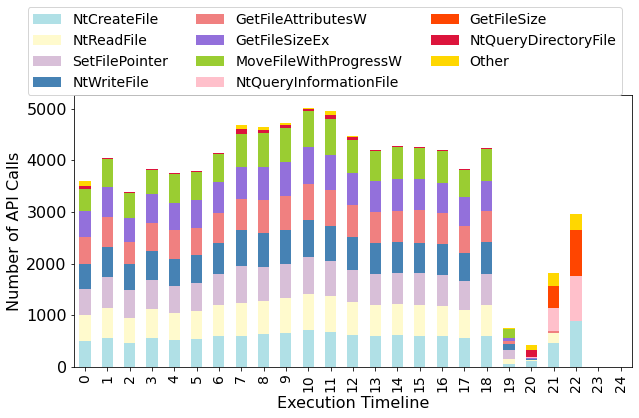}
         \vspace*{-6mm}
         \caption{Avoslocker's file API frequency} 
         \label{fig:avoslocker-file-calls}
     \end{subfigure}
     \begin{subfigure}[b]{0.32\textwidth}
         \centering
         \includegraphics[width=\textwidth]{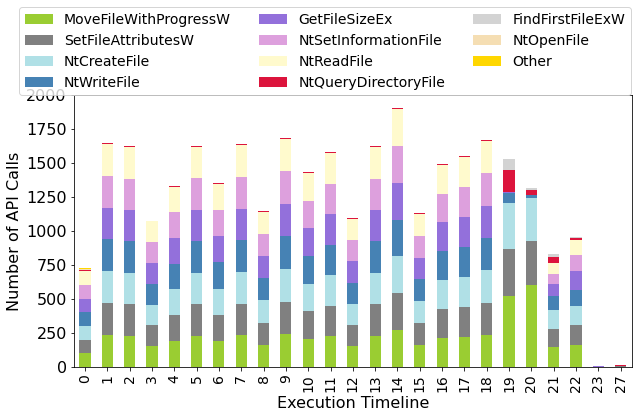}
         \vspace*{-6mm}
         \caption{SunCrypt's file API frequency}
         \label{fig:suncrypt-file-calls}
     \end{subfigure}
     
     \begin{subfigure}[b]{0.32\textwidth}
         \includegraphics[width=\textwidth]{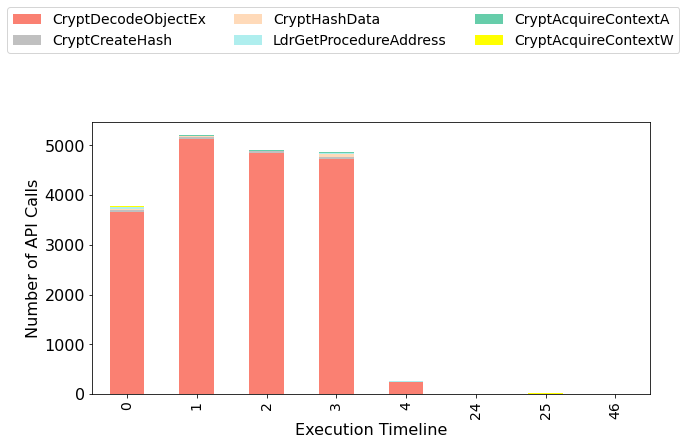}
         \vspace*{-6mm}
         \caption{Dropbox's crypto API frequency} 
         \label{fig:dropbox-crypto-calls}
     \end{subfigure}
      \begin{subfigure}[b]{0.32\textwidth}
         \includegraphics[width=\textwidth]{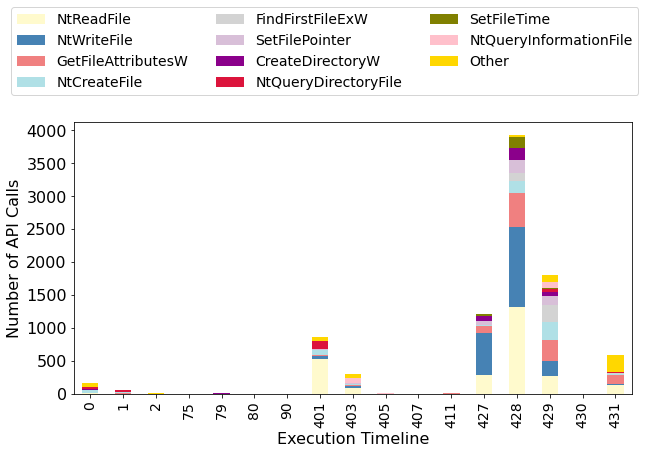}
         \vspace*{-6mm}
         \caption{Notepad++'s file API frequency} 
         \label{fig:notepadpp-file-calls}
     \end{subfigure}
     \begin{subfigure}[b]{0.32\textwidth}
         \includegraphics[width=\textwidth]{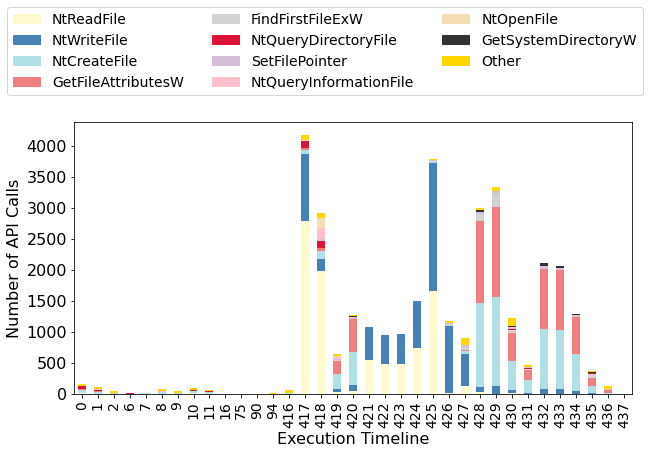}
         \vspace*{-6mm}
         \caption{TeamViewer's file API frequency} 
         \label{fig:teamviewer-file-calls}
     \end{subfigure}

\caption{API call statistics of executing ransomware samples (top) and benign software (bottom).  The x-axis is the timestamp during execution and the y-axis shows the number of calls. If no call is made during a second, then it is not shown in the figures. Ransomware uses intensive crypto and file API calls during execution with a distinctive pattern. The same colors represent the same APIs across subfigures.}
\label{fig:crypto-and-file}
\end{figure*}

To take a close look at the file encryption behaviors, we analyze cryptographic and file-related API usage and call frequency of ransomware samples. 

\subsubsection{Experimental Setup 1}

\noindent\textbf{Ransomware samples.} We collect 262 ransomware samples from MalwareBarzaar\footnote{https://bazaar.abuse.ch/browse/} and VirusShare\footnote{https://virusshare.com/}. Among these samples, we find 54 of the samples (from 35 distinct families) provide more meaningful traces for manual analysis. The SHA256 hashes of the 54 samples are shown in Table~\ref{tab:sha256} in supplemental materials, which can help find the exact samples. 

\noindent\textbf{Analysis environment.} We set up an isolated environment for safely executing ransomware samples using Cuckoo Sandbox (v2.0.7)\footnote{https://cuckoosandbox.org/}, with VirtualBox (v6.1)\footnote{https://www.virtualbox.org/} as the hypervisor and Windows 7 as the guest system. We use 4 CPUs and 4096 MB memory for the VM. We install several applications, including Chrome, Adobe PDF Reader, NotePad++, and LibreOffice, to make the environment more realistic. We also put random files under several directories, such as disk C, Documents, and Downloads for ransomware to encrypt. Each execution starts with a clean system image. To avoid VM escape, we use a different host system (Ubuntu 21.04). We also set up a fake internet service using INetSim\footnote{https://www.inetsim.org/}. Lastly, to minimize the risk of spreading, we disconnect the machine from the internet.

\noindent\textbf{Report and API analyses.} For each successful execution, the Cuckoo sandbox generates a report in JSON format, from which we extract data for analysis. We conduct API call frequency analyses based on the reported API calls. 

\noindent\textbf{Comparison with benign software samples.} We download the top popular Windows applications from a range of different categories from software.informer\footnote{https://software.informer.com/System-Tools/} or the software’s official website. We use 38 samples from 12 categories for manual analysis (Table~\ref{tab:benign} in supplemental materials). We run the benign samples in the Cuckoo sandbox the same way we run ransomware samples and collect the execution reports for analysis. Among those samples, we use the ones with intensive file access behaviors in the evaluation of our classification (Section~\ref{sec:evaluation}).


\subsubsection{Characterization Findings on Encryption Behaviors} 

Our characterization study identifies interesting encryption behaviors.

\noindent\textbf{CryptoAPI frequency.} Some ransomware samples make intensive CryptoAPI calls, in the order of thousands, during execution. We show AvosLocker as an example in Figure \ref{fig:avoslocker-crypto-calls}. It uses the API sequence \texttt{CryptAcquireContextA}, \texttt{CryptDecodeObjectEx},  \texttt{CryptEncrypt} for encryption. The number of calls to each of the three APIs is similar, adding up to around 4,000 calls per second.

\noindent\textbf{Ransomware file access.} Similar to crypto-related calls, file access frequency is also in the order of thousands per second. The peaks of calls are around 5000 and 1900 calls per second for AvosLocker and SunCrypt (Figures \ref{fig:avoslocker-file-calls} and \ref{fig:suncrypt-file-calls}), respectively. By examining the directory paths touched through calls to \texttt{NtCreateFile}, \texttt{NtWriteFile}, and \texttt{NtOpenFile}, we notice that these samples start traversing the directory at C:// and then go into child directories. MountLocker even searches disks through a:// to z://. AvosLocker accesses 780 unique directories during the execution, SunCrypt accesses 180, and MountLocker accesses 2,236 in our testbed. The three families use a similar combination of 6 to 7 file-related APIs repetitively during the encryption process. The combination includes APIs to create files, read files, write files, query file size or information, and set file pointers. The number of each API being called every second is somewhat evenly distributed, with no absolute dominance.

\noindent\textbf{Comparison with benign samples.} To contrast with ransomware behavior, we also analyze the execution of 38 benign software samples. The top API categories of the majority of benign samples include system, registry, and miscellaneous, varying from ransomware behavior. However, there are a few exceptions. Dropbox (Figure \ref{fig:dropbox-crypto-calls}) makes over 3500 crypto-related calls per second in the first four seconds of execution, with a peak of over 5000 calls per second. The difference is in the composition of the API calls made. While AvosLocker uses a combination of three CryptoAPIs with relatively even distribution, the vast majority of calls made by Dropbox are to a single CryptoAPI, i.e., \texttt{CryptDecodeObjectEx}. 

Moreover, some benign samples also make a notable amount of file-related calls, as shown in Figures \ref{fig:notepadpp-file-calls} (Notepad++) and \ref{fig:teamviewer-file-calls} (TeamViewer). By examining the composition and call patterns, one can easily tell the benign usage pattern is rather random, varying from program to program. 

In summary, ransomware shows a distinguishable repetitive file-access API pattern throughout the execution. We further analyze the distinction between ransomware behavior and intensive benign file access in-depth. The results show that this unique pattern can be quantified to help classify malicious from benign execution. Our classification method is presented in Section~\ref{sec:classification}.

\subsection{API Occurrence Contrast Analysis}
\label{sec:api_stats}
In another characterization study, we conduct a contrast analysis of 288 APIs, comparing their occurrences and usage frequencies in ransomware and benign programs. 

\subsubsection{Experimental Setup 2} We collect 348 ransomware samples from MalwareBarzaar. The samples are from 37 families. For benign samples, we collect 330 of them from software.informer. We use the same sandbox setting in this experiment and the JSON report for API analysis. The virtual machine has 4 CPUs and 8192 MB of memory. We refer to this setting as setup 2.

\subsubsection{Characterization Findings on Contrast Analysis}
We observe some APIs are used more commonly by ransomware than benign software. For instance, \texttt{WriteConsoleW} is used by 50\% of ransomware samples we measure, while only occurs in the execution traces of 5\% of benign samples. We show a list of such APIs in Table~\ref{tab:api_stats_occur}. Moreover, even if some APIs occur in a similar number of ransomware and benign software, the call frequency could vary substantially. For example, \texttt{NtWriteFile} is used by a comparable number of ransomware (333 samples) and benign programs (328 samples). However, ransomware samples make, on average, 40554 calls during execution, which is around 8 times compared to benign (5031 on average). More examples of such APIs are shown in Table~\ref{tab:apt_stats_freq}. Comparably, there are also a set of APIs that are more commonly observed during benign executions. A list is presented in Table~\ref{tab:api_stats_benign}. 

The invocation patterns of specific APIs vary between ransomware and benign programs. This API usage contrast is useful for building new detection. Later, we show how they aid detection in Section~\ref{sec:detect-algo}.

\begin{table}[]
\centering
\caption{List of APIs that significantly more prevalent in ransomware when compared to benign. The percentage is calculated by (number of ransomware that calls this API / 348) for ransomware and (number of benign programs call this API / 330) for benign programs.}
\vspace{-2mm}
\small \begin{tabular}{
>{\columncolor[HTML]{EFEFEF}}c c
>{\columncolor[HTML]{EFEFEF}}c c}
\hline
\cellcolor[HTML]{C0C0C0}\textbf{} & \cellcolor[HTML]{C0C0C0}\textbf{API} & \cellcolor[HTML]{C0C0C0}\textbf{\% RW} & \cellcolor[HTML]{C0C0C0}\textbf{\% Benign} \\ \hline
1                                 & NtOpenDirectoryObject                & 65\%                                   & 32\%                                       \\ \hline
2                                 & CoInitializeSecurity                 & 59\%                                   & 26\%                                       \\ \hline
3                                 & MoveFileWithProgressW                & 57\%                                   & 29\%                                       \\ \hline
4                                 & WriteConsoleW                        & 50\%                                   & 5\%                                        \\ \hline
5                                 & Process32NextW                       & 49\%                                   & 11\%                                       \\ \hline
6                                 & CreateToolhelp32Snapshot             & 49\%                                   & 12\%                                       \\ \hline
7                                 & Process32FirstW                      & 48\%                                   & 11\%                                       \\ \hline
8                                 & CryptEncrypt                         & 18\%                                   & 0\%                                        \\ \hline
9                                 & CryptExportKey                       & 15\%                                   & 3\%                                        \\ \hline
10                                & CryptGenKey                          & 10\%                                   & 0\%                                        \\ \hline
\end{tabular}

\label{tab:api_stats_occur}
\vspace{-1mm}
\end{table}

\begin{table}[]
\centering
\caption{List of APIs that are comparably prevalent in ransomware and benign but have a much higher frequency in ransomware execution. RW stands for ransomware. RW freq mean is the average call frequency based on all ransomware samples that use this API. \# RW is the number of ransomware samples that used this API and \# benign is the number of benign programs that used this API. The call frequency is collected during a 300-second execution period for each sample.}
\vspace{-1mm}

\small \begin{tabular}{
>{\columncolor[HTML]{EFEFEF}}c c
>{\columncolor[HTML]{EFEFEF}}c c
>{\columncolor[HTML]{EFEFEF}}c }
\hline
\multicolumn{1}{l}{\cellcolor[HTML]{C0C0C0}} & \cellcolor[HTML]{C0C0C0}\textbf{API} & \cellcolor[HTML]{C0C0C0}\textbf{\begin{tabular}[c]{@{}c@{}}RW freq\\ mean\end{tabular}} & \cellcolor[HTML]{C0C0C0}\textbf{\begin{tabular}[c]{@{}c@{}}Benign \\ freq\\ mean\end{tabular}} & \cellcolor[HTML]{C0C0C0}\textbf{\begin{tabular}[c]{@{}c@{}}\# RW / \\ \# Benign\end{tabular}} \\ \hline
1                                            & CryptCreateHash                      & 159568.5                                                                                & 12.6                                                                                           & 48 / 63                                                                                       \\ \hline
2                                            & NtWriteFile                          & 40554.4                                                                                 & 5031.2                                                                                         & 333 / 328                                                                                     \\ \hline
3                                            & NtReadFile                           & 33883                                                                                   & 11532.1                                                                                        & 324 / 329                                                                                     \\ \hline
4                                            & SetFilePointerEx                     & 7472                                                                                    & 179.8                                                                                          & 240 / 247                                                                                     \\ \hline
5                                            & NtAllocateVirtualMemory              & 4839.2                                                                                  & 1602.5                                                                                         & 342 / 330                                                                                     \\ \hline
6                                            & NtFreeVirtualMemory                  & 4599.4                                                                                  & 579.9                                                                                          & 331 / 330                                                                                     \\ \hline
7                                            & NtCreateFile                         & 4564.3                                                                                  & 1292.2                                                                                         & 340 / 329                                                                                     \\ \hline
8                                            & FindFirstFileExW                     & 2666.1                                                                                  & 963.6                                                                                          & 281 / 308                                                                                     \\ \hline
9                                            & CryptAcquireContextA                 & 2170.6                                                                                  & 16.1                                                                                           & 50 / 60                                                                                       \\ \hline
10                                           & GetFileType                          & 1149.9                                                                                  & 226.7                                                                                          & 240 / 257                                                                                     \\ \hline
11                                           & SetFileAttributesW                   & 1105.6                                                                                  & 103.1                                                                                          & 152 / 115                                                                                     \\ \hline
12                                           & NtDeviceIoControlFile                & 682.1                                                                                   & 71.9                                                                                           & 243 / 193                                                                                     \\ \hline
13                                           & RegDeleteValueW                      & 245.3                                                                                   & 13.8                                                                                           & 137 / 150                                                                                     \\ \hline
14                                           & OpenSCManagerW                       & 90.8                                                                                    & 4.5                                                                                            & 191 / 180                                                                                     \\ \hline
15                                           & GetUserNameExW                       & 34.3                                                                                    & 6.5                                                                                            & 133 / 166                                                                                     \\ \hline
16                                           & OpenServiceW                         & 23.6                                                                                    & 7.6                                                                                            & 174 / 140                                                                                     \\ \hline
17                                           & NtOpenThread                         & 14.3                                                                                    & 4                                                                                              & 170 / 164                                                                                     \\ \hline
18                                           & CoCreateInstanceEx                   & 6                                                                                       & 2                                                                                              & 126 / 96                                                                                      \\ \hline
\end{tabular}

\label{tab:apt_stats_freq}
\end{table}
\begin{table}[]
\centering
\caption{List of APIs that significantly more prevalent in benign software when compared to ransomware. The percentage is calculated by (number of ransomware that calls this API / 348) for ransomware and (number of benign programs that call this API / 330) for benign programs.}
\vspace{-2mm}
\small \begin{tabular}{
>{\columncolor[HTML]{EFEFEF}}c c
>{\columncolor[HTML]{EFEFEF}}c c}
\hline
\cellcolor[HTML]{C0C0C0} & \cellcolor[HTML]{C0C0C0}\textbf{API}               & \cellcolor[HTML]{C0C0C0}\textbf{\% RW} & \cellcolor[HTML]{C0C0C0}\textbf{\% Benign} \\ \hline
1                        & RemoveDirectoryA                                   & 0\%                                    & 53\%                                       \\ \hline
2                        & \cellcolor[HTML]{FFFFFF}NtDeleteKey                & 1\%                                    & 94\%                                       \\ \hline
3                        & GetSystemDirectoryA                                & 3\%                                    & 57\%                                       \\ \hline
4                        & \cellcolor[HTML]{FFFFFF}FindResourceA              & 5\%                                    & 63\%                                       \\ \hline
5                        & NtReadVirtualMemory                                & 9\%                                    & 74\%                                       \\ \hline
6                        & \cellcolor[HTML]{FFFFFF}SendNotifyMessageW         & 10\%                                   & 96\%                                       \\ \hline
7                        & \cellcolor[HTML]{FFFFFF}FindWindowW                & 12\%                                   & 90\%                                       \\ \hline
8                        & \cellcolor[HTML]{FFFFFF}SetEndOfFile               & 12\%                                   & 81\%                                       \\ \hline
9                        & GetKeyState                                        & 13\%                                   & 97\%                                       \\ \hline
10                       & GetTempPathW                                       & 13\%                                   & 76\%                                       \\ \hline
11                       & \cellcolor[HTML]{FFFFFF}GetFileVersionInfoW        & 14\%                                   & 88\%                                       \\ \hline
12                       & \cellcolor[HTML]{FFFFFF}GetFileVersionInfoSizeW    & 15\%                                   & 88\%                                       \\ \hline
13                       & \cellcolor[HTML]{FFFFFF}RegCreateKeyExA            & 15\%                                   & 65\%                                       \\ \hline
14                       & \cellcolor[HTML]{FFFFFF}FindResourceW              & 16\%                                   & 98\%                                       \\ \hline
15                       & \cellcolor[HTML]{FFFFFF}GetFileInformationByHandle & 16\%                                   & 82\%                                       \\ \hline
16                       & \cellcolor[HTML]{FFFFFF}DrawTextExW                & 17\%                                   & 100\%                                      \\ \hline
17                       & \cellcolor[HTML]{FFFFFF}GetCursorPos               & 17\%                                   & 98\%                                       \\ \hline
18                       & \cellcolor[HTML]{FFFFFF}SearchPathW                & 17\%                                   & 98\%                                       \\ \hline
19                       & SetFileTime                                        & 18\%                                   & 97\%                                       \\ \hline
20                       & \cellcolor[HTML]{FFFFFF}GetVolumePathNameW         & 18\%                                   & 80\%                                       \\ \hline
21                       & \cellcolor[HTML]{FFFFFF}SizeofResource             & 20\%                                   & 99\%                                       \\ \hline
22                       & \cellcolor[HTML]{FFFFFF}NtCreateKey                & 20\%                                   & 97\%                                       \\ \hline
23                       & \cellcolor[HTML]{FFFFFF}FindResourceExW            & 21\%                                   & 100\%                                      \\ \hline
24                       & OleInitialize                                      & 23\%                                   & 87\%                                       \\ \hline
25                       & \cellcolor[HTML]{FFFFFF}GetForegroundWindow        & 24\%                                   & 100\%                                      \\ \hline
26                       & \cellcolor[HTML]{FFFFFF}EnumWindows                & 24\%                                   & 96\%                                       \\ \hline
\end{tabular}

\label{tab:api_stats_benign}
\end{table}

\subsection{Dynamic resolution of API and Windows modules} 
\begin{figure}
    \centering
    \includegraphics[width=0.7\textwidth]{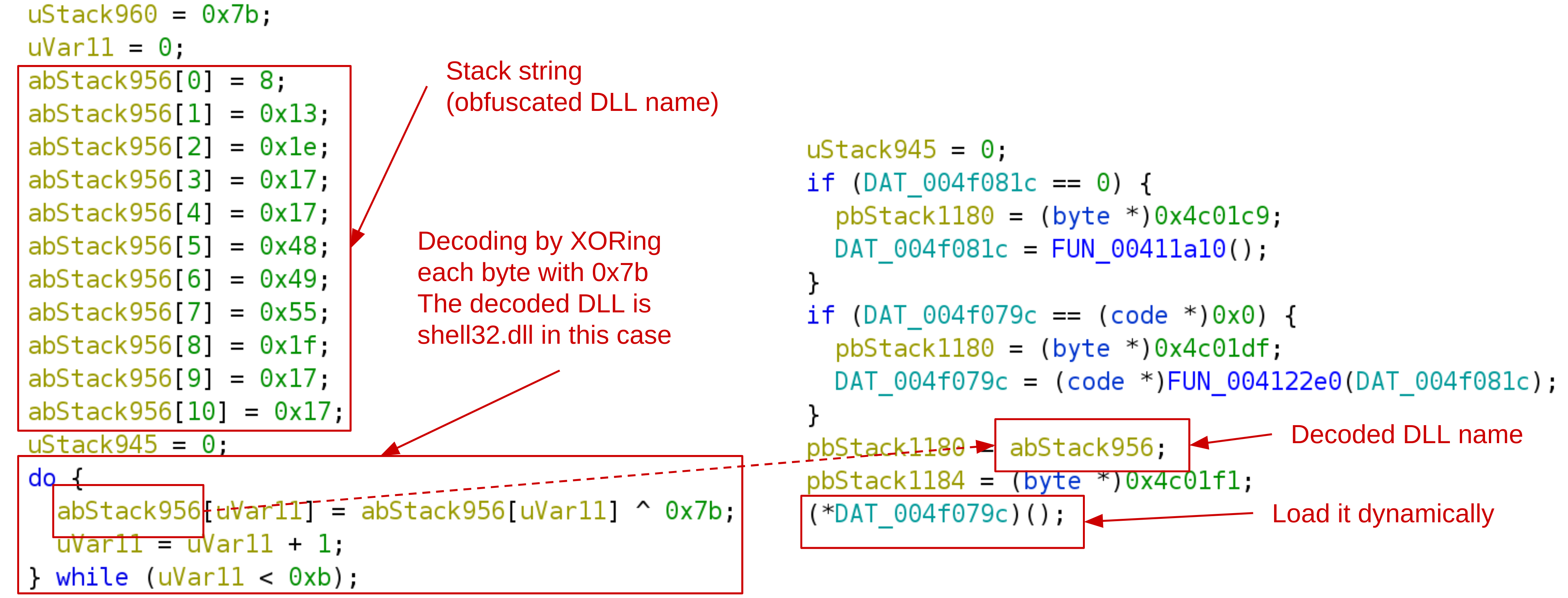}
    \caption{\small Example of LockBit code that dynamically resolves libraries.}
    \label{fig:lockbit-shell32}
\end{figure}

From reverse engineering, we found some samples obfuscate strings and Windows library names, including crypto-related modules, and resolve them at runtime. Dynamic resolution is to load a module or an API using its address during execution. In this way, the usage of a module remains unknown until being executed. For instance, LockBit obfuscates strings into random text and only statically imports five modules, preventing critical information from being extracted easily. Each string and library name is encoded by XORing or subtracting a unique value, and is decoded individually during execution. Figure \ref{fig:lockbit-shell32} shows how LockBit decodes and loads \texttt{shell32.dll}. 28 such strings were found in the entry function of the LockBit. 21 of them turn out to be Windows 32 libraries, which are a much longer list than the imports. Four of them are crypto-related, namely \texttt{advapi32.dll}, \texttt{bcrypt.dll}, \texttt{crypt32.dll}, and \texttt{cryptbase.dll}. 

To further inspect this feature, we analyzed the execution report and found that the samples use \texttt{LdrLoadDll} and \texttt{LdrGetProcedureAddress} to dynamically load DLLs and APIs. By looking at the loaded modules, we found 7 crypto-related modules, namely \texttt{advapi32}, \texttt{cryptsp}, \texttt{cryptbase}, \texttt{bcryptimitives}, \texttt{crypt32}, \texttt{rsaenh}, and \texttt{bcrypt}. In the import table of the 54 ransomware samples (experimental setup 1), only \texttt{advapi32} and \texttt{crypt32} show, in 38 and 13 samples, respectively. However, all 7 crypto-related modules are dynamically loaded through \texttt{LdrLoadDll}, by 48, 33, 22, 19, 15, 8, and 3 samples, respectively. Other non-crypto modules that are frequently loaded by ransomware samples include \texttt{kernel32} (48 samples), \texttt{ole32} (42 samples), \texttt{rpcrt4} (39 samples), and \texttt{user32} (36 samples).

A short import table with few crypto-related APIs may put detection off guard and let the malware through. Encryption is one of the most distinctive features of ransomware. Hiding it increases their chance of escaping detection. One step further, although not observed in our analysis, the hacker could write dead code where the execution would never reach and put irrelevant APIs in the import table for distraction, hindering the efficiency of static analysis.

\noindent\textbf{Comparison between newer and older ransomware samples.} To find out if there exist any trends of evolution, we analyzed the usage of top dynamic resolved APIs in the newer sample group (samples from 2020 and after) and the older sample group (2019 and before). The numbers and percentage of changes are summarized in Table \ref{tab:dynamic-short}. Overall, the usage of dynamic resolutions has increased in recent years. The \texttt{rsaenh} module becomes 23.36\% more prevalent in newer samples. The usage of \texttt{CryptAquireContentW} and \texttt{CryptGenRandom}, which are APIs from \texttt{cryptsp}, increased by 26.67\% and 27.63\%, respectively. None of the samples before 2019 dynamically loaded the \texttt{brcyrptprimitives} module but its APIs are being used by up to 31.58\% of recent samples.

\noindent\textbf{Comparison with benign samples.} This comparison between ransomware and benign samples help identify the distinctions in run-time API usages, possibly. Benign samples also dynamically resolve Windows libraries, including crypto-related modules. Top modules include \texttt{advapi32} (33 samples out of 38 dynamically load it) and \texttt{cryptsp} (23 samples). While the same crypto-related modules are used, the most frequent APIs loaded by benign samples are different. Most top benign loaded APIs are from \texttt{advapi32}, which stands for Advanced Windows 32 Base API. This module also contains basic Windows API other than crypto functions, such as \texttt{EventRegister} (Table~\ref{tab:dynamic-short}).

\begin{table}[]
\centering
\caption{API examples in crypto-related modules dynamically resolved by ransomware and benign samples. Total number: number of samples that dynamically load this API.  \# New /\# Old: numbers of new samples (2020 or after) and old samples (2019 or earlier) that load this API. Change \%: the difference between the newer and older sample group, (\# newer sample / 38 - \# older sample / 16). Positive values suggest an increase in usage in recent years.}

\small \begin{tabular}{
>{\columncolor[HTML]{EFEFEF}}l l
>{\columncolor[HTML]{EFEFEF}}l c
>{\columncolor[HTML]{EFEFEF}}c c}
\hline
\cellcolor[HTML]{C0C0C0}                                                                                                 & \cellcolor[HTML]{C0C0C0}\textbf{\begin{tabular}[c]{@{}l@{}}Top \\ modules\end{tabular}} & \cellcolor[HTML]{C0C0C0}\textbf{API examples}                               & \cellcolor[HTML]{C0C0C0}\textbf{\begin{tabular}[c]{@{}c@{}}Total\\ count\end{tabular}} & \cellcolor[HTML]{C0C0C0}\textbf{\begin{tabular}[c]{@{}c@{}} \# New /\\\# Old\end{tabular}} & \cellcolor[HTML]{C0C0C0}\textbf{\begin{tabular}[c]{@{}c@{}}Change\%\end{tabular}} \\ \hline
\cellcolor[HTML]{EFEFEF}                                                                                                 &                                                                                         & CryptReleaseContext                                                         & 42                                                                                     & 29 / 13                                                                              & -4.93\%                                                                               \\ \hhline{|>{\arrayrulecolor{tgray}}->{\arrayrulecolor{white}}->{\arrayrulecolor{tblack}}|-|-|-|-|} 
\cellcolor[HTML]{EFEFEF}                                                                                                 &                                                                                         & CryptAcquireContextW                                                        & 27                                                                                     & 22 / 5                                                                               & \cellcolor[HTML]{FFFFFF}26.64\%                                                       \\ \hhline{|>{\arrayrulecolor{tgray}}->{\arrayrulecolor{white}}->{\arrayrulecolor{tblack}}|-|-|-|-|} 
\cellcolor[HTML]{EFEFEF}                                                                                                 & \multirow{-3}{*}{cryptsp}                                                               & CryptGenRandom                                                              & 24                                                                                     & 20 / 4                                                                               & 27.63\%                                                                               \\ \hhline{|>{\arrayrulecolor{tgray}}->{\arrayrulecolor{tblack}}|-|-|-|-|-|}  
\cellcolor[HTML]{EFEFEF}                                                                                                 &                                                                                         & CryptDestroyKey                                                             & 10                                                                                     & 7 / 3                                                                                & -0.33\%                                                                               \\ \hhline{|>{\arrayrulecolor{tgray}}->{\arrayrulecolor{white}}->{\arrayrulecolor{tblack}}|-|-|-|-|} 
\cellcolor[HTML]{EFEFEF}                                                                                                 &                                                                                         & CryptGenKey                                                                 & 10                                                                                     & 7 / 3                                                                                & \cellcolor[HTML]{FFFFFF}-0.33\%                                                       \\ \hhline{|>{\arrayrulecolor{tgray}}->{\arrayrulecolor{white}}->{\arrayrulecolor{tblack}}|-|-|-|-|} 
\cellcolor[HTML]{EFEFEF}                                                                                                 & \multirow{-3}{*}{advapi32}                                                              & CryptImportKey                                                              & 10                                                                                     & 7 / 3                                                                                & \cellcolor[HTML]{FFFFFF}-0.33\%                                                       \\ \hhline{|>{\arrayrulecolor{tgray}}->{\arrayrulecolor{tblack}}|-|-|-|-|-|} 
\cellcolor[HTML]{EFEFEF}                                                                                                 &                                                                                         & CPGenRandom                                                                 & 19                                                                                     & 16 / 3                                                                               & 23.36\%                                                                               \\ \hhline{|>{\arrayrulecolor{tgray}}->{\arrayrulecolor{white}}->{\arrayrulecolor{tblack}}|-|-|-|-|} 
\cellcolor[HTML]{EFEFEF}                                                                                                 &                                                                                         & CPGenKey                                                                    & 19                                                                                     & 16 / 3                                                                               & \cellcolor[HTML]{FFFFFF}23.36\%                                                       \\ \hhline{|>{\arrayrulecolor{tgray}}->{\arrayrulecolor{white}}->{\arrayrulecolor{tblack}}|-|-|-|-|} 
\cellcolor[HTML]{EFEFEF}                                                                                                 & \multirow{-3}{*}{rsaenh}                                                                & CPEncrypt                                                                   & 19                                                                                     & 16 / 3                                                                               & \cellcolor[HTML]{FFFFFF}23.36\%                                                       \\ \hhline{|>{\arrayrulecolor{tgray}}->{\arrayrulecolor{tblack}}|-|-|-|-|-|}  
\cellcolor[HTML]{EFEFEF}                                                                                                 & cryptbase                                                                               & \begin{tabular}[c]{@{}l@{}}RtlGenRandom \\ (SystemFunction036)\end{tabular} & 19                                                                                     & 12 / 7                                                                               & -12.17\%                                                                              \\ \hhline{|>{\arrayrulecolor{tgray}}->{\arrayrulecolor{tblack}}|-|-|-|-|-|}  
\cellcolor[HTML]{EFEFEF}                                                                                                 &                                                                                         & GetHashInterface                                                            & 12                                                                                     & 12 / 0                                                                               & 31.58\%                                                                               \\ \hhline{|>{\arrayrulecolor{tgray}}->{\arrayrulecolor{white}}->{\arrayrulecolor{tblack}}|-|-|-|-|} 
\multirow{-12}{*}{\cellcolor[HTML]{EFEFEF}\begin{tabular}[c]{@{}l@{}}R\\ a\\ n\\ s\\ o\\ m\\ w\\ a\\ r\\ e\end{tabular}} & \multirow{-2}{*}{\begin{tabular}[c]{@{}l@{}}bcrypt-\\ primitives\end{tabular}}          & GetCipherInterface                                                          & 7                                                                                      & 7 / 0                                                                                & 18.42\%                                                                               \\ \hline
\cellcolor[HTML]{EFEFEF}                                                                                                 &                                                                                         & EventUnregister                                                             & 26                                                                                     & -                                                                                    & -                                                                                     \\ \hhline{|>{\arrayrulecolor{tgray}}->{\arrayrulecolor{white}}->{\arrayrulecolor{tblack}}|-|-|-|-|} 
\cellcolor[HTML]{EFEFEF}                                                                                                 &                                                                                         & EventRegister                                                               & 26                                                                                     & -                                                                                    & -                                                                                     \\ \hhline{|>{\arrayrulecolor{tgray}}->{\arrayrulecolor{white}}->{\arrayrulecolor{tblack}}|-|-|-|-|} 
\cellcolor[HTML]{EFEFEF}                                                                                                 & \multirow{-3}{*}{advapi32}                                                              & EventWrite                                                                  & 26                                                                                     & -                                                                                    & -                                                                                     \\ \hhline{|>{\arrayrulecolor{tgray}}->{\arrayrulecolor{tblack}}|-|-|-|-|-|} 
\multirow{-4}{*}{\cellcolor[HTML]{EFEFEF}\begin{tabular}[c]{@{}l@{}}B\\ e\\ n\\ i\end{tabular}}                          & cryptsp                                                                                 & CryptReleaseContext                                                         & 23                                                                                     & -                                                                                    & -                                                                                     \\ \hline
\end{tabular}

\vspace{-2mm}
\label{tab:dynamic-short}
\end{table}

\subsection{Other Behavioral Signatures}
\begin{table*}[]
\centering
\caption{Non-encryption behavioral signatures of ransomware samples. \# Samples / \# Families: numbers of ransomware samples and families have this signature. \# New / \# Old: numbers of new samples (2020 or after) and old samples (2019 or earlier) that have this signature. Change \%: the difference between the newer and older sample group, (\# newer sample / 38 - \# older sample / 16). Positive values suggest an increase in usage in recent years. \# Benign: number of benign samples also have this signature. Severity level is based on its severity score defined by the Cuckoo community. Diff \%: difference in this signature's prevalence between ransomware and benign samples (\# Samples / 54 - \# Benign Samples / 38). The numbers in () are the number of samples in this group. The behavioral signatures more prevalent in ransomware are highlighted in bold.}

\scriptsize \begin{tabular}{
>{\columncolor[HTML]{EFEFEF}}l l
>{\columncolor[HTML]{EFEFEF}}c c
>{\columncolor[HTML]{EFEFEF}}c c
>{\columncolor[HTML]{EFEFEF}}c c}
\hline
\cellcolor[HTML]{C0C0C0}\textbf{Category}                                                                                                & \cellcolor[HTML]{C0C0C0}\textbf{Signature}                                                                               & \cellcolor[HTML]{C0C0C0}\textbf{\begin{tabular}[c]{@{}c@{}}Seve-\\ rity\end{tabular}} & \cellcolor[HTML]{C0C0C0}\textbf{\begin{tabular}[c]{@{}c@{}}\# Samples (54)\\ / \# Families (35)\end{tabular}} & \cellcolor[HTML]{C0C0C0}\textbf{\begin{tabular}[c]{@{}c@{}}\# New (38) /\\ \# Old (16)\end{tabular}} & \cellcolor[HTML]{C0C0C0}\textbf{Change \%} & \cellcolor[HTML]{C0C0C0}\textbf{\begin{tabular}[c]{@{}c@{}}\# Benign\\ (38)\end{tabular}} & \cellcolor[HTML]{C0C0C0}\textbf{Diff \%} \\ \hline
\cellcolor[HTML]{EFEFEF}                                                                                                                 & \textbf{\begin{tabular}[c]{@{}l@{}}Attempts to detect sandbox through the presence \\ of a file or process\end{tabular}} & High                                                                                  & 24 / 17                                                                                                       & 18 / 6                                                                                               & 9.87\%                                     & 1                                                                                       & 41.8\%                                   \\ \hhline{|>{\arrayrulecolor{tgray}}->{\arrayrulecolor{tblack}}|-|-|-|-|-|-|-|}   
\cellcolor[HTML]{EFEFEF}                                                                                                                 & \textbf{Attempts to delay analysis (a process tried to sleep)}                                                           & Med                                                                                   & 17 / 12                                                                                                       & 13 / 4                                                                                               & 9.21\%                                     & 11                                                                                      & 2.5\%                                    \\ \hhline{|>{\arrayrulecolor{tgray}}->{\arrayrulecolor{tblack}}|-|-|-|-|-|-|-|} 
\cellcolor[HTML]{EFEFEF}                                                                                                                 & Checks foreground human activities                                                                                       & Med                                                                                   & 16 / 10                                                                                                       & 13 / 3                                                                                               & 15.46\%                                    & 21                                                                                      & -25.6\%                                  \\ \hhline{|>{\arrayrulecolor{tgray}}->{\arrayrulecolor{tblack}}|-|-|-|-|-|-|-|} 
\cellcolor[HTML]{EFEFEF}                                                                                                                 & Creates a process named as a common system process                                                                       & Med                                                                                   & 4 / 3                                                                                                         & 3 / 1                                                                                                & 1.64\%                                     & 3                                                                                       & -0.5\%                                   \\ \hhline{|>{\arrayrulecolor{tgray}}->{\arrayrulecolor{tblack}}|-|-|-|-|-|-|-|} 
\multirow{-5}{*}{\cellcolor[HTML]{EFEFEF}\begin{tabular}[c]{@{}l@{}}Anti-\\ detection \\ and \\ anti-\\ analysis\end{tabular}}           & \begin{tabular}[c]{@{}l@{}}Checks amount of memory or disk size in system to \\ detect virtual machine\end{tabular}      & Med                                                                                   & 31 / 20                                                                                                       & 24 / 7                                                                                               & 19.41\%                                    & 27                                                                                      & -13.6\%                                  \\ \hline
\cellcolor[HTML]{EFEFEF}                                                                                                                 & \textbf{Operates on local firewall's policies and settings}                                                              & High                                                                                  & 7 / 4                                                                                                         & 7 / 0                                                                                                & 18.42\%                                    & 0                                                                                       & 13.0\%                                   \\ \hhline{|>{\arrayrulecolor{tgray}}->{\arrayrulecolor{tblack}}|-|-|-|-|-|-|-|} 
\cellcolor[HTML]{EFEFEF}                                                                                                                 & \textbf{Attempts to modify desktop wallpaper}                                                                            & High                                                                                  & 6 / 3                                                                                                         & 5 / 1                                                                                                & 6.91\%                                     & 0                                                                                       & 11.1\%                                   \\ \hhline{|>{\arrayrulecolor{tgray}}->{\arrayrulecolor{tblack}}|-|-|-|-|-|-|-|} 
\cellcolor[HTML]{EFEFEF}                                                                                                                 & \textbf{\begin{tabular}[c]{@{}l@{}}Disables Windows Security features \\ (e.g., Windows defender)\end{tabular}}          & High                                                                                  & 3 / 2                                                                                                         & 1 / 2                                                                                                & -9.87\%                                    & 0                                                                                       & 5.6\%                                    \\ \hhline{|>{\arrayrulecolor{tgray}}->{\arrayrulecolor{tblack}}|-|-|-|-|-|-|-|} 
\cellcolor[HTML]{EFEFEF}                                                                                                                 & \textbf{Clears Windows event logs}                                                                                       & High                                                                                  & 1 / 1                                                                                                         & 1 / 0                                                                                                & 2.63\%                                     & 0                                                                                       & 1.9\%                                    \\ \hhline{|>{\arrayrulecolor{tgray}}->{\arrayrulecolor{tblack}}|-|-|-|-|-|-|-|} 
\multirow{-5}{*}{\cellcolor[HTML]{EFEFEF}\begin{tabular}[c]{@{}l@{}}Modification\\ of system \\ and security \\ setting\end{tabular}} & \textbf{Installs itself for autorun at Windows startup}                                                                  & Low                                                                                   & 22 / 13                                                                                                       & 14 / 8                                                                                               & -13.16\%                                   & 13                                                                                      & 6.5\%                                    \\ \hline
\cellcolor[HTML]{EFEFEF}                                                                                                                 & \textbf{Deletes the Shadow Copy}                                                                                         & High                                                                                  & 16 / 12                                                                                                       & 12 / 4                                                                                               & 6.58\%                                     & 0                                                                                       & 29.6\%                                   \\ \hhline{|>{\arrayrulecolor{tgray}}->{\arrayrulecolor{tblack}}|-|-|-|-|-|-|-|} 
\cellcolor[HTML]{EFEFEF}                                                                                                                 & \textbf{Modifies boot configuration settings}                                                                            & High                                                                                  & 11 / 7                                                                                                        & 10 / 1                                                                                               & 20.07\%                                    & 0                                                                                       & 20.4\%                                   \\ \hhline{|>{\arrayrulecolor{tgray}}->{\arrayrulecolor{tblack}}|-|-|-|-|-|-|-|} 
\multirow{-3}{*}{\cellcolor[HTML]{EFEFEF}\begin{tabular}[c]{@{}l@{}}Prevention \\ of recovery\end{tabular}}                           & \textbf{Uses wbadmin utility to delete backups}                                                                          & High                                                                                  & 5 / 2                                                                                                         & 5 / 0                                                                                                & 13.16\%                                    & 0                                                                                       & 9.3\%                                    \\ \hline
\cellcolor[HTML]{EFEFEF}                                                                                                                 & \textbf{Appends a new file extension to files}                                                                           & High                                                                                  & 51 / 33                                                                                                       & 36 / 15                                                                                              & 0.99\%                                     & 5                                                                                       & 81.3\%                                   \\ \hhline{|>{\arrayrulecolor{tgray}}->{\arrayrulecolor{tblack}}|-|-|-|-|-|-|-|} 
\cellcolor[HTML]{EFEFEF}                                                                                                                 & \textbf{Moves a number of files}                                                                                         & High                                                                                  & 13 / 10                                                                                                       & 10 / 3                                                                                               & 7.57\%                                     & 2                                                                                       & 18.8\%                                   \\ \hhline{|>{\arrayrulecolor{tgray}}->{\arrayrulecolor{tblack}}|-|-|-|-|-|-|-|} 
\multirow{-3}{*}{\cellcolor[HTML]{EFEFEF}\begin{tabular}[c]{@{}l@{}}Modification \\ of file\end{tabular}}                                & Deletes a number of files                                                                                                & High                                                                                  & 6 / 5                                                                                                         & 1 / 5                                                                                                & -28.62\%                                   & 6                                                                                       & -4.7\%                                   \\ \hline
\end{tabular}

\label{tab:signature}
\end{table*}

The previous sections focus on the encryption and file access API usage during ransomware execution. In this section, we examine other ransomware behaviors. We summarized 16 common behavioral signatures in 4 categories (Table \ref{tab:signature}). We derived the severity of each based on its severity score defined by the Cuckoo community. Ransomware samples are further divided into newer and older groups based on the year for analysis of evolution. We also calculate the difference between the prevalence of each signature in ransomware and benign software. The higher the value, the more indicative of ransomware the signature is. The ones that are more prevalent in ransomware are marked in bold in Table \ref{tab:signature}.

\noindent\textbf{Prevention of recovery.} The nature of ransomware is to hold the victim’s data hostage. Therefore, they use multiple techniques to avoid data or system recovery. To reduce the possibility of recovery from the volume shadow copy, which is a snapshot that duplicates data stored in a volume, many ransomware attempts to delete it. 16 samples use commands such as \texttt{vssadmin delete shadows /all /quiet} and \texttt{wmic shadowcopy delete} to do so. We also observed that 11 samples modify the boot configuration of the system to prevent the Windows system from going into the recovery environment (Windows RE). This is often done by using bcedit commands (e.g., \texttt{bcdedit  /set {default} recoveryenabled no} and \texttt{bcdedit  /set {default} bootstatuspolicy ignoreallfailures}). Yet another technique to prevent the usage of backups is to delete the global catalog that contains system backup information. 5 samples do it through the command \texttt{wbadmin delete catalog -quiet}.

\noindent\textbf{Modification of file.} File type changing is often a signal of data being encrypted. We notice that 51 out of 54 ransomware samples append a new extension to the files in the system, some are the same as the ransomware family name (e.g., .alcatraz by Alcatraz) and some look random (e.g., .36Q62M by Sodinokibi). Moving and deleting a large number of files (observed in 9 and 3 samples, respectively) are also signs of file modification.

\noindent\textbf{Modification of system and security setting.} Before launching the attack, ransomware often changes Windows system setting to their convenience, especially security settings. 7 samples modify local firewall rules and 3 samples disable several Windows Defender features. After the attack, one sample (i.e., Hive) also clears the Windows event log to hide its trace. 22 samples install themselves to run at system startup, possibly launching the attack again every time the system reboots in case of recovery. In addition, 6 samples attempt to modify the desktop wallpaper to make the message more attention-calling and threatening. 

\noindent\textbf{Anti-analysis and anti-detection.} The majority of ransomware samples attempt to hinder analysis and detection in order to increase the success rate of attacks. Because of the extensive use of VMs and sandboxes in malware analysis, many ransomware samples try to distinguish a VM from a physical machine \cite{Shi2017, inside-look}. Out of the 54 samples examined, 31 check the amount of system memory or disk size, 24 try to detect the presence of a sandbox, and 16 check foreground human activities. Some dynamic detection or analysis run a program for a few minutes to evaluate the risk. To evade detection, 17 samples delay the execution. In the case of finding clues of being analyzed in a VM, the malware would likely end execution or hang indefinitely, leaving no trace for analysis.

\begin{table}[]
\centering
\caption{Top processes used by ransomware samples. \# Samples / \# Families: numbers of ransomware samples and families initiate this process. \# New / \# Old: numbers of new samples (2020 or after) and old samples (2019 or earlier) that use this process. Change \%: the difference between the newer and older sample group. Positive values suggest an increase in usage in recent years. \# Benign: number of benign samples. Diff \%: difference in this process' prevalence between ransomware and benign samples. The numbers in () are the number of samples in this group. The processes more prevalent in ransomware are in bold. All processes have a .exe suffix.}

\small \begin{tabular}{
>{\columncolor[HTML]{EFEFEF}}c c
>{\columncolor[HTML]{EFEFEF}}c c
>{\columncolor[HTML]{EFEFEF}}c c
>{\columncolor[HTML]{EFEFEF}}c c}
\hline
\cellcolor[HTML]{C0C0C0}\textbf{Process} & \cellcolor[HTML]{C0C0C0}\textbf{\begin{tabular}[c]{@{}c@{}}\#Samples\\(54)\end{tabular}} & \cellcolor[HTML]{C0C0C0}\textbf{\begin{tabular}[c]{@{}c@{}}\#Families\\(35)\end{tabular}} & \cellcolor[HTML]{C0C0C0}\textbf{\begin{tabular}[c]{@{}c@{}}\#New \\ (38)\end{tabular}} & \cellcolor[HTML]{C0C0C0}\textbf{\begin{tabular}[c]{@{}c@{}}\#Old \\ (16)\end{tabular}} & \cellcolor[HTML]{C0C0C0}\textbf{\begin{tabular}[c]{@{}c@{}}Change\\ \% \end{tabular}} & \cellcolor[HTML]{C0C0C0}\textbf{\begin{tabular}[c]{@{}c@{}}\#Benign\\(38)\end{tabular}} & \cellcolor[HTML]{C0C0C0}\textbf{\begin{tabular}[c]{@{}c@{}}Diff\\ \% \end{tabular}} \\ \hline
lsass                                & 54                                                                                                & 35                                                                                                 & 38                                                                                                   & 16                                                                                                    & 0.00\%                                     & 38                                                                                                & 0.0\%                                    \\ \hline
\textbf{cmd}                         & 36                                                                                                & 23                                                                                                 & 24                                                                                                   & 12                                                                                                    & -11.84\%                                   & 1                                                                                                 & 95.0\%                                   \\ \hline
\textbf{vssadmin}                    & 20                                                                                                & 15                                                                                                 & 15                                                                                                   & 5                                                                                                     & 8.22\%                                     & 0                                                                                                 & 55.6\%                                   \\ \hline
explorer                            & 15                                                                                                & 9                                                                                                  & 13                                                                                                   & 2                                                                                                     & 21.71\%                                    & 22                                                                                                & -68.3\%                                  \\ \hline
\textbf{reg}                         & 13                                                                                                & 7                                                                                                  & 7                                                                                                    & 6                                                                                                     & -19.08\%                                   & 1                                                                                                 & 31.1\%                                   \\ \hline
\textbf{WMIC}                        & 12                                                                                                & 8                                                                                                  & 11                                                                                                   & 1                                                                                                     & 22.70\%                                    & 0                                                                                                 & 33.3\%                                   \\ \hline
\textbf{bcdedit}                     & 10                                                                                                & 6                                                                                                  & 9                                                                                                    & 1                                                                                                     & 17.43\%                                    & 0                                                                                                 & 27.8\%                                   \\ \hline
\textbf{netsh}                       & 7                                                                                                 & 4                                                                                                  & 7                                                                                                    & 0                                                                                                     & 18.42\%                                    & 0                                                                                                 & 19.4\%                                   \\ \hline
\textbf{PING}                        & 6                                                                                                 & 4                                                                                                  & 4                                                                                                    & 2                                                                                                     & -1.97\%                                    & 0                                                                                                 & 16.7\%                                   \\ \hline
\textbf{taskkill}                    & 6                                                                                                 & 4                                                                                                  & 4                                                                                                    & 2                                                                                                     & -1.97\%                                    & 1                                                                                                 & 11.7\%                                   \\ \hline
\textbf{net}                         & 5                                                                                                 & 4                                                                                                  & 5                                                                                                    & 0                                                                                                     & 13.16\%                                    & 1                                                                                                 & 8.9\%                                    \\ \hline
\textbf{net1}                        & 5                                                                                                 & 4                                                                                                  & 5                                                                                                    & 0                                                                                                     & 13.16\%                                    & 1                                                                                                 & 8.9\%                                    \\ \hline
\textbf{wbadmin}                     & 5                                                                                                 & 2                                                                                                  & 5                                                                                                    & 0                                                                                                     & 13.16\%                                    & 0                                                                                                 & 13.9\%                                   \\ \hline
\textbf{notepad}                     & 4                                                                                                 & 3                                                                                                  & 3                                                                                                    & 1                                                                                                     & 1.64\%                                     & 0                                                                                                 & 11.1\%                                   \\ \hline
\textbf{chrome}                      & 4                                                                                                 & 4                                                                                                  & 2                                                                                                    & 2                                                                                                     & -7.24\%                                    & 1                                                                                                 & 6.1\%                                    \\ \hline
\textbf{powershell}                  & 4                                                                                                 & 3                                                                                                  & 3                                                                                                    & 1                                                                                                     & 1.64\%                                     & 0                                                                                                 & 11.1\%                                   \\ \hline
\textbf{icacls}                      & 3                                                                                                 & 2                                                                                                  & 2                                                                                                    & 1                                                                                                     & -0.99\%                                    & 0                                                                                                 & 8.3\%                                    \\ \hline
\textbf{schtasks}                    & 3                                                                                                 & 3                                                                                                  & 1                                                                                                    & 2                                                                                                     & -9.87\%                                    & 1                                                                                                 & 3.3\%                                    \\ \hline
\textbf{cscript}                     & 3                                                                                                 & 2                                                                                                  & 0                                                                                                    & 3                                                                                                     & -18.75\%                                   & 0                                                                                                 & 8.3\%                                    \\ \hline
\end{tabular}

\vspace{-3mm}
\label{tab:process}
\end{table}
\noindent\textbf{Processes.} Additionally, we summarize the processes initiated by the 54 samples during execution (Table \ref{tab:process}). Consistent with the usage of prevention-of-recovery commands, we find 20, 12, 10, and 5 samples utilize vssadmin, WMIC, bcedit, and wbadmin, respectively. Frequent usage of cmd (36 samples) and powershell (4 samples) are also observed, likely for executing the commands. 

\noindent\textbf{Comparison between newer and older ransomware samples.} We do the compare between newer and older samples to identify the trends of evolution (Table \ref{tab:signature}). One notable trend is the increased usage of recovery prevention. 20.07\% more newer samples modify the boot configurations, 13.16\% more delete backups using wbadmin, and 6.58\% more deletes the volume shadow copy. Moreover, the usage of evasive techniques has a trend of growing as well, with up to a 19.41\% increase. 

\noindent\textbf{Comparison with benign samples.} We further inspected the benign sample to see the similarities and differences with ransomware. Shared behavioral signatures include checking foreground human activities (observed in 21 out of 38 benign samples) and checking system memory and disk size (27 samples). Although these features look suspicious, they would not separate malicious and benign well. Modification of security settings and prevention system recovery are rarely seen in regular applications. On the other hand, except for lsass and explorer, the processes used by ransomware are not common in regular software execution, including command line-related processes (e.g., cmd and powershell) and system backup-related processes (e.g., vssadmin, WMIC, bcdedit, and wbadmin).
\section{New API-profiling Based Classification Method}
\label{sec:classification}

This section presents our new API-based classification algorithm for identifying ransomware threats (\textbf{RQ3}).

\subsection{Algorithms}
\label{sec:detect-algo}

Our method consists of two main operations: {\em i)} {\bf consistency-based classification} and {\em ii)} {\bf refinement using API contrast score}. In consistency-based classification, we quantify the ransomware execution patterns according to their API repetition patterns. We build several consistency-based mathematical models for classification. Then, in the refinement operation, we compute API contrast scores to further improve the detection accuracy.

\subsubsection{Consistency-based classification.}

We present multiple computational methods, with varying complexity, for summarizing ransomware's API invocation behavioral patterns. These methods are for the first stage of our detection.

\noindent\textit{Consistency-based detection.} We design four consistency metrics to quantify the variation during the execution. We treat the API composition for a short execution time period as a vector and compare the vector with a previous period. Small variations between vectors suggest more consistency. Specifically, the metrics are Cosine-based consistency (equation \ref{eq:cos-consistency}), Manhattan-based consistency (equation \ref{eq:manh-consistency}), frequency-weighted consistency (equation \ref{eq:weight-consistency}), and Euclidean-based consistency (equation \ref{eq:euc-consistency}):

\noindent\begin{minipage}{.45\linewidth}
\begin{equation}
 1-\frac{c\bullet p}{\left\| c \right\|\left\| p \right\|}
 \label{eq:cos-consistency}
\end{equation}
\end{minipage}%
\begin{minipage}{.45\linewidth}
\begin{equation}
 \sum_{i=1}^{n}\left| c_i-p_i \right|
 \label{eq:manh-consistency}
\end{equation}
\end{minipage}

\noindent\begin{minipage}{.45\linewidth}
\begin{equation}
 \sum_{i=1}^{n}f_i\left| c_i-p_i \right|
 \label{eq:weight-consistency}
\end{equation}
\end{minipage}
\begin{minipage}{.45\linewidth}
\begin{equation}
 \sqrt{\sum_{i=1}^{n}(c_i-p_i)^2}
 \label{eq:euc-consistency}
\end{equation}
\end{minipage}

\noindent $c$ is the vector representing the current execution window, $p$ is the vector representing the previous execution window, and $f$ is the vector representing the frequency of top APIs. When calculating the scores, we consider the top 10 file API and use 3-second and 1-second windows for the previous and current, respectively. All of those parameters can be adjusted at the time of application.

As the names suggest, Manhattan-based and Euclidean-based consistency use Manhattan and Euclidean distance to compute the difference between execution periods. The smaller the score, the more consistent the execution. Frequency-weighted consistency is a variation of the Manhattan distance that each element is weighted by the frequency. That is, the top frequent API takes a larger part in the score. Lastly, cosine-based consistency is calculated based on the cosine similarity between the two vectors representing previous and current execution. Because the more similar the vectors are, the closer to 1 the cosine similarity is, we use 1 minus cosine similarity here to be consistent with other metrics (i.e., smaller values represent more consistency). For all four consistency algorithms, the smaller the value, the more malicious the program is.

\noindent\textit{Evenness-based detection.} In this method, we analyze the frequency distribution of multiple top APIs together. This method aims to capture the ransomware feature that the API usage composition of each epoch is relatively evenly distributed during its execution. We develop two evenness-related metrics to quantify the execution pattern, namely normalized evenness (equation \ref{eq:normal-even}) and squared evenness (equation \ref{eq:square-even}):

\vspace{2mm}
\noindent\begin{minipage}{.45\linewidth}
\begin{equation}
\sum_{i=1}^{n}\frac{\left| API_i-avg \right|}{avg}
 \label{eq:normal-even}
\end{equation}
\end{minipage}%
\begin{minipage}{.45\linewidth}
\begin{equation}
\sum_{i=1}^{n}(API_i-avg)^2
 \label{eq:square-even}
\end{equation}
\end{minipage}
\vspace{2mm}

\noindent $avg$ is the average of top API usage, $n$ is the number of top APIs used. The evenness is first calculated for each small execution period and then averaged to present the whole execution. 

\noindent\textit{Changepoint-based detection.} In this method, we count notable changes using Bayesian Online Changepoint Detection (BOCD)~\cite{changepoint} to separate malicious and benign traces. BOCD is designed to identify abrupt changes in sequential data. Ideally, ransomware execution should have fewer changepoints due to the constant pattern.

As a baseline, we also implement a single API distribution approach, whose detection is based on the frequency distributions of top file APIs, such as \texttt{NtWriteFile}, \texttt{NtReadFile}, and \texttt{NtCreateFile}, using the Poisson distribution, Wilcoxon rank sum test, and Jensen-Shannon (JS) divergence. The equation for computing JS divergence is as follows:
 \begin{equation}
    JS(P\parallel Q) = \frac{1}{2}*KL(P\parallel M) + \frac{1}{2}*KL(M\parallel P)
\end{equation}
 \noindent where
 \begin{equation}
    M = \frac{1}{2} * (P + Q)
\end{equation}
 \noindent and KL is the KL divergence that 
 \begin{equation}
    KL(P\parallel Q) = \sum_{x\in \chi}^{}p(x)\log(\frac{p(x)}{q(x)})
\end{equation}
 \noindent When computing a divergence score for a ransomware sample, $P$ represents the API distribution of the specific sample and $Q$ represents the overall benign distribution. Vice versa for benign.

\subsubsection{Refinement using API contrast score.}
\label{sec:contrast-score}

The refinement operation -- the second stage of our detection -- builds on leveraging the API occurrence variation between ransomware and benign programs. In refinement, our screening is centered on API contrast scores, explained next. {\bf First}, with the labeled dataset, we perform a comparative counting analysis to compute a contrast score for each distinct API. This training process also organizes the APIs into three distinct sets, based on their occurrences and invocation frequencies in ransomware and benign execution. The three sets are {\em i)} likely ransomware API set $\mathbb{R}$, which contains APIs that often occur in ransomware, but rarely in benign programs, {\em ii)} likely benign API set $\mathbb{B}$, which contains APIs that commonly appear in benign, but less often in ransomware, and {\em iii)} co-occurring API set $\mathbb{O}$, which consists of APIs used by both types of samples, but with a much lower call frequency in benign execution. This assignment is based on relative or pre-defined thresholds, as shown in Equation~\ref{eqn:API-contrast-score}. {\bf Second}, we compute the single contrast score $C_i$ for each API$_i$ following Equation~\ref{eqn:API-contrast-score}. {\bf Finally}, during the testing phase, given the profile of an unknown execution, we compute the total contrast score for all $n$ occurrences of APIs in the profile, i.e., $\sum_{i=1}^n C_i$, with deduplication (each distinct API will only be counted once).

\begin{equation}
    \label{eqn:API-contrast-score}
    C_i = 
    \begin{cases}
      1,~\mbox{API}_i~\rightarrow~\mathbb{R},  & \text{if}~\frac{\textit{occr}^R_i}{\textit{occr}_i^B} \geq \tau_1\\
      
      -1,~\mbox{API}_i~\rightarrow~\mathbb{B}, &  \text{if}~\frac{\textit{occr}^R_i}{\textit{occr}_i^B} \leq \tau_2\\
      1,~\mbox{API}_i~\rightarrow~\mathbb{O},  & \text{if}~\frac{\textit{occr}^R_i}{\textit{occr}_i^B} \in (\tau_2, \tau_1) \wedge~\frac{\textit{freq}^R_i}{\textit{freq}_i^B} \geq \tau_3 \\
      0,  & \text{otherwise}
    \end{cases}
\end{equation}

In Equation~\ref{eqn:API-contrast-score}, set $\mathbb{R}$ contains APIs occurring highly frequently in ransomware, but rarely in benign, $\mathbb{B}$ contains APIs associated with benign software, but not with ransomware, and set $\mathbb{O}$ contains APIs that occur in both, $\textit{occr}_i^R$ is the count of ransomware samples in which the execution traces include the occurrence of API$_i$, and $\textit{occr}_i^B$ is the count of benign samples whose traces include the occurrence of API$_i$.

$\textit{freq}_i^R$ is the average call frequency of API$_i$ in $\mathbb{O}$'s ransomware execution, $\textit{freq}_i^B$ is the average call frequency of API$_i$ in $\mathbb{O}$'s benign samples. In our implementation, $\tau_1$ is set to 2, $\tau_2$ is set to 3, and $\tau_3$ is set to 2 (i.e., the call frequency exceeds the average benign call frequency by at least two times). The frequency limit used for each API in set $\mathbb{O}$ can be found in Table~\ref{tab:rw_score_freq_list} in the appendix. Because we already consider file-related APIs in the previous classification stage, we only include non-file APIs in the refinement. The lists of APIs in each set are in Tables~\ref{tab:rw_score_api_list}, \ref{tab:rw_score_freq_list}, and \ref{tab:be_score_api_list}, respectively.

\begin{table}
\centering
\caption{List of top ransomware APIs used in API contrast score (set $\mathbb{R}$).}
\small 
\begin{tabular}{
>{\columncolor[HTML]{EFEFEF}}c c
>{\columncolor[HTML]{EFEFEF}}c c}
\hline
\multicolumn{4}{c}{\cellcolor[HTML]{C0C0C0}\textbf{API}}                    \\ \hline
CoInitializeSecurity     & Process32FirstW & WriteConsoleW & CryptEncrypt   \\ \hline
CreateToolhelp32Snapshot & Process32NextW  & CryptGenKey   & CryptExportKey \\ \hline
\end{tabular}

\label{tab:rw_score_api_list}
\end{table}

\begin{table}
\centering
\caption{List of co-occurring APIs used in calculating the API contrast score (set $\mathbb{O}$). The frequency limits are calculated based on the observed frequency and the duration of execution. We set them to be double the average benign frequency. Because the execution time is 600s for setup 1 samples and 300s for setup 2 samples, we further double the value when applied on setup 1 samples.}
\small \begin{tabular}{
>{\columncolor[HTML]{EFEFEF}}c c
>{\columncolor[HTML]{EFEFEF}}c c}
\hline
\cellcolor[HTML]{C0C0C0}\textbf{API} & \cellcolor[HTML]{C0C0C0}\textbf{Threshold} & \cellcolor[HTML]{C0C0C0}\textbf{API} & \cellcolor[HTML]{C0C0C0}\textbf{Threshold} \\ \hline
NtAllocateVirtualMemory              & 6412                                       & RegDeleteValueW                      & 56                                         \\ \hline
NtFreeVirtualMemory                  & 2320                                       & GetUserNameExW                       & 28                                         \\ \hline
OpenSCManagerW                       & 20                                         & CoCreateInstanceEx                   & 8                                          \\ \hline
OpenServiceW                         & 32                                         & CryptAcquireContextA                 & 64                                         \\ \hline
NtOpenThread                         & 16                                         & CryptCreateHash                      & 52                                         \\ \hline
\end{tabular}

\label{tab:rw_score_freq_list}
\end{table}

\begin{table}
\centering
\caption{List of top benign APIs used in calculating the API contrast score (set $\mathbb{B}$).}

\small 
\begin{tabular}{
>{\columncolor[HTML]{EFEFEF}}c c
>{\columncolor[HTML]{EFEFEF}}c c}
\hline
\multicolumn{4}{c}{\cellcolor[HTML]{C0C0C0}\textbf{API}}                                                                   \\ \hline
NtDeleteKey   & \cellcolor[HTML]{FFFFFF}GetCursorPos   & GetForegroundWindow     & \cellcolor[HTML]{FFFFFF}FindResourceExW \\ \hline
EnumWindows   & \cellcolor[HTML]{FFFFFF}SizeofResource & GetFileVersionInfoSizeW & \cellcolor[HTML]{FFFFFF}FindResourceA   \\ \hline
GetKeyState   & OleInitialize                          & GetFileVersionInfoW     & \cellcolor[HTML]{FFFFFF}RegCreateKeyExA \\ \hline
DrawTextExW   & \cellcolor[HTML]{FFFFFF}FindWindowW    & SendNotifyMessageW      & \multicolumn{1}{l}{}                    \\ \hline
FindResourceW & \cellcolor[HTML]{FFFFFF}NtCreateKey    & NtReadVirtualMemory     & \multicolumn{1}{l}{}                    \\ \hline
\end{tabular}

\label{tab:be_score_api_list}
\end{table}

\begin{figure*}
     \centering
     \begin{subfigure}[b]{0.35\textwidth}
         \centering
         \includegraphics[width=\textwidth]{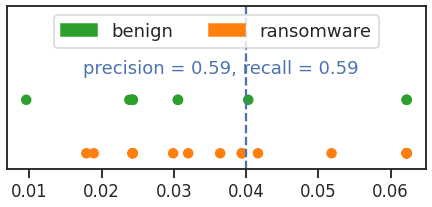}
         \vspace*{-6mm}
         \caption{JS Divergence (NtWriteFile)} 
         \label{fig:js-diver}
     \end{subfigure}
      \begin{subfigure}[b]{0.35\textwidth}
         \centering
         \includegraphics[width=\textwidth]{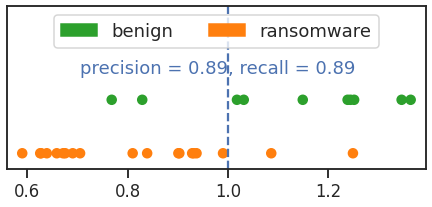}
         \vspace*{-6mm}
         \caption{Normalized evenness} 
         \label{fig:normal-even}
     \end{subfigure}
     
     \begin{subfigure}[b]{0.35\textwidth}
         \centering
         \includegraphics[width=\textwidth]{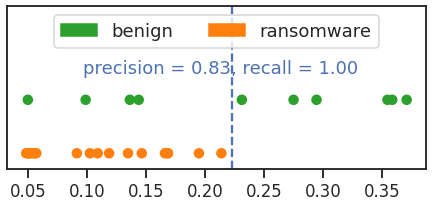}
         \vspace*{-6mm}
         \caption{Squared evenness}
         \label{fig:square-even}
     \end{subfigure}
     \begin{subfigure}[b]{0.35\textwidth}
         \centering
         \includegraphics[width=\textwidth]{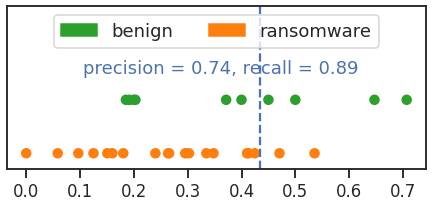}
         \vspace*{-6mm}
         \caption{Number of changepoints}
         \label{fig:changepoint}
     \end{subfigure}

     \begin{subfigure}[b]{0.35\textwidth}
         \includegraphics[width=\textwidth]{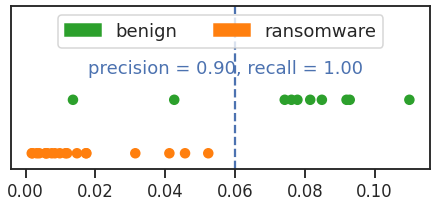}
         \vspace*{-6mm}
         \caption{Cosine-based consistency} 
         \label{fig:cos-consistency}
     \end{subfigure}
      \begin{subfigure}[b]{0.35\textwidth}
         \includegraphics[width=\textwidth]{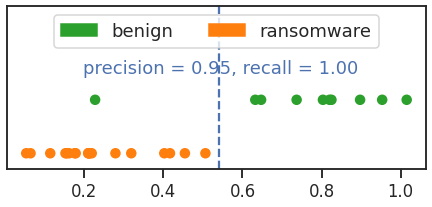}
         \vspace*{-6mm}
         \caption{ Manhattan-based consistency} 
         \label{fig:manh-consistency}
     \end{subfigure}
     
     \begin{subfigure}[b]{0.35\textwidth}
         \includegraphics[width=\textwidth]{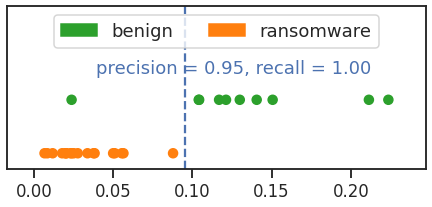}
         \vspace*{-6mm}
         \caption{Frequency-weighted consistency} 
         \label{fig:weight-consistency}
     \end{subfigure}
     \begin{subfigure}[b]{0.35\textwidth}
         \includegraphics[width=\textwidth]{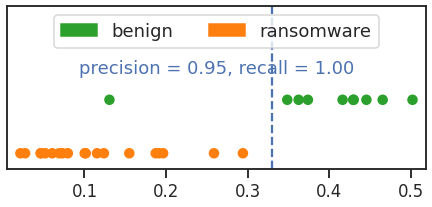}
         \vspace*{-6mm}
         \caption{ Euclidean-based consistency} 
         \label{fig:euc-consistency}
     \end{subfigure}
     
\caption{Classification results using different methods and models. Each green dot (top row) represents a benign software sample and each orange dot (bottom row) represents a ransomware sample. The dotted blue line is a boundary for helping understand how well the separation is. The precision and recall shown in the figures are based on the chosen boundary and are in terms of the ransomware class. }

\label{fig:file-freq-models}
\end{figure*}

\subsection{Classification Results}
\label{sec:evaluation}

In this section, we present the evaluation results of our two-stage approach, first on consistency-based classification, then on API contrast-based refinement. 
The test is conducted on real-world samples from 15 distinct ransomware families.
Then, we provide an in-depth analysis of the top important APIs with attack context as case studies. Finally, we demonstrate the performance of the commercial defense, identifying the security gap.

\subsection{Results of Various Classification Models}

We first present the classification results based on various mathematical models. Consistency-based algorithms show the best performance. The results of classifying 19 ransomware samples and 10 benign programs with notable amounts of file activities are shown in Figure~\ref{fig:file-freq-models}.

\noindent\textbf{Single API distribution.} Using only the distribution of a single API (i.e., baseline), we observe the scores for ransomware and benign samples are highly overlapping, implying the inadequacy of this model. An example of JS divergence of NtWriteFile’s distribution is shown in Figure \ref{fig:js-diver}.

\noindent\textbf{Evenness and the number of changepoints.} As shown in Figures \ref{fig:normal-even}, \ref{fig:square-even}, and \ref{fig:changepoint}, the evenness and changepoint metrics worked better than relying on single API distribution. We can see the trend that most ransomware samples are on the left side in the figure while benign samples have relatively larger values in both cases. However, there are still no clear boundaries between them and there exists room for improvement.

\begin{figure}

\centering
\begin{minipage}{.5\textwidth}
  \centering
  \captionsetup{width=0.85\textwidth}
  \includegraphics[width=.9\linewidth]{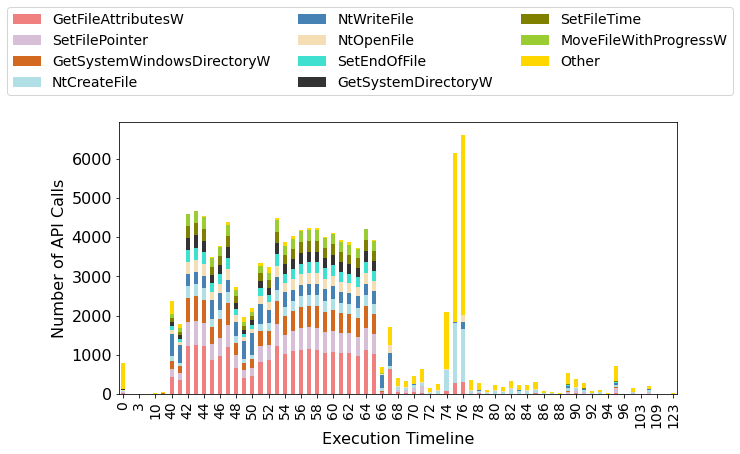}
  \vspace{-2mm}
  \captionof{figure}{Git's file API frequency (false positive)}
  \label{fig:git-file-access}
\end{minipage}\hfill %
\begin{minipage}{.5\textwidth}
  \centering
  \captionsetup{width=0.9\textwidth}
  \includegraphics[width=.85\linewidth]{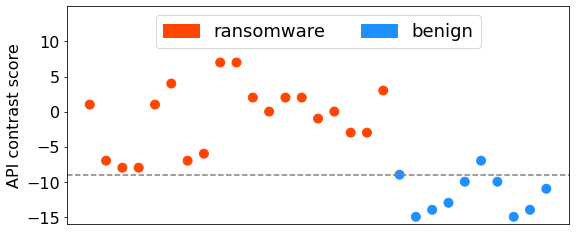}
  \captionof{figure}{API contrast scores of ransomware and benign samples.}
  \label{fig:api_score_diff}
\end{minipage}
\end{figure}

\noindent\textbf{Consistency.} When considering the consistent execution pattern, there is a clear separation between ransomware and benign samples. All ransomware samples have a relatively small value, gathering on the left side of the figure (Figures \ref{fig:cos-consistency}, \ref{fig:manh-consistency}, \ref{fig:weight-consistency}, and \ref{fig:euc-consistency}). Specifically, Manhattan-based consistency, frequency-weighted consistency, and Euclidean-based consistency show optimal performance. With a proper threshold, they accurately catch all ransomware cases with only one false positive. Cosine-based consistency has slightly a lower precision of 0.9. The false positive case that appeared in all four settings is Git, which has a period of execution with high consistency, shown in Figure~\ref{fig:git-file-access}. In summary, the consistency in ransomware file-access API usage helps to identify the threats from benign intensive file accesses.

\subsection{Results of API Contrast-based Refinement}
\label{sec:api-score}

We evaluate the effectiveness of API-contrast refinement, specifically how much it help the classification.
The API patterns are distilled from samples from setup 2 and tested on setup 1.
When considering only the top ransomware API score (i.e., only using APIs in sets $\mathbb{R}$ and $\mathbb{O}$, as described in Section~\ref{sec:contrast-score}), all benign samples have a lower score, ranging from 0 to 4. Git, the false positive case generated by all four consistency algorithms, only has a score of 2, helping reduce the possibility of being malicious. Another false positive case produced by the cosine-based consistency is PaintNet, which hits none of these rules and gets a 0. On the other hand, ransomware samples have relatively high scores with a maximum of 11. The scores for all samples are listed in Table~\ref{tab:api_score}.

\begin{table}[]
\centering
\caption{API scores calculated based on the usage of top ransomware APIs and top benign APIs. The calculation process is described in Section~\ref{sec:detect-algo}. RW stands for ransomware. Ransomware API score is calculated based on APIs in sets $\mathbb{R}$ and $\mathbb{O}$. Benign API score is calculated based on APIs in set $\mathbb{B}$. API contrast score is the final score, shown in Figure~\ref{fig:api_score_diff}, from adding benign score to ransomware score. The lower the final score, the more likely the sample is benign.}
\small \begin{tabular}{
>{\columncolor[HTML]{EFEFEF}}c c
>{\columncolor[HTML]{EFEFEF}}c c
>{\columncolor[HTML]{EFEFEF}}c }
\hline
\cellcolor[HTML]{C0C0C0}                     & \cellcolor[HTML]{C0C0C0}\textbf{Sample} & \cellcolor[HTML]{C0C0C0}\textbf{\begin{tabular}[c]{@{}c@{}}RW API\\ Score\end{tabular}} & \cellcolor[HTML]{C0C0C0}\textbf{\begin{tabular}[c]{@{}c@{}}Benign\\ API\\ Score\end{tabular}} & \cellcolor[HTML]{C0C0C0}\textbf{\begin{tabular}[c]{@{}c@{}}API \\ Contrast\\ Score\end{tabular}} \\ \hline
\cellcolor[HTML]{EFEFEF}                     & Hive (47db)                             & 8                                                                                       & -7                                                                                            & 1                                                                                                \\ \hhline{|>{\arrayrulecolor{tgray}}->{\arrayrulecolor{tblack}}|-|-|-|-|} 
\cellcolor[HTML]{EFEFEF}                     & LockBit (a2ad)                          & 8                                                                                       & -15                                                                                           & -7                                                                                               \\ \hhline{|>{\arrayrulecolor{tgray}}->{\arrayrulecolor{tblack}}|-|-|-|-|} 
\cellcolor[HTML]{EFEFEF}                     & LockBit (dec4)                          & 7                                                                                       & -15                                                                                           & -8                                                                                               \\ \hhline{|>{\arrayrulecolor{tgray}}->{\arrayrulecolor{tblack}}|-|-|-|-|} 
\cellcolor[HTML]{EFEFEF}                     & LockFile (2a23)                         & 3                                                                                       & -11                                                                                           & -8                                                                                               \\ \hhline{|>{\arrayrulecolor{tgray}}->{\arrayrulecolor{tblack}}|-|-|-|-|} 
\cellcolor[HTML]{EFEFEF}                     & Ryuk (9eb7)                             & 1                                                                                       & 0                                                                                             & 1                                                                                                \\ \hhline{|>{\arrayrulecolor{tgray}}->{\arrayrulecolor{tblack}}|-|-|-|-|} 
\cellcolor[HTML]{EFEFEF}                     & Ryuk (40b8)                             & 7                                                                                       & -3                                                                                            & 4                                                                                                \\ \hhline{|>{\arrayrulecolor{tgray}}->{\arrayrulecolor{tblack}}|-|-|-|-|} 
\cellcolor[HTML]{EFEFEF}                     & Sodinokibi (9b11)                       & 5                                                                                       & -12                                                                                           & -7                                                                                               \\ \hhline{|>{\arrayrulecolor{tgray}}->{\arrayrulecolor{tblack}}|-|-|-|-|} 
\cellcolor[HTML]{EFEFEF}                     & Sodinokibi (fd16)                       & 6                                                                                       & -12                                                                                           & -6                                                                                               \\ \hhline{|>{\arrayrulecolor{tgray}}->{\arrayrulecolor{tblack}}|-|-|-|-|} 
\cellcolor[HTML]{EFEFEF}                     & VirLock (7a92)                          & 9                                                                                       & -2                                                                                            & 7                                                                                                \\ \hhline{|>{\arrayrulecolor{tgray}}->{\arrayrulecolor{tblack}}|-|-|-|-|} 
\cellcolor[HTML]{EFEFEF}                     & VirLock (f4b1)                          & 11                                                                                      & -4                                                                                            & 7                                                                                                \\ \hhline{|>{\arrayrulecolor{tgray}}->{\arrayrulecolor{tblack}}|-|-|-|-|} 
\cellcolor[HTML]{EFEFEF}                     & MountLocker (5eae)                      & 3                                                                                       & -1                                                                                            & 2                                                                                                \\ \hhline{|>{\arrayrulecolor{tgray}}->{\arrayrulecolor{tblack}}|-|-|-|-|} 
\cellcolor[HTML]{EFEFEF}                     & Karma (6c98)                            & 0                                                                                       & 0                                                                                             & 0                                                                                                \\ \hhline{|>{\arrayrulecolor{tgray}}->{\arrayrulecolor{tblack}}|-|-|-|-|} 
\cellcolor[HTML]{EFEFEF}                     & AvosLocker (7188)                       & 6                                                                                       & -4                                                                                            & 2                                                                                                \\ \hhline{|>{\arrayrulecolor{tgray}}->{\arrayrulecolor{tblack}}|-|-|-|-|} 
\cellcolor[HTML]{EFEFEF}                     & AvosLocker (f810)                       & 6                                                                                       & -4                                                                                            & 2                                                                                                \\ \hhline{|>{\arrayrulecolor{tgray}}->{\arrayrulecolor{tblack}}|-|-|-|-|} 
\cellcolor[HTML]{EFEFEF}                     & Dharma (dc5b)                           & 9                                                                                       & -10                                                                                           & -1                                                                                               \\ \hhline{|>{\arrayrulecolor{tgray}}->{\arrayrulecolor{tblack}}|-|-|-|-|} 
\cellcolor[HTML]{EFEFEF}                     & DoejoCrypt (e044)                       & 0                                                                                       & 0                                                                                             & 0                                                                                                \\ \hhline{|>{\arrayrulecolor{tgray}}->{\arrayrulecolor{tblack}}|-|-|-|-|} 
\cellcolor[HTML]{EFEFEF}                     & Mydoom (dd28)                           & 0                                                                                       & -3                                                                                            & -3                                                                                               \\ \hhline{|>{\arrayrulecolor{tgray}}->{\arrayrulecolor{tblack}}|-|-|-|-|} 
\cellcolor[HTML]{EFEFEF}                     & Sage (ac27)                             & 2                                                                                       & -5                                                                                            & -3                                                                                               \\ \hhline{|>{\arrayrulecolor{tgray}}->{\arrayrulecolor{tblack}}|-|-|-|-|} 
\multirow{-19}{*}{\cellcolor[HTML]{EFEFEF}R} & SunCrypt (759f)                         & 3                                                                                       & 0                                                                                             & 3                                                                                                \\ \hline
\cellcolor[HTML]{EFEFEF}                     & Chrome                                  & 1                                                                                       & -10                                                                                           & -9                                                                                               \\ \hhline{|>{\arrayrulecolor{tgray}}->{\arrayrulecolor{tblack}}|-|-|-|-|} 
\cellcolor[HTML]{EFEFEF}                     & Git                                     & 2                                                                                       & -17                                                                                           & -15                                                                                              \\ \hhline{|>{\arrayrulecolor{tgray}}->{\arrayrulecolor{tblack}}|-|-|-|-|} 
\cellcolor[HTML]{EFEFEF}                     & Notepad++                               & 0                                                                                       & -14                                                                                           & -14                                                                                              \\ \hhline{|>{\arrayrulecolor{tgray}}->{\arrayrulecolor{tblack}}|-|-|-|-|} 
\cellcolor[HTML]{EFEFEF}                     & TeamViewer                              & 4                                                                                       & -17                                                                                           & -13                                                                                              \\ \hhline{|>{\arrayrulecolor{tgray}}->{\arrayrulecolor{tblack}}|-|-|-|-|} 
\cellcolor[HTML]{EFEFEF}                     & Bitdefender                             & 2                                                                                       & -12                                                                                           & -10                                                                                              \\ \hhline{|>{\arrayrulecolor{tgray}}->{\arrayrulecolor{tblack}}|-|-|-|-|} 
\cellcolor[HTML]{EFEFEF}                     & PaintNet                                & 0                                                                                       & -7                                                                                            & -7                                                                                               \\ \hhline{|>{\arrayrulecolor{tgray}}->{\arrayrulecolor{tblack}}|-|-|-|-|} 
\cellcolor[HTML]{EFEFEF}                     & iCloud                                  & 0                                                                                       & -10                                                                                           & -10                                                                                              \\ \hhline{|>{\arrayrulecolor{tgray}}->{\arrayrulecolor{tblack}}|-|-|-|-|} 
\cellcolor[HTML]{EFEFEF}                     & OneDrive                                & 1                                                                                       & -16                                                                                           & -15                                                                                              \\ \hhline{|>{\arrayrulecolor{tgray}}->{\arrayrulecolor{tblack}}|-|-|-|-|} 
\cellcolor[HTML]{EFEFEF}                     & Skype                                   & 2                                                                                       & -16                                                                                           & -14                                                                                              \\ \hhline{|>{\arrayrulecolor{tgray}}->{\arrayrulecolor{tblack}}|-|-|-|-|} 
\multirow{-10}{*}{\cellcolor[HTML]{EFEFEF}B} & ScreenSplit                             & 3                                                                                       & -14                                                                                           & -11                                                                                              \\ \hline
\end{tabular}
\label{tab:api_score}
\end{table}

Comparably, for the top benign API score (i.e., only using APIs in set $\mathbb{B}$), the benign samples tend to cluster within the range of -7 to -17 (Table~\ref{tab:api_score}). The false positive sample, Git, has a score of -17, which is the lowest among all samples, implying benignness. However, the scores of ransomware are rather spread out, with a lower bound of -15, overlapping with the benign range.

The API contrast score, which takes advantage of both scores discussed above, demonstrates the most promising performance. All benign samples have a score of at most -7, gathering at the lower area (in blue) in Figure~\ref{fig:api_score_diff}. Git has a score of -15, falling in the range of benign. Filtering all positive predictions from Manhattan-based with a threshold of -10, the benign case Git is separated out while the decision on all ransomware cases remains unchanged, helping further boost the precision. The refinement outcome is the same for frequency-weighted and Euclidean-based consistency metrics. 

On top of the consistency-based classification, the API contrast score helps further evaluate the risk and reduce false positives. However, this stage of detection could be evaded by sophisticated attacks. We acknowledge and discuss the limitations of it in Section~\ref{sec:discussion}.

\subsection{Case Study of Most Informative APIs}
\label{sec:cases}

To further investigate the most informative APIs for revealing ransomware behaviors, we conduct a feature importance analysis based on a random forest model (using setup 2). The trained model achieves 0.99 ransomware class recall and precision using API call frequencies during the whole execution. We then manually analyze the top 29 APIs (Table~\ref{tab:top29} in the appendix) with their usage context and compare with benign application statistics. Next, we present a few detailed case studies of the APIs, differentiating the usage between ransomware and benign samples.

\noindent \textbf{Persistence (\texttt{NtDeleteKey})}. We observe that the lack of registry key deletion is a feature of ransomware maintaining persistence after the attack. Windows registry is a database that keeps important information related to the operation of the system and services running in the system.

\begin{itemize}
    \item \textbf{Ransomware}: Ransomware creates a registry value in the “Run” subkey for auto-launching after the system reboots. In our experiments, we observed that 343 ransomware samples called \texttt{NtOpenKey} and 71 called \texttt{NtCreateKey}, whereas only 4 called \texttt{NtDeleteKey}.
    \item \textbf{Benign}: On the opposite, most benign software calls \texttt{NtDeleteKey} at the end of execution to delete any registry keys they opened (via \texttt{NtOpenKey}) or created (via \texttt{NtCreateKey}). Among 330 benign applications that called \texttt{NtCreateKey}, 319 of them called \texttt{NtOpenKey}, and 310 called \texttt{NtDeleteKey}. A few exceptions of benign applications also exist, such as Norton and Viber, which launch at Windows startup and thus keep their registry keys.
\end{itemize}

\noindent \textbf{Kernel security driver access (\texttt{DeviceIoControl})}. Our experiments show that ransomware often uses control code “3735560”, which is related to the kernel security driver, when calling \texttt{DeviceIoControl} API. \texttt{DeviceIoControl} is used by programs to interact with device drivers in the system. Sending a specific control code will cause the corresponding driver to perform corresponding actions.

\begin{itemize}
    \begin{sloppypar}
    \item \textbf{Ransomware}: 149 out of 348 ransomware samples invoked \texttt{DeviceIoControl} during execution. Among those samples, 79\% samples used 3735560 (0x390008 in hexadecimal) for control code, in which 0x39 corresponds to the macros “\texttt{IOCTL\_KSEC\_RANDOM\_FILL\_BUFFER}”, or “\texttt{IOCTL\_KSEC\_RNG\_REKEY}”.  “KSEC” stands for Kernal SECurity. The driver contains security and crypto-related functions, which are potentially used by ransomware samples for encryption key generation.
    \end{sloppypar}
    \item \textbf{Benign}: 321 out of 330 benign software invoked \texttt{DeviceIoControl}. Only 40\% of them used control code 3735560. The majority of benign samples invoked the API with code 589916 (0x9005C) and 590016 (0x900C0), with 0x9 referring to the file system (FILE\_DEVICE\_FILE\_SYSTEM).
\end{itemize}

Besides, foreground-related API invocation differences may also serve as useful features for classification. For example, ransomware checks currently active program less frequently than benign ones, mainly for anti-analysis purposes. Only 82 ransomware samples (24\%) used the \texttt{GetForegroundWindow} API, among which around 70\% of samples belong to only a few families (i.e., LockBit, Stop, Ryuk, and Venus). In benign scenarios, We observe 329 benign applications invoked \texttt{GetForegroundWindow}. Another helpful API is \texttt{DrawTextExW}, which is for front-end formatting. Only 59 of 348 ransomware samples (17\%) call \texttt{DrawTextExW} with low frequency. On contrast, 329 out of 330 benign samples make use of it, with an average call of 2256 times and a high of 49000.

\subsection{Comparison with Benign File Operations} 
In this section, we compare several intensive file operations for benign purposes with ransomware. Benign software can also be designed to handle a significant amount of file operations, such as backup. To investigate how to differentiate such benign behaviors from ransomware, we manually run a set of file operations using the 7zip file manager. The operations we perform include compressing, copying, moving, encrypting, extracting, and deleting. Each operation is perform on 1000 to 3000 files in at least 3 distinct directories. The encryption algorithm used is AES-256.

While some file operations have a repetitive execution period, they can be distinguished from malicious behaviors in a few ways (Figure~\ref{fig:benign-file-access} in the appendix). Copying (Figure~\ref{fig:copy}) and moving (Figure~\ref{fig:move}) a large number of files show a few repetitive execution windows. However, the API composition is simpler during these periods, consisting of only 3 APIs, while typical ransomware execution uses 6 to 10. For extracting files (Figure~\ref{fig:extract}), the vast majority of calls were made to a single API, namely \texttt{NtWriteFile}. The execution process of compression (Figure~\ref{fig:compress}) and encryption (Figure~\ref{fig:encrypt}) has a 2-stage pattern, calling \texttt{NtOpenFile} and \texttt{NtQueryDirectoryFile} first for preparation and then using \texttt{NtReadFile}, \texttt{NtCreateFile}, and \texttt{GetFileInformationByHandle} for the operations. Deleting also has this 2-stage feature with the usage of a different set of APIs (Figure~\ref{fig:delete}). The repetitive period lasts longer in the case of deletion, but the call frequency is also lower for each second.

On the other hand, when looking at the top ransomware and benign API usage scores (described in Sections~\ref{sec:detect-algo} and \ref{sec:api-score}), we can also distinguish the file operations from malicious behaviors. The highest ransomware API score among all 6 operations is 1 (out of 18), suggesting the benignness. None of the operations used any of the top prevalent ransomware APIs. Only 2 of them exceed the frequency threshold of \texttt{RegDeleteValueW}. Furthermore, when looking at the top benign APIs, the scores range from -13 to -14, falling in the cluster of benign programs (benign scores are shown in Table~\ref{tab:api_score}).

In summary, with further inspection, it is possible to separate benign file accesses from malicious ransomware behaviors. However, execution patterns depend on the implementation of specific programs and ransomware could evolve to mimic benign behaviors. Therefore, while being helpful, those observations might not generalize to all cases.

\begin{figure*}

     \centering
     \begin{subfigure}[b]{0.32\textwidth}
         \centering
         \includegraphics[width=\textwidth]{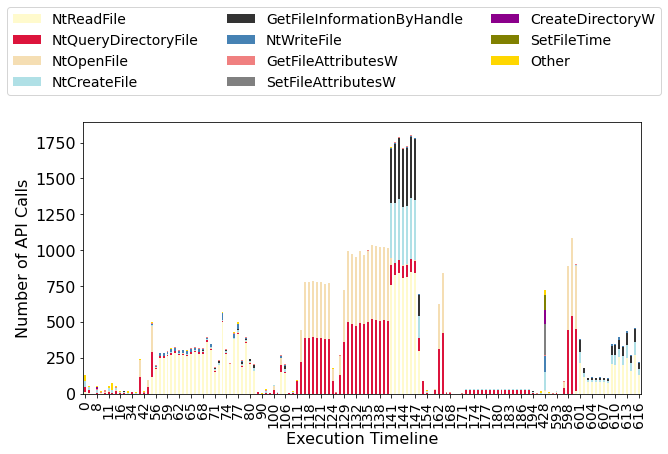}
         \vspace*{-4mm}
         \caption{compress} 
         \label{fig:compress}
     \end{subfigure}
      \begin{subfigure}[b]{0.3\textwidth}
         \centering
         \includegraphics[width=\textwidth]{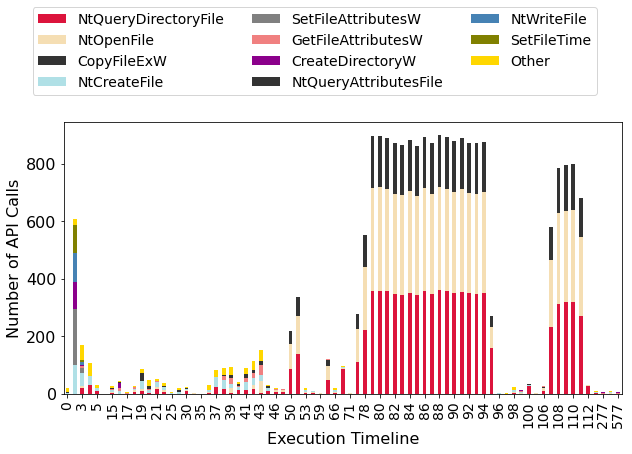}
         \vspace*{-4mm}
         \caption{copy} 
         \label{fig:copy}
     \end{subfigure}
     
     \begin{subfigure}[b]{0.3\textwidth}
         \centering
         \includegraphics[width=\textwidth]{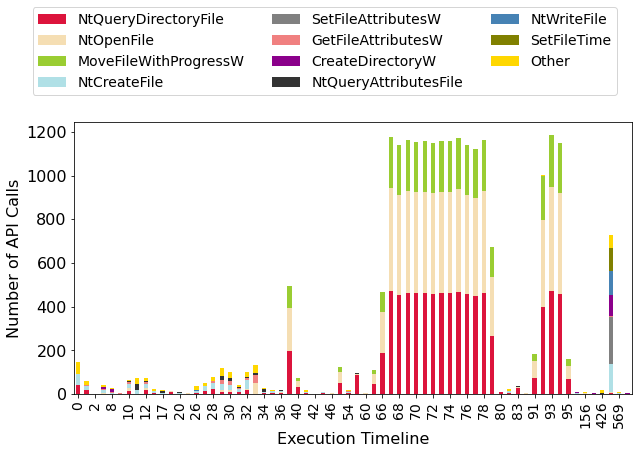}
         \vspace*{-4mm}
         \caption{move}
         \label{fig:move}
     \end{subfigure}
     \begin{subfigure}[b]{0.3\textwidth}
         \includegraphics[width=\textwidth]{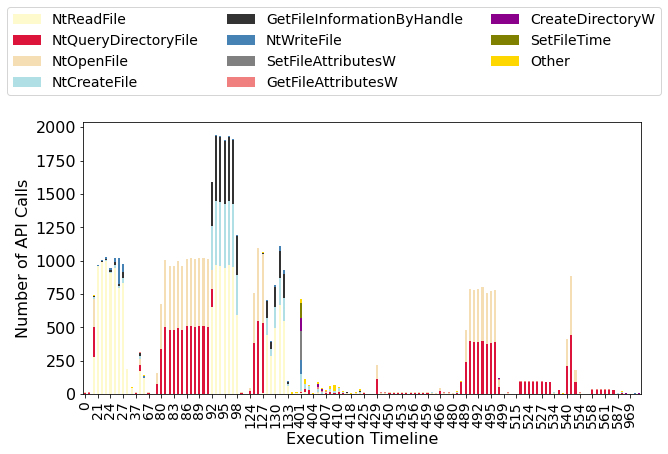}
         \vspace*{-4mm}
         \caption{encrypt} 
         \label{fig:encrypt}
     \end{subfigure}
     
      \begin{subfigure}[b]{0.3\textwidth}
         \includegraphics[width=\textwidth]{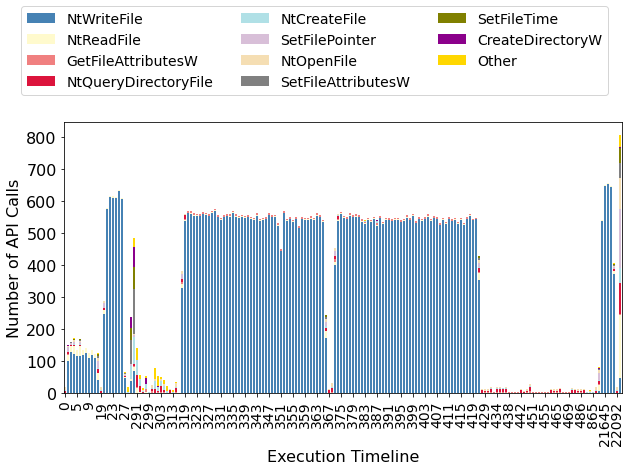}
         \vspace*{-4mm}
         \caption{extract} 
         \label{fig:extract}
     \end{subfigure}
     \begin{subfigure}[b]{0.33\textwidth}
         \includegraphics[width=\textwidth]{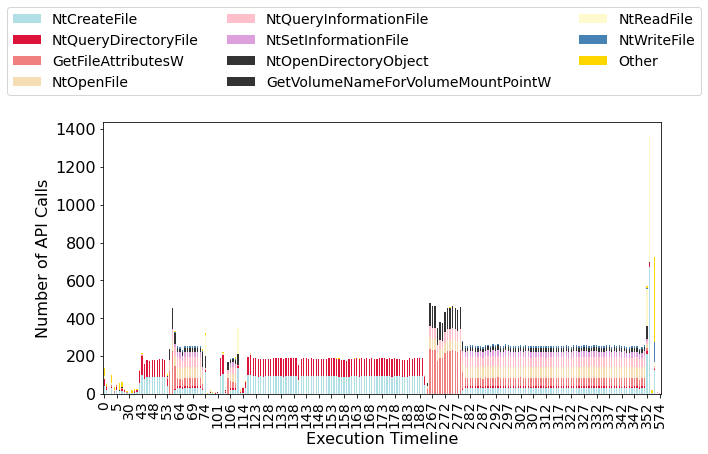}
         \vspace*{-4mm}
         \caption{delete} 
         \label{fig:delete}
     \end{subfigure}

\caption{File-related API call statistics of benign file operations (performed using the 7zip file manager). The same colors represent the same APIs across subfigures.}
\label{fig:benign-file-access}
\end{figure*}

\hfill \break 
\textbf{We summarize our experimental findings as follows}:
\begin{itemize}
    \item Ransomware has extremely high crypto API and file API usage frequency during execution, up to 5,000 times a second, with a unique high repetition pattern throughout the execution process, which we leverage for detection.
   \item The usage patterns of specific APIs exhibit disparities between ransomware and benign programs. This variation helps our classification further evaluate the risk.
   \begin{sloppypar}
 \item Our feasibility study shows that our two-stage API-based classification achieves perfect precision and recall in differentiating ransomware execution from benign. Our consistency-based detection, capturing the unique ransomware file API usage features, effectively recognizes all malicious execution during evaluation, with only 1 false alarm. The API contrast score, as a second step, successfully filters the false positives.
 \end{sloppypar}
  \item For commercial defense, the efficacy and comprehensiveness of decryptors are very limited, with only 1 out of 6 tested decryptors successfully recovering infected files. Malware scanners use static signature-based detection. 69\% complex password-protected ransomware samples evade the detection of all 70 malware scanners tested. Anti-virus software detects generic malware behaviors, overlooking the most essential ransomware encryption actions.
\end{itemize}

\section{Discussion and Limitations}
\label{sec:discussion}

\noindent\textbf{Deployability of API-based solutions.} While API-based solutions do not require low-level system modifications, one challenge is the runtime overhead caused by API instrumentation. Previous work~\cite{maffia2021longitudinal} reported that instrumenting over 300 APIs causes an average of 115\% runtime overhead compared to uninstrumented baseline. This notable slowdown might demotivate users from applying such protection. Possible worksarounds include selective API instrumentation and selective monitoring. Selective API instrumentation refers to only instrument a small set of the most important APIs. Research has shown that some APIs take more significant roles in the classification task~\cite{api-2023}. We also discuss some top informative APIs in Section~\ref{sec:cases}. Selective monitoring refers to only monitor a program exclusively during predetermined time intervals, such as only in the evenings. The length and frequency of monitoring time could also be determined based on the program’s trust level.

\noindent\textbf{Defense evasion.} Newly emerged evasion strategies~\cite{de2022evading,fadadu2020evading} render many existing ransomware detections less effective. They split the malicious program into several subprocesses and make each process only do the small portion of the tasks. In this way, the traces of each process will contain less information about the attack and thus less likely to be detected. However, spliting and creating too many process could also be a flag. Monitoring the process hierarchy and analyzing subprocesses created by the same parent could also help identify malicious patterns. Imitation-based attack against kernel-level I/O analysis has also been crafted in recent years~\cite{zhou2023limits}. Multi-layer protection, i.e., incorporating API monitoring into I/O based detection, was purposed by the authors as a possible counter-measurement.

\noindent\textbf{Detecting core ransomware-specific features.} 
Static signature-based detection, while effective in catching known malware samples, falls short in detecting any variants. For behavioral detection, although there are many prevalent features in ransomware executions, such as prevention of recovery, none of them is required for launching an attack and keeping the data hostage, i.e., not the core behaviors of ransomware and can evolve at any time. However, in our experiments, none of the antivirus software reacts to intensive encryption or file modification. SonicWall reports that~\cite{9attempts}, on average, 9.7 ransomware attacks are attempted every business day for a company running in the US. Although generic detection catches many threats, with such a huge base number, even a small escape rate could cause big problems. 

Our classification focuses on identifying the unique execution pattern during the rapid encryption of data. This feature holds for the majority of current ransomware, as encryption is the core of ransomware and slower procedures increase the likelihood of detection and termination. Our method may miss newly emerged samples that exhibit significantly different functionalities, which is a limitation shared by feature-based detection approaches. However, because encryption is the fundamental nature of crypto-ransomware and is not easily discardable, we believe that it will remain a key feature of ransomware attacks.

Moreover, it is worth noting that different techniques and implementations can be used to achieve a single function, resulting in different traces for analysis. Strategies built using specific traces may not generalize well to other malicious samples or families. In our consistency analysis, we address this issue by distilling the core repetitive patterns from the execution. This enables the identification of the encryption process utilizing diverse API combinations.

\noindent\textbf{Limitations.} Our work has several limitations due to the usage of sandbox and VM. The sandbox cannot catch the dynamically resolved APIs and I/O through memory mapping, resulting in several reports with only a few records, even when we observe numerous infected files in the system. We were also unable to retrieve the memory access behaviors, which will be our future work. In many reports, hundreds of records have exactly the same timestamp, making it impossible to calculate the inter-arrival time of some API sequences. Thus, we only report the call frequencies per second. In some cases, the time interval between the first and last API call in the report exceeds the total execution time. In these cases, we omit the exceeding part. 

Our second limitation involves the possibility of new attack strategies evading our rule- and statistic-based detection. Our second stage API contrast score, despite being helpful, considers only non-file APIs and should not be used alone. The reason is that some behaviors, while common, are not essential for launching a ransomware attack. The hackers can selectively discard or incorporate specific functions to bypass this detection stage. Like other detection works relying on specific API usage, the effect of system updates and environment changes needs to be carefully handled. 

Additionally, in our implementation, we manually set the thresholds for selecting APIs. This poses another challenge to maintain detection accuracy against evolving attack patterns. Similar to various machine learning models that require periodic retraining to address concept drift, our decision thresholds will require future adjustments to adapt to emerging attacks. Alternative thresholds or additional selection criteria should also be explored in the future to enhance the effectiveness.

\section{Conclusion}
\label{sec:conclusion}


Development of deployable ransomware protection is crucial for future research. Our work focuses on discussing the deployability of existing solutions and providing insights on API detections based on ransomware behaviors profiling. We also report new insights from our in-depth API case studies and the evaluation of commercial defenses, which previously have not been reported in the literature. Ongoing work is focused on addressing overhead-related deployment challenges. 

\begin{acks}
This work has been supported by the Office of Naval Research under Grant N00014-22-1-2057.
\end{acks}

\bibliographystyle{ACM-Reference-Format}
\bibliography{ref}


\begin{thebibliography}{170}


\ifx \showCODEN    \undefined \def \showCODEN     #1{\unskip}     \fi
\ifx \showDOI      \undefined \def \showDOI       #1{#1}\fi
\ifx \showISBNx    \undefined \def \showISBNx     #1{\unskip}     \fi
\ifx \showISBNxiii \undefined \def \showISBNxiii  #1{\unskip}     \fi
\ifx \showISSN     \undefined \def \showISSN      #1{\unskip}     \fi
\ifx \showLCCN     \undefined \def \showLCCN      #1{\unskip}     \fi
\ifx \shownote     \undefined \def \shownote      #1{#1}          \fi
\ifx \showarticletitle \undefined \def \showarticletitle #1{#1}   \fi
\ifx \showURL      \undefined \def \showURL       {\relax}        \fi
\providecommand\bibfield[2]{#2}
\providecommand\bibinfo[2]{#2}
\providecommand\natexlab[1]{#1}
\providecommand\showeprint[2][]{arXiv:#2}

\bibitem[Abraham and George(2019)]%
        {abraham2019survey}
\bibfield{author}{\bibinfo{person}{Jitti~Annie Abraham} {and} \bibinfo{person}{Susan~M George}.} \bibinfo{year}{2019}\natexlab{}.
\newblock \showarticletitle{A survey on preventing crypto ransomware using machine learning}. In \bibinfo{booktitle}{\emph{2019 2nd International Conference on Intelligent Computing, Instrumentation and Control Technologies (ICICICT)}}, Vol.~\bibinfo{volume}{1}. IEEE, \bibinfo{pages}{259--263}.
\newblock


\bibitem[Adams and MacKay(2007)]%
        {changepoint}
\bibfield{author}{\bibinfo{person}{Ryan~Prescott Adams} {and} \bibinfo{person}{David J.~C. MacKay}.} \bibinfo{year}{2007}\natexlab{}.
\newblock \bibinfo{title}{Bayesian Online Changepoint Detection}.
\newblock \bibinfo{howpublished}{\url{https://arxiv.org/abs/0710.3742}}.
\newblock


\bibitem[Ahmadian and Shahriari(2016)]%
        {ahmadian20162entfox}
\bibfield{author}{\bibinfo{person}{Mohammad~Mehdi Ahmadian} {and} \bibinfo{person}{Hamid~Reza Shahriari}.} \bibinfo{year}{2016}\natexlab{}.
\newblock \showarticletitle{2entFOX: A framework for high survivable ransomwares detection}. In \bibinfo{booktitle}{\emph{2016 13th international iranian society of cryptology conference on information security and cryptology (ISCISC)}}. IEEE, \bibinfo{pages}{79--84}.
\newblock


\bibitem[Ahmadian et~al\mbox{.}(2015)]%
        {ahmadian2015connection}
\bibfield{author}{\bibinfo{person}{Mohammad~Mehdi Ahmadian}, \bibinfo{person}{Hamid~Reza Shahriari}, {and} \bibinfo{person}{Seyed~Mohammad Ghaffarian}.} \bibinfo{year}{2015}\natexlab{}.
\newblock \showarticletitle{Connection-monitor \& connection-breaker: A novel approach for prevention and detection of high survivable ransomwares}. In \bibinfo{booktitle}{\emph{2015 12th International iranian society of cryptology conference on information security and cryptology (ISCISC)}}. IEEE, \bibinfo{pages}{79--84}.
\newblock


\bibitem[Ahmed et~al\mbox{.}(2021)]%
        {peeler}
\bibfield{author}{\bibinfo{person}{Muhammad~Ejaz Ahmed}, \bibinfo{person}{Hyoungshick Kim}, \bibinfo{person}{Seyit Camtepe}, {and} \bibinfo{person}{Surya Nepal}.} \bibinfo{year}{2021}\natexlab{}.
\newblock \showarticletitle{Peeler: Profiling Kernel-Level Events to Detect Ransomware}. In \bibinfo{booktitle}{\emph{Computer Security -- ESORICS 2021}}, \bibfield{editor}{\bibinfo{person}{Elisa Bertino}, \bibinfo{person}{Haya Shulman}, {and} \bibinfo{person}{Michael Waidner}} (Eds.). \bibinfo{address}{Cham}, \bibinfo{pages}{240--260}.
\newblock
\showISBNx{978-3-030-88418-5}


\bibitem[Ahmed et~al\mbox{.}(2022)]%
        {ahmed2022mitigating}
\bibfield{author}{\bibinfo{person}{Usman Ahmed}, \bibinfo{person}{Jerry Chun-Wei Lin}, {and} \bibinfo{person}{Gautam Srivastava}.} \bibinfo{year}{2022}\natexlab{}.
\newblock \showarticletitle{Mitigating adversarial evasion attacks of ransomware using ensemble learning}.
\newblock \bibinfo{journal}{\emph{Computers and Electrical Engineering}}  \bibinfo{volume}{100} (\bibinfo{year}{2022}), \bibinfo{pages}{107903}.
\newblock


\bibitem[Ahmed et~al\mbox{.}(2020a)]%
        {ahmed2020automated}
\bibfield{author}{\bibinfo{person}{Yahye~Abukar Ahmed}, \bibinfo{person}{Baris Kocer}, {and} \bibinfo{person}{Bander Ali~Saleh Al-rimy}.} \bibinfo{year}{2020}\natexlab{a}.
\newblock \showarticletitle{Automated analysis approach for the detection of high survivable ransomware}.
\newblock \bibinfo{journal}{\emph{KSII Transactions on Internet and Information Systems (TIIS)}} \bibinfo{volume}{14}, \bibinfo{number}{5} (\bibinfo{year}{2020}), \bibinfo{pages}{2236--2257}.
\newblock


\bibitem[Ahmed et~al\mbox{.}(2020b)]%
        {ahmed2020system}
\bibfield{author}{\bibinfo{person}{Yahye~Abukar Ahmed}, \bibinfo{person}{Bar{\i}{\c{s}} Ko{\c{c}}er}, \bibinfo{person}{Shamsul Huda}, \bibinfo{person}{Bander Ali~Saleh Al-rimy}, {and} \bibinfo{person}{Mohammad~Mehedi Hassan}.} \bibinfo{year}{2020}\natexlab{b}.
\newblock \showarticletitle{A system call refinement-based enhanced Minimum Redundancy Maximum Relevance method for ransomware early detection}.
\newblock \bibinfo{journal}{\emph{Journal of Network and Computer Applications}}  \bibinfo{volume}{167} (\bibinfo{year}{2020}), \bibinfo{pages}{102753}.
\newblock


\bibitem[Al-Rimy et~al\mbox{.}(2020)]%
        {al2020pseudo}
\bibfield{author}{\bibinfo{person}{Bander Ali~Saleh Al-Rimy}, \bibinfo{person}{Mohd~Aiziani Maarof}, \bibinfo{person}{Mamoun Alazab}, \bibinfo{person}{Fawaz Alsolami}, \bibinfo{person}{Syed Zainudeen~Mohd Shaid}, \bibinfo{person}{Fuad~A Ghaleb}, \bibinfo{person}{Tawfik Al-Hadhrami}, {and} \bibinfo{person}{Abdullah~Marish Ali}.} \bibinfo{year}{2020}\natexlab{}.
\newblock \showarticletitle{A pseudo feedback-based annotated TF-IDF technique for dynamic crypto-ransomware pre-encryption boundary delineation and features extraction}.
\newblock \bibinfo{journal}{\emph{IEEE Access}}  \bibinfo{volume}{8} (\bibinfo{year}{2020}), \bibinfo{pages}{140586--140598}.
\newblock


\bibitem[Al-Rimy et~al\mbox{.}(2021)]%
        {al2021redundancy}
\bibfield{author}{\bibinfo{person}{Bander Ali~Saleh Al-Rimy}, \bibinfo{person}{Mohd~Aizaini Maarof}, \bibinfo{person}{Mamoun Alazab}, \bibinfo{person}{Syed Zainudeen~Mohd Shaid}, \bibinfo{person}{Fuad~A Ghaleb}, \bibinfo{person}{Abdulmohsen Almalawi}, \bibinfo{person}{Abdullah~Marish Ali}, {and} \bibinfo{person}{Tawfik Al-Hadhrami}.} \bibinfo{year}{2021}\natexlab{}.
\newblock \showarticletitle{Redundancy coefficient gradual up-weighting-based mutual information feature selection technique for crypto-ransomware early detection}.
\newblock \bibinfo{journal}{\emph{Future Generation Computer Systems}}  \bibinfo{volume}{115} (\bibinfo{year}{2021}), \bibinfo{pages}{641--658}.
\newblock


\bibitem[Al-rimy et~al\mbox{.}(2018b)]%
        {al2018zero}
\bibfield{author}{\bibinfo{person}{Bander Ali~Saleh Al-rimy}, \bibinfo{person}{Mohd~Aizaini Maarof}, \bibinfo{person}{Yuli~Adam Prasetyo}, \bibinfo{person}{Syed Zainudeen~Mohd Shaid}, {and} \bibinfo{person}{Asmawi Fadillah~Mohd Ariffin}.} \bibinfo{year}{2018}\natexlab{b}.
\newblock \showarticletitle{Zero-day aware decision fusion-based model for crypto-ransomware early detection}.
\newblock \bibinfo{journal}{\emph{International Journal of Integrated Engineering}} \bibinfo{volume}{10}, \bibinfo{number}{6} (\bibinfo{year}{2018}).
\newblock


\bibitem[Al-rimy et~al\mbox{.}(2018a)]%
        {bander2018}
\bibfield{author}{\bibinfo{person}{Bander Ali~Saleh Al-rimy}, \bibinfo{person}{Mohd~Aizaini Maarof}, {and} \bibinfo{person}{Syed Zainudeen~Mohd Shaid}.} \bibinfo{year}{2018}\natexlab{a}.
\newblock \showarticletitle{Ransomware threat success factors, taxonomy, and countermeasures: A survey and research directions}.
\newblock \bibinfo{journal}{\emph{Computers and Security}}  \bibinfo{volume}{74} (\bibinfo{date}{5} \bibinfo{year}{2018}), \bibinfo{pages}{144--166}.
\newblock
\showISSN{0167-4048}


\bibitem[Al-rimy et~al\mbox{.}(2019)]%
        {ALRIMY2019476}
\bibfield{author}{\bibinfo{person}{Bander Ali~Saleh Al-rimy}, \bibinfo{person}{Mohd~Aizaini Maarof}, {and} \bibinfo{person}{Syed Zainudeen~Mohd Shaid}.} \bibinfo{year}{2019}\natexlab{}.
\newblock \showarticletitle{Crypto-ransomware early detection model using novel incremental bagging with enhanced semi-random subspace selection}.
\newblock \bibinfo{journal}{\emph{Future Generation Computer Systems}}  \bibinfo{volume}{101} (\bibinfo{year}{2019}), \bibinfo{pages}{476--491}.
\newblock
\showISSN{0167-739X}
\urldef\tempurl%
\url{https://doi.org/10.1016/j.future.2019.06.005}
\showDOI{\tempurl}


\bibitem[Alam et~al\mbox{.}(2020)]%
        {alam2020rapper}
\bibfield{author}{\bibinfo{person}{Manaar Alam}, \bibinfo{person}{Sayan Sinha}, \bibinfo{person}{Sarani Bhattacharya}, \bibinfo{person}{Swastika Dutta}, \bibinfo{person}{Debdeep Mukhopadhyay}, {and} \bibinfo{person}{Anupam Chattopadhyay}.} \bibinfo{year}{2020}\natexlab{}.
\newblock \showarticletitle{Rapper: Ransomware prevention via performance counters}.
\newblock \bibinfo{journal}{\emph{arXiv preprint arXiv:2004.01712}} (\bibinfo{year}{2020}).
\newblock


\bibitem[Aldauiji et~al\mbox{.}(2022)]%
        {aldauiji2022utilizing}
\bibfield{author}{\bibinfo{person}{Fatimah Aldauiji}, \bibinfo{person}{Omar Batarfi}, {and} \bibinfo{person}{Manal Bayousef}.} \bibinfo{year}{2022}\natexlab{}.
\newblock \showarticletitle{Utilizing cyber threat hunting techniques to find ransomware attacks: A survey of the state of the art}.
\newblock \bibinfo{journal}{\emph{IEEE Access}}  \bibinfo{volume}{10} (\bibinfo{year}{2022}), \bibinfo{pages}{61695--61706}.
\newblock


\bibitem[Alhawi et~al\mbox{.}(2018)]%
        {Alhawi2018NetCoverse}
\bibfield{author}{\bibinfo{person}{Omar~M.K. Alhawi}, \bibinfo{person}{James Baldwin}, {and} \bibinfo{person}{Ali Dehghantanha}.} \bibinfo{year}{2018}\natexlab{}.
\newblock \showarticletitle{Leveraging machine learning techniques for windows ransomware network traffic detection}.
\newblock \bibinfo{journal}{\emph{Advances in Information Security}}  \bibinfo{volume}{70} (\bibinfo{year}{2018}), \bibinfo{pages}{93--106}.
\newblock
\showISSN{15682633}


\bibitem[Almashhadani et~al\mbox{.}(2019)]%
        {Almashhadani2019multi-classifier}
\bibfield{author}{\bibinfo{person}{Ahmad~O. Almashhadani}, \bibinfo{person}{Mustafa Kaiiali}, \bibinfo{person}{Sakir Sezer}, {and} \bibinfo{person}{Philip O'Kane}.} \bibinfo{year}{2019}\natexlab{}.
\newblock \showarticletitle{A Multi-Classifier Network-Based Crypto Ransomware Detection System: A Case Study of {L}ocky Ransomware}.
\newblock \bibinfo{journal}{\emph{IEEE Access}}  \bibinfo{volume}{7} (\bibinfo{year}{2019}), \bibinfo{pages}{47053--47067}.
\newblock
\showISSN{21693536}


\bibitem[Almousa et~al\mbox{.}(2021)]%
        {Api-based}
\bibfield{author}{\bibinfo{person}{May Almousa}, \bibinfo{person}{Sai Basavaraju}, {and} \bibinfo{person}{Mohd Anwar}.} \bibinfo{year}{2021}\natexlab{}.
\newblock \showarticletitle{{API}-Based Ransomware Detection Using Machine Learning-Based Threat Detection Models}. In \bibinfo{booktitle}{\emph{2021 18th International Conference on Privacy, Security and Trust (PST)}}. \bibinfo{pages}{1--7}.
\newblock


\bibitem[Alqahtani and Sheldon(2022)]%
        {alqahtani2022survey}
\bibfield{author}{\bibinfo{person}{Abdullah Alqahtani} {and} \bibinfo{person}{Frederick~T Sheldon}.} \bibinfo{year}{2022}\natexlab{}.
\newblock \showarticletitle{A survey of crypto ransomware attack detection methodologies: an evolving outlook}.
\newblock \bibinfo{journal}{\emph{Sensors}} \bibinfo{volume}{22}, \bibinfo{number}{5} (\bibinfo{year}{2022}), \bibinfo{pages}{1837}.
\newblock


\bibitem[Alsoghyer and Almomani(2019)]%
        {electronics8080868}
\bibfield{author}{\bibinfo{person}{Samah Alsoghyer} {and} \bibinfo{person}{Iman Almomani}.} \bibinfo{year}{2019}\natexlab{}.
\newblock \showarticletitle{Ransomware Detection System for Android Applications}.
\newblock \bibinfo{journal}{\emph{Electronics}} \bibinfo{volume}{8}, \bibinfo{number}{8} (\bibinfo{year}{2019}).
\newblock
\showISSN{2079-9292}
\urldef\tempurl%
\url{https://doi.org/10.3390/electronics8080868}
\showDOI{\tempurl}


\bibitem[Alsoghyer and Almomani(2020)]%
        {alsoghyer2020effectiveness}
\bibfield{author}{\bibinfo{person}{Samah Alsoghyer} {and} \bibinfo{person}{Iman Almomani}.} \bibinfo{year}{2020}\natexlab{}.
\newblock \showarticletitle{On the effectiveness of application permissions for Android ransomware detection}. In \bibinfo{booktitle}{\emph{2020 6th conference on data science and machine learning applications (CDMA)}}. IEEE, \bibinfo{pages}{94--99}.
\newblock


\bibitem[Alwashali et~al\mbox{.}(2021)]%
        {ali2021}
\bibfield{author}{\bibinfo{person}{Ali Ahmed Mohammed~Ali Alwashali}, \bibinfo{person}{Nor Azlina~Abd Rahman}, {and} \bibinfo{person}{Noris Ismail}.} \bibinfo{year}{2021}\natexlab{}.
\newblock \showarticletitle{A Survey of Ransomware as a Service ({RaaS}) and Methods to Mitigate the Attack}. In \bibinfo{booktitle}{\emph{Proceedings of the International Conference on Developments in eSystems Engineering, DeSE}}, Vol.~\bibinfo{volume}{2021-December}. \bibinfo{pages}{92--96}.
\newblock
\showISBNx{9781665408882}
\showISSN{21611343}


\bibitem[Alzahrani et~al\mbox{.}(2019)]%
        {randetector}
\bibfield{author}{\bibinfo{person}{Abdulrahman Alzahrani}, \bibinfo{person}{Hani Alshahrani}, \bibinfo{person}{Ali Alshehri}, {and} \bibinfo{person}{Huirong Fu}.} \bibinfo{year}{2019}\natexlab{}.
\newblock \showarticletitle{An Intelligent Behavior-Based Ransomware Detection System For Android Platform}. In \bibinfo{booktitle}{\emph{2019 First IEEE International Conference on Trust, Privacy and Security in Intelligent Systems and Applications (TPS-ISA)}}. \bibinfo{pages}{28--35}.
\newblock
\urldef\tempurl%
\url{https://doi.org/10.1109/TPS-ISA48467.2019.00013}
\showDOI{\tempurl}


\bibitem[Alzahrani et~al\mbox{.}(2018)]%
        {alzahrani2018randroid}
\bibfield{author}{\bibinfo{person}{Abdulrahman Alzahrani}, \bibinfo{person}{Ali Alshehri}, \bibinfo{person}{Hani Alshahrani}, \bibinfo{person}{Raed Alharthi}, \bibinfo{person}{Huirong Fu}, \bibinfo{person}{Anyi Liu}, {and} \bibinfo{person}{Ye Zhu}.} \bibinfo{year}{2018}\natexlab{}.
\newblock \showarticletitle{RanDroid: Structural similarity approach for detecting ransomware applications in Android platform}. In \bibinfo{booktitle}{\emph{2018 IEEE International Conference on Electro/Information Technology (EIT)}}. IEEE, \bibinfo{pages}{0892--0897}.
\newblock


\bibitem[Alzahrani and Alghazzawi(2019)]%
        {alzahrani2019review}
\bibfield{author}{\bibinfo{person}{Nisreen Alzahrani} {and} \bibinfo{person}{Daniyal Alghazzawi}.} \bibinfo{year}{2019}\natexlab{}.
\newblock \showarticletitle{A review on android ransomware detection using deep learning techniques}. In \bibinfo{booktitle}{\emph{Proceedings of the 11th international conference on management of digital EcoSystems}}. \bibinfo{pages}{330--335}.
\newblock


\bibitem[Andronio et~al\mbox{.}(2015)]%
        {andronio2015heldroid}
\bibfield{author}{\bibinfo{person}{Nicol{\'o} Andronio}, \bibinfo{person}{Stefano Zanero}, {and} \bibinfo{person}{Federico Maggi}.} \bibinfo{year}{2015}\natexlab{}.
\newblock \showarticletitle{Heldroid: Dissecting and detecting mobile ransomware}. In \bibinfo{booktitle}{\emph{Research in Attacks, Intrusions, and Defenses: 18th International Symposium, RAID 2015, Kyoto, Japan, November 2-4, 2015. Proceedings 18}}. Springer, \bibinfo{pages}{382--404}.
\newblock


\bibitem[Aurangzeb et~al\mbox{.}(2017)]%
        {aurangzeb2017ransomware}
\bibfield{author}{\bibinfo{person}{Sana Aurangzeb}, \bibinfo{person}{Muhammad Aleem}, \bibinfo{person}{Muhammad~Azhar Iqbal}, \bibinfo{person}{Muhammad~Arshad Islam}, {et~al\mbox{.}}} \bibinfo{year}{2017}\natexlab{}.
\newblock \showarticletitle{Ransomware: a survey and trends}.
\newblock \bibinfo{journal}{\emph{Journal of Information Assurance \& Security}} \bibinfo{volume}{6}, \bibinfo{number}{2} (\bibinfo{year}{2017}), \bibinfo{pages}{48--58}.
\newblock


\bibitem[Aurangzeb et~al\mbox{.}(2022)]%
        {aurangzeb2022bigrc}
\bibfield{author}{\bibinfo{person}{Sana Aurangzeb}, \bibinfo{person}{Haris Anwar}, \bibinfo{person}{Muhammad~Asif Naeem}, {and} \bibinfo{person}{Muhammad Aleem}.} \bibinfo{year}{2022}\natexlab{}.
\newblock \showarticletitle{BigRC-EML: big-data based ransomware classification using ensemble machine learning}.
\newblock \bibinfo{journal}{\emph{Cluster Computing}} \bibinfo{volume}{25}, \bibinfo{number}{5} (\bibinfo{year}{2022}), \bibinfo{pages}{3405--3422}.
\newblock


\bibitem[Aurangzeb et~al\mbox{.}(2021)]%
        {hardware-profile}
\bibfield{author}{\bibinfo{person}{Sana Aurangzeb}, \bibinfo{person}{Rao Naveed~Bin Rais}, \bibinfo{person}{Muhammad Aleem}, \bibinfo{person}{Muhammad~Arshad Islam}, {and} \bibinfo{person}{Muhammad~Azhar Iqbal}.} \bibinfo{year}{2021}\natexlab{}.
\newblock \showarticletitle{On the classification of {M}icrosoft-{W}indows ransomware using hardware profile}.
\newblock \bibinfo{journal}{\emph{PeerJ Computer Science}}  \bibinfo{volume}{7} (\bibinfo{date}{2} \bibinfo{year}{2021}).
\newblock
\showISSN{23765992}


\bibitem[Ayub et~al\mbox{.}(2020)]%
        {ayub2020request}
\bibfield{author}{\bibinfo{person}{Md~Ahsan Ayub}, \bibinfo{person}{Andrea Continella}, {and} \bibinfo{person}{Ambareen Siraj}.} \bibinfo{year}{2020}\natexlab{}.
\newblock \showarticletitle{An {I/O} request packet ({IRP}) driven effective ransomware detection scheme using artificial neural network}. In \bibinfo{booktitle}{\emph{2020 IEEE 21st International Conference on Information Reuse and Integration for Data Science (IRI)}}. IEEE, \bibinfo{pages}{319--324}.
\newblock


\bibitem[Ayub et~al\mbox{.}(2023)]%
        {ayub2023rwarmor}
\bibfield{author}{\bibinfo{person}{Md~Ahsan Ayub}, \bibinfo{person}{Ambareen Siraj}, \bibinfo{person}{Bobby Filar}, {and} \bibinfo{person}{Maanak Gupta}.} \bibinfo{year}{2023}\natexlab{}.
\newblock \showarticletitle{RWArmor: a static-informed dynamic analysis approach for early detection of cryptographic windows ransomware}.
\newblock \bibinfo{journal}{\emph{International Journal of Information Security}} (\bibinfo{year}{2023}), \bibinfo{pages}{1--24}.
\newblock


\bibitem[Bae et~al\mbox{.}(2020)]%
        {bae2020ransomware}
\bibfield{author}{\bibinfo{person}{Seong~Il Bae}, \bibinfo{person}{Gyu~Bin Lee}, {and} \bibinfo{person}{Eul~Gyu Im}.} \bibinfo{year}{2020}\natexlab{}.
\newblock \showarticletitle{Ransomware detection using machine learning algorithms}.
\newblock \bibinfo{journal}{\emph{Concurrency and Computation: Practice and Experience}} \bibinfo{volume}{32}, \bibinfo{number}{18} (\bibinfo{year}{2020}), \bibinfo{pages}{e5422}.
\newblock


\bibitem[Baek et~al\mbox{.}(2018)]%
        {ssd-insider}
\bibfield{author}{\bibinfo{person}{Sungha Baek}, \bibinfo{person}{Youngdon Jung}, \bibinfo{person}{Aziz Mohaisen}, \bibinfo{person}{Sungjin Lee}, {and} \bibinfo{person}{Daehun Nyang}.} \bibinfo{year}{2018}\natexlab{}.
\newblock \showarticletitle{{SSD-Insider}: Internal defense of solid-state drive against ransomware with perfect data recovery}. In \bibinfo{booktitle}{\emph{Proceedings of International Conference on Distributed Computing Systems}} (Vienna, Austria). \bibinfo{pages}{875--884}.
\newblock
\showISBNx{9781538668719}


\bibitem[Bajpai et~al\mbox{.}(2018)]%
        {bajpai2018key}
\bibfield{author}{\bibinfo{person}{Pranshu Bajpai}, \bibinfo{person}{Aditya~K Sood}, {and} \bibinfo{person}{Richard Enbody}.} \bibinfo{year}{2018}\natexlab{}.
\newblock \showarticletitle{A key-management-based taxonomy for ransomware}. In \bibinfo{booktitle}{\emph{2018 APWG Symposium on Electronic Crime Research (eCrime)}}. IEEE, \bibinfo{pages}{1--12}.
\newblock


\bibitem[Beaman et~al\mbox{.}(2021)]%
        {Beaman2021}
\bibfield{author}{\bibinfo{person}{Craig Beaman}, \bibinfo{person}{Ashley Barkworth}, \bibinfo{person}{Toluwalope~David Akande}, \bibinfo{person}{Saqib Hakak}, {and} \bibinfo{person}{Muhammad~Khurram Khan}.} \bibinfo{year}{2021}\natexlab{}.
\newblock \showarticletitle{Ransomware: Recent advances, analysis, challenges and future research directions}.
\newblock \bibinfo{journal}{\emph{Computers \& Security}}  \bibinfo{volume}{111} (\bibinfo{date}{12} \bibinfo{year}{2021}).
\newblock
\showISSN{01674048}


\bibitem[Begovic et~al\mbox{.}(2023)]%
        {begovic2023cryptographic}
\bibfield{author}{\bibinfo{person}{Kenan Begovic}, \bibinfo{person}{Abdulaziz Al-Ali}, {and} \bibinfo{person}{Qutaibah Malluhi}.} \bibinfo{year}{2023}\natexlab{}.
\newblock \showarticletitle{Cryptographic ransomware encryption detection: Survey}.
\newblock \bibinfo{journal}{\emph{Computers \& Security}} (\bibinfo{year}{2023}), \bibinfo{pages}{103349}.
\newblock


\bibitem[Benmalek(2024)]%
        {benmalek2024ransomware}
\bibfield{author}{\bibinfo{person}{Mourad Benmalek}.} \bibinfo{year}{2024}\natexlab{}.
\newblock \showarticletitle{Ransomware on cyber-physical systems: Taxonomies, case studies, security gaps, and open challenges}.
\newblock \bibinfo{journal}{\emph{Internet of Things and Cyber-Physical Systems}} (\bibinfo{year}{2024}).
\newblock


\bibitem[Berrueta et~al\mbox{.}(2019)]%
        {berrueta2019survey}
\bibfield{author}{\bibinfo{person}{Eduardo Berrueta}, \bibinfo{person}{Daniel Morato}, \bibinfo{person}{Eduardo Maga{\~n}a}, {and} \bibinfo{person}{Mikel Izal}.} \bibinfo{year}{2019}\natexlab{}.
\newblock \showarticletitle{A survey on detection techniques for cryptographic ransomware}.
\newblock \bibinfo{journal}{\emph{IEEE Access}}  \bibinfo{volume}{7} (\bibinfo{year}{2019}), \bibinfo{pages}{144925--144944}.
\newblock


\bibitem[Bijitha et~al\mbox{.}(2020)]%
        {bijitha2020survey}
\bibfield{author}{\bibinfo{person}{CV Bijitha}, \bibinfo{person}{Rohit Sukumaran}, {and} \bibinfo{person}{Hiran~V Nath}.} \bibinfo{year}{2020}\natexlab{}.
\newblock \showarticletitle{A survey on ransomware detection techniques}. In \bibinfo{booktitle}{\emph{Secure Knowledge Management In Artificial Intelligence Era: 8th International Conference, SKM 2019, Goa, India, December 21--22, 2019, Proceedings 8}}. Springer, \bibinfo{pages}{55--68}.
\newblock


\bibitem[Bortolameotti et~al\mbox{.}(2017)]%
        {bortolameotti2017decanter}
\bibfield{author}{\bibinfo{person}{Riccardo Bortolameotti}, \bibinfo{person}{Thijs van Ede}, \bibinfo{person}{Marco Caselli}, \bibinfo{person}{Maarten~H Everts}, \bibinfo{person}{Pieter Hartel}, \bibinfo{person}{Rick Hofstede}, \bibinfo{person}{Willem Jonker}, {and} \bibinfo{person}{Andreas Peter}.} \bibinfo{year}{2017}\natexlab{}.
\newblock \showarticletitle{Decanter: Detection of anomalous outbound http traffic by passive application fingerprinting}. In \bibinfo{booktitle}{\emph{Proceedings of the 33rd Annual computer security applications Conference}}. \bibinfo{pages}{373--386}.
\newblock


\bibitem[Cabaj et~al\mbox{.}(2018)]%
        {cabaj2018software}
\bibfield{author}{\bibinfo{person}{Krzysztof Cabaj}, \bibinfo{person}{Marcin Gregorczyk}, {and} \bibinfo{person}{Wojciech Mazurczyk}.} \bibinfo{year}{2018}\natexlab{}.
\newblock \showarticletitle{Software-defined networking-based crypto ransomware detection using HTTP traffic characteristics}.
\newblock \bibinfo{journal}{\emph{Computers \& Electrical Engineering}}  \bibinfo{volume}{66} (\bibinfo{year}{2018}), \bibinfo{pages}{353--368}.
\newblock


\bibitem[Cabaj and Mazurczyk(2016)]%
        {Cabaj2016cryptowall}
\bibfield{author}{\bibinfo{person}{Krzysztof Cabaj} {and} \bibinfo{person}{Wojciech Mazurczyk}.} \bibinfo{year}{2016}\natexlab{}.
\newblock \showarticletitle{Using software-defined networking for ransomware mitigation: The case of cryptowall}.
\newblock \bibinfo{journal}{\emph{IEEE Network}}  \bibinfo{volume}{30} (\bibinfo{date}{11} \bibinfo{year}{2016}), \bibinfo{pages}{14--20}.
\newblock
Issue 6.
\showISSN{1558156X}


\bibitem[Castiglione and Pavlovic(2019)]%
        {castiglione2019dynamic}
\bibfield{author}{\bibinfo{person}{Jason Castiglione} {and} \bibinfo{person}{Dusko Pavlovic}.} \bibinfo{year}{2019}\natexlab{}.
\newblock \showarticletitle{Dynamic distributed secure storage against ransomware}.
\newblock \bibinfo{journal}{\emph{IEEE Transactions on computational social systems}} \bibinfo{volume}{7}, \bibinfo{number}{6} (\bibinfo{year}{2019}), \bibinfo{pages}{1469--1475}.
\newblock


\bibitem[Celdr{\'a}n et~al\mbox{.}(2023)]%
        {celdran2023behavioral}
\bibfield{author}{\bibinfo{person}{Alberto~Huertas Celdr{\'a}n}, \bibinfo{person}{Pedro Miguel~S{\'a}nchez S{\'a}nchez}, \bibinfo{person}{Jan von~der Assen}, \bibinfo{person}{Dennis Shushack}, \bibinfo{person}{{\'A}ngel Luis~Perales G{\'o}mez}, \bibinfo{person}{G{\'e}r{\^o}me Bovet}, \bibinfo{person}{Gregorio~Mart{\'\i}nez P{\'e}rez}, {and} \bibinfo{person}{Burkhard Stiller}.} \bibinfo{year}{2023}\natexlab{}.
\newblock \showarticletitle{Behavioral fingerprinting to detect ransomware in resource-constrained devices}.
\newblock \bibinfo{journal}{\emph{Computers \& Security}}  \bibinfo{volume}{135} (\bibinfo{year}{2023}), \bibinfo{pages}{103510}.
\newblock


\bibitem[Cen et~al\mbox{.}(2024)]%
        {cen2024ransomware}
\bibfield{author}{\bibinfo{person}{Mingcan Cen}, \bibinfo{person}{Frank Jiang}, \bibinfo{person}{Xingsheng Qin}, \bibinfo{person}{Qinghong Jiang}, {and} \bibinfo{person}{Robin Doss}.} \bibinfo{year}{2024}\natexlab{}.
\newblock \showarticletitle{Ransomware early detection: A survey}.
\newblock \bibinfo{journal}{\emph{Computer Networks}}  \bibinfo{volume}{239} (\bibinfo{year}{2024}), \bibinfo{pages}{110138}.
\newblock


\bibitem[Chen et~al\mbox{.}(2017b)]%
        {chen2017uncovering}
\bibfield{author}{\bibinfo{person}{Jing Chen}, \bibinfo{person}{Chiheng Wang}, \bibinfo{person}{Ziming Zhao}, \bibinfo{person}{Kai Chen}, \bibinfo{person}{Ruiying Du}, {and} \bibinfo{person}{Gail-Joon Ahn}.} \bibinfo{year}{2017}\natexlab{b}.
\newblock \showarticletitle{Uncovering the face of android ransomware: Characterization and real-time detection}.
\newblock \bibinfo{journal}{\emph{IEEE Transactions on Information Forensics and Security}} \bibinfo{volume}{13}, \bibinfo{number}{5} (\bibinfo{year}{2017}), \bibinfo{pages}{1286--1300}.
\newblock


\bibitem[Chen and Bridges(2017)]%
        {chen2017automated}
\bibfield{author}{\bibinfo{person}{Qian Chen} {and} \bibinfo{person}{Robert~A Bridges}.} \bibinfo{year}{2017}\natexlab{}.
\newblock \showarticletitle{Automated behavioral analysis of malware: A case study of wannacry ransomware}. In \bibinfo{booktitle}{\emph{2017 16th IEEE International Conference on machine learning and applications (ICMLA)}}. IEEE, \bibinfo{pages}{454--460}.
\newblock


\bibitem[Chen et~al\mbox{.}(2019)]%
        {chen2019automated}
\bibfield{author}{\bibinfo{person}{Qian Chen}, \bibinfo{person}{Sheikh~Rabiul Islam}, \bibinfo{person}{Henry Haswell}, {and} \bibinfo{person}{Robert~A Bridges}.} \bibinfo{year}{2019}\natexlab{}.
\newblock \showarticletitle{Automated ransomware behavior analysis: Pattern extraction and early detection}. In \bibinfo{booktitle}{\emph{Science of Cyber Security: Second International Conference, SciSec 2019, Nanjing, China, August 9--11, 2019, Revised Selected Papers 2}}. Springer, \bibinfo{pages}{199--214}.
\newblock


\bibitem[Chen et~al\mbox{.}(2017a)]%
        {call-graph}
\bibfield{author}{\bibinfo{person}{Zhi-Guo Chen}, \bibinfo{person}{Ho-Seok Kang}, \bibinfo{person}{Shang-Nan Yin}, {and} \bibinfo{person}{Sung-Ryul Kim}.} \bibinfo{year}{2017}\natexlab{a}.
\newblock \showarticletitle{Automatic Ransomware Detection and Analysis Based on Dynamic API Calls Flow Graph} \emph{(\bibinfo{series}{RACS '17})}. \bibinfo{pages}{196–201}.
\newblock
\showISBNx{9781450350273}


\bibitem[Cheng et~al\mbox{.}(2019)]%
        {Cheng2019dptCry}
\bibfield{author}{\bibinfo{person}{Guang Cheng}, \bibinfo{person}{Chunsheng Guo}, {and} \bibinfo{person}{Yongning Tang}.} \bibinfo{year}{2019}\natexlab{}.
\newblock \showarticletitle{dpt{C}ry: an approach to decrypting ransomware {W}anna{C}ry based on {API} hooking}.
\newblock \bibinfo{journal}{\emph{CCF Transactions on Networking 2019 2:3}}  \bibinfo{volume}{2} (\bibinfo{date}{12} \bibinfo{year}{2019}), \bibinfo{pages}{207--216}.
\newblock
Issue 3.
\showISSN{2520-8470}


\bibitem[Cimitile et~al\mbox{.}(2018)]%
        {cimitile2018talos}
\bibfield{author}{\bibinfo{person}{Aniello Cimitile}, \bibinfo{person}{Francesco Mercaldo}, \bibinfo{person}{Vittoria Nardone}, \bibinfo{person}{Antonella Santone}, {and} \bibinfo{person}{Corrado~Aaron Visaggio}.} \bibinfo{year}{2018}\natexlab{}.
\newblock \showarticletitle{Talos: no more ransomware victims with formal methods}.
\newblock \bibinfo{journal}{\emph{International Journal of Information Security}}  \bibinfo{volume}{17} (\bibinfo{year}{2018}), \bibinfo{pages}{719--738}.
\newblock


\bibitem[Coglio et~al\mbox{.}(2023)]%
        {api-2023}
\bibfield{author}{\bibinfo{person}{Filippo Coglio}, \bibinfo{person}{Ahmed Lekssays}, \bibinfo{person}{Barbara Carminati}, {and} \bibinfo{person}{Elena Ferrari}.} \bibinfo{year}{2023}\natexlab{}.
\newblock \showarticletitle{Early-Stage Ransomware Detection Based on Pre-attack Internal API Calls}. In \bibinfo{booktitle}{\emph{Advanced Information Networking and Applications}}, \bibfield{editor}{\bibinfo{person}{Leonard Barolli}} (Ed.). \bibinfo{publisher}{Springer International Publishing}, \bibinfo{address}{Cham}, \bibinfo{pages}{417--429}.
\newblock
\showISBNx{978-3-031-28451-9}


\bibitem[Cohen and Nissim(2018)]%
        {cohen2018}
\bibfield{author}{\bibinfo{person}{Aviad Cohen} {and} \bibinfo{person}{Nir Nissim}.} \bibinfo{year}{2018}\natexlab{}.
\newblock \showarticletitle{Trusted detection of ransomware in a private cloud using machine learning methods leveraging meta-features from volatile memory}.
\newblock \bibinfo{journal}{\emph{Expert Systems with Applications}}  \bibinfo{volume}{102} (\bibinfo{year}{2018}), \bibinfo{pages}{158--178}.
\newblock


\bibitem[Continella et~al\mbox{.}(2016)]%
        {ShieldFS}
\bibfield{author}{\bibinfo{person}{Andrea Continella}, \bibinfo{person}{Alessandro Guagnelli}, \bibinfo{person}{Giovanni Zingaro}, \bibinfo{person}{Giulio De~Pasquale}, \bibinfo{person}{Alessandro Barenghi}, \bibinfo{person}{Stefano Zanero}, {and} \bibinfo{person}{Federico Maggi}.} \bibinfo{year}{2016}\natexlab{}.
\newblock \showarticletitle{Shield{FS}: A Self-Healing, Ransomware-Aware Filesystem}. In \bibinfo{booktitle}{\emph{Proceedings of the 32nd Annual Conference on Computer Security Applications}} (Los Angeles, California, USA). \bibinfo{pages}{336–347}.
\newblock
\showISBNx{9781450347716}


\bibitem[Cusack et~al\mbox{.}(2018)]%
        {cusack2018machine}
\bibfield{author}{\bibinfo{person}{Greg Cusack}, \bibinfo{person}{Oliver Michel}, {and} \bibinfo{person}{Eric Keller}.} \bibinfo{year}{2018}\natexlab{}.
\newblock \showarticletitle{Machine learning-based detection of ransomware using SDN}. In \bibinfo{booktitle}{\emph{Proceedings of the 2018 ACM international workshop on security in software defined networks \& network function virtualization}}. \bibinfo{pages}{1--6}.
\newblock


\bibitem[Cuzzocrea et~al\mbox{.}(2018)]%
        {cuzzocrea2018novel}
\bibfield{author}{\bibinfo{person}{Alfredo Cuzzocrea}, \bibinfo{person}{Fabio Martinelli}, {and} \bibinfo{person}{Francesco Mercaldo}.} \bibinfo{year}{2018}\natexlab{}.
\newblock \showarticletitle{A novel structural-entropy-based classification technique for supporting android ransomware detection and analysis}. In \bibinfo{booktitle}{\emph{2018 IEEE International Conference on Fuzzy Systems (FUZZ-IEEE)}}. IEEE, \bibinfo{pages}{1--7}.
\newblock


\bibitem[Dargahi et~al\mbox{.}(2019)]%
        {dargahi2019cyber}
\bibfield{author}{\bibinfo{person}{Tooska Dargahi}, \bibinfo{person}{Ali Dehghantanha}, \bibinfo{person}{Pooneh~Nikkhah Bahrami}, \bibinfo{person}{Mauro Conti}, \bibinfo{person}{Giuseppe Bianchi}, {and} \bibinfo{person}{Loris Benedetto}.} \bibinfo{year}{2019}\natexlab{}.
\newblock \showarticletitle{A Cyber-Kill-Chain based taxonomy of crypto-ransomware features}.
\newblock \bibinfo{journal}{\emph{Journal of Computer Virology and Hacking Techniques}}  \bibinfo{volume}{15} (\bibinfo{year}{2019}), \bibinfo{pages}{277--305}.
\newblock


\bibitem[De~Gaspari et~al\mbox{.}(2022)]%
        {de2022evading}
\bibfield{author}{\bibinfo{person}{Fabio De~Gaspari}, \bibinfo{person}{Dorjan Hitaj}, \bibinfo{person}{Giulio Pagnotta}, \bibinfo{person}{Lorenzo De~Carli}, {and} \bibinfo{person}{Luigi~V Mancini}.} \bibinfo{year}{2022}\natexlab{}.
\newblock \showarticletitle{Evading behavioral classifiers: a comprehensive analysis on evading ransomware detection techniques}.
\newblock \bibinfo{journal}{\emph{Neural Computing and Applications}} \bibinfo{volume}{34}, \bibinfo{number}{14} (\bibinfo{year}{2022}), \bibinfo{pages}{12077--12096}.
\newblock


\bibitem[Desai(2019)]%
        {desai2019survey}
\bibfield{author}{\bibinfo{person}{Usama Desai}.} \bibinfo{year}{2019}\natexlab{}.
\newblock \showarticletitle{A survey on Android ransomware and its detection methods}.
\newblock \bibinfo{journal}{\emph{International Research Journal of Engineering and Technology}} \bibinfo{volume}{6}, \bibinfo{number}{03} (\bibinfo{year}{2019}), \bibinfo{pages}{3081--3087}.
\newblock


\bibitem[Elkhail et~al\mbox{.}(2023)]%
        {elkhail2023seamlessly}
\bibfield{author}{\bibinfo{person}{Abdulrahman~Abu Elkhail}, \bibinfo{person}{Nada Lachtar}, \bibinfo{person}{Duha Ibdah}, \bibinfo{person}{Rustam Aslam}, \bibinfo{person}{Hamza Khan}, \bibinfo{person}{Anys Bacha}, {and} \bibinfo{person}{Hafiz Malik}.} \bibinfo{year}{2023}\natexlab{}.
\newblock \showarticletitle{Seamlessly Safeguarding Data Against Ransomware Attacks}.
\newblock \bibinfo{journal}{\emph{IEEE Transactions on Dependable and Secure Computing}} \bibinfo{volume}{20}, \bibinfo{number}{1} (\bibinfo{year}{2023}), \bibinfo{pages}{1--16}.
\newblock


\bibitem[Fadadu et~al\mbox{.}(2020)]%
        {fadadu2020evading}
\bibfield{author}{\bibinfo{person}{Fenil Fadadu}, \bibinfo{person}{Anand Handa}, \bibinfo{person}{Nitesh Kumar}, {and} \bibinfo{person}{Sandeep~Kumar Shukla}.} \bibinfo{year}{2020}\natexlab{}.
\newblock \showarticletitle{Evading API call sequence based malware classifiers}. In \bibinfo{booktitle}{\emph{Information and Communications Security: 21st International Conference, ICICS 2019, Beijing, China, December 15--17, 2019, Revised Selected Papers 21}}. Springer, \bibinfo{pages}{18--33}.
\newblock


\bibitem[Faris et~al\mbox{.}(2020)]%
        {faris2020optimizing}
\bibfield{author}{\bibinfo{person}{Hossam Faris}, \bibinfo{person}{Maria Habib}, \bibinfo{person}{Iman Almomani}, \bibinfo{person}{Mohammed Eshtay}, {and} \bibinfo{person}{Ibrahim Aljarah}.} \bibinfo{year}{2020}\natexlab{}.
\newblock \showarticletitle{Optimizing extreme learning machines using chains of salps for efficient Android ransomware detection}.
\newblock \bibinfo{journal}{\emph{Applied Sciences}} \bibinfo{volume}{10}, \bibinfo{number}{11} (\bibinfo{year}{2020}), \bibinfo{pages}{3706}.
\newblock


\bibitem[Fernandez~Maimo et~al\mbox{.}(2019)]%
        {fernandez2019intelligent}
\bibfield{author}{\bibinfo{person}{Lorenzo Fernandez~Maimo}, \bibinfo{person}{Alberto Huertas~Celdran}, \bibinfo{person}{Angel~L Perales~Gomez}, \bibinfo{person}{Felix~J Garcia~Clemente}, \bibinfo{person}{James Weimer}, {and} \bibinfo{person}{Insup Lee}.} \bibinfo{year}{2019}\natexlab{}.
\newblock \showarticletitle{Intelligent and dynamic ransomware spread detection and mitigation in integrated clinical environments}.
\newblock \bibinfo{journal}{\emph{Sensors}} \bibinfo{volume}{19}, \bibinfo{number}{5} (\bibinfo{year}{2019}), \bibinfo{pages}{1114}.
\newblock


\bibitem[Fernando and Komninos(2024)]%
        {fernando2024fesad}
\bibfield{author}{\bibinfo{person}{Damien~Warren Fernando} {and} \bibinfo{person}{Nikos Komninos}.} \bibinfo{year}{2024}\natexlab{}.
\newblock \showarticletitle{FeSAD ransomware detection framework with machine learning using adaption to concept drift}.
\newblock \bibinfo{journal}{\emph{Computers \& Security}}  \bibinfo{volume}{137} (\bibinfo{year}{2024}), \bibinfo{pages}{103629}.
\newblock


\bibitem[Ferrante et~al\mbox{.}(2018)]%
        {ferrante2018extinguishing}
\bibfield{author}{\bibinfo{person}{Alberto Ferrante}, \bibinfo{person}{Miroslaw Malek}, \bibinfo{person}{Fabio Martinelli}, \bibinfo{person}{Francesco Mercaldo}, {and} \bibinfo{person}{Jelena Milosevic}.} \bibinfo{year}{2018}\natexlab{}.
\newblock \showarticletitle{Extinguishing ransomware-a hybrid approach to android ransomware detection}. In \bibinfo{booktitle}{\emph{Foundations and Practice of Security: 10th International Symposium, FPS 2017, Nancy, France, October 23-25, 2017, Revised Selected Papers 10}}. Springer, \bibinfo{pages}{242--258}.
\newblock


\bibitem[Garg et~al\mbox{.}(2018)]%
        {garg2018past}
\bibfield{author}{\bibinfo{person}{Dhruv Garg}, \bibinfo{person}{Abha Thakral}, \bibinfo{person}{Tarun Nalwa}, {and} \bibinfo{person}{Tanupriya Choudhury}.} \bibinfo{year}{2018}\natexlab{}.
\newblock \showarticletitle{A past examination and future expectation: Ransomware}. In \bibinfo{booktitle}{\emph{2018 International Conference on Advances in Computing and Communication Engineering (ICACCE)}}. IEEE, \bibinfo{pages}{243--247}.
\newblock


\bibitem[Gazzan and Sheldon(2023)]%
        {gazzan2023enhanced}
\bibfield{author}{\bibinfo{person}{Mazen Gazzan} {and} \bibinfo{person}{Frederick~T Sheldon}.} \bibinfo{year}{2023}\natexlab{}.
\newblock \showarticletitle{An enhanced minimax loss function technique in generative adversarial network for ransomware behavior prediction}.
\newblock \bibinfo{journal}{\emph{Future Internet}} \bibinfo{volume}{15}, \bibinfo{number}{10} (\bibinfo{year}{2023}), \bibinfo{pages}{318}.
\newblock


\bibitem[Gharib and Ghorbani(2017)]%
        {gharib2017dna}
\bibfield{author}{\bibinfo{person}{Amirhossein Gharib} {and} \bibinfo{person}{Ali Ghorbani}.} \bibinfo{year}{2017}\natexlab{}.
\newblock \showarticletitle{Dna-droid: A real-time android ransomware detection framework}. In \bibinfo{booktitle}{\emph{Network and System Security: 11th International Conference, NSS 2017, Helsinki, Finland, August 21--23, 2017, Proceedings 11}}. Springer, \bibinfo{pages}{184--198}.
\newblock


\bibitem[Gonzalez and Hayajneh(2017)]%
        {gonzalez2017detection}
\bibfield{author}{\bibinfo{person}{Daniel Gonzalez} {and} \bibinfo{person}{Thaier Hayajneh}.} \bibinfo{year}{2017}\natexlab{}.
\newblock \showarticletitle{Detection and prevention of crypto-ransomware}. In \bibinfo{booktitle}{\emph{2017 IEEE 8th Annual Ubiquitous Computing, Electronics and Mobile Communication Conference (UEMCON)}}. IEEE, \bibinfo{pages}{472--478}.
\newblock


\bibitem[Gómez-Hernández et~al\mbox{.}(2018)]%
        {r-locker}
\bibfield{author}{\bibinfo{person}{J.~A. Gómez-Hernández}, \bibinfo{person}{L. Álvarez González}, {and} \bibinfo{person}{P. García-Teodoro}.} \bibinfo{year}{2018}\natexlab{}.
\newblock \showarticletitle{R-{L}ocker: Thwarting ransomware action through a honeyfile-based approach}.
\newblock \bibinfo{journal}{\emph{Computers \& Security}}  \bibinfo{volume}{73} (\bibinfo{date}{3} \bibinfo{year}{2018}), \bibinfo{pages}{389--398}.
\newblock
\showISSN{0167-4048}


\bibitem[Hasan and Rahman(2017)]%
        {hasan2017ranshunt}
\bibfield{author}{\bibinfo{person}{Md~Mahbub Hasan} {and} \bibinfo{person}{Md~Mahbubur Rahman}.} \bibinfo{year}{2017}\natexlab{}.
\newblock \showarticletitle{RansHunt: A support vector machines based ransomware analysis framework with integrated feature set}. In \bibinfo{booktitle}{\emph{2017 20th international conference of computer and information technology (ICCIT)}}. IEEE, \bibinfo{pages}{1--7}.
\newblock


\bibitem[Hill and Bellekens(2018)]%
        {hill2018cryptoknight}
\bibfield{author}{\bibinfo{person}{Gregory Hill} {and} \bibinfo{person}{Xavier Bellekens}.} \bibinfo{year}{2018}\natexlab{}.
\newblock \showarticletitle{CryptoKnight: generating and modelling compiled cryptographic primitives}.
\newblock \bibinfo{journal}{\emph{Information}} \bibinfo{volume}{9}, \bibinfo{number}{9} (\bibinfo{year}{2018}), \bibinfo{pages}{231}.
\newblock


\bibitem[Hirano and Kobayashi(2019)]%
        {hirano2019machine}
\bibfield{author}{\bibinfo{person}{Manabu Hirano} {and} \bibinfo{person}{Ryotaro Kobayashi}.} \bibinfo{year}{2019}\natexlab{}.
\newblock \showarticletitle{Machine learning based ransomware detection using storage access patterns obtained from live-forensic hypervisor}. In \bibinfo{booktitle}{\emph{2019 sixth international conference on internet of things: Systems, Management and security (IOTSMS)}}. IEEE, \bibinfo{pages}{1--6}.
\newblock


\bibitem[Homayoun et~al\mbox{.}(2017)]%
        {homayoun2017know}
\bibfield{author}{\bibinfo{person}{Sajad Homayoun}, \bibinfo{person}{Ali Dehghantanha}, \bibinfo{person}{Marzieh Ahmadzadeh}, \bibinfo{person}{Sattar Hashemi}, {and} \bibinfo{person}{Raouf Khayami}.} \bibinfo{year}{2017}\natexlab{}.
\newblock \showarticletitle{Know abnormal, find evil: frequent pattern mining for ransomware threat hunting and intelligence}.
\newblock \bibinfo{journal}{\emph{IEEE transactions on emerging topics in computing}} \bibinfo{volume}{8}, \bibinfo{number}{2} (\bibinfo{year}{2017}), \bibinfo{pages}{341--351}.
\newblock


\bibitem[Homayoun et~al\mbox{.}(2019)]%
        {homayoun2019drthis}
\bibfield{author}{\bibinfo{person}{Sajad Homayoun}, \bibinfo{person}{Ali Dehghantanha}, \bibinfo{person}{Marzieh Ahmadzadeh}, \bibinfo{person}{Sattar Hashemi}, \bibinfo{person}{Raouf Khayami}, \bibinfo{person}{Kim-Kwang~Raymond Choo}, {and} \bibinfo{person}{David~Ellis Newton}.} \bibinfo{year}{2019}\natexlab{}.
\newblock \showarticletitle{DRTHIS: Deep ransomware threat hunting and intelligence system at the fog layer}.
\newblock \bibinfo{journal}{\emph{Future Generation Computer Systems}}  \bibinfo{volume}{90} (\bibinfo{year}{2019}), \bibinfo{pages}{94--104}.
\newblock


\bibitem[Honda et~al\mbox{.}(2018)]%
        {honda2018ransomware}
\bibfield{author}{\bibinfo{person}{Toshiki Honda}, \bibinfo{person}{Kohei Mukaiyama}, \bibinfo{person}{Takeharu Shirai}, \bibinfo{person}{Tetsushi Ohki}, {and} \bibinfo{person}{Masakatsu Nishigaki}.} \bibinfo{year}{2018}\natexlab{}.
\newblock \showarticletitle{Ransomware Detection Considering User's Document Editing}. In \bibinfo{booktitle}{\emph{2018 IEEE 32nd International Conference on Advanced Information Networking and Applications (AINA)}}. IEEE, \bibinfo{pages}{907--914}.
\newblock


\bibitem[Huang et~al\mbox{.}(2017)]%
        {flashguard}
\bibfield{author}{\bibinfo{person}{Jian Huang}, \bibinfo{person}{Jun Xu}, \bibinfo{person}{Xinyu Xing}, \bibinfo{person}{Peng Liu}, {and} \bibinfo{person}{Moinuddin~K. Qureshi}.} \bibinfo{year}{2017}\natexlab{}.
\newblock \showarticletitle{{FlashGuard}: Leveraging Intrinsic Flash Properties to Defend Against Encryption Ransomware}. In \bibinfo{booktitle}{\emph{Proceedings of the 2017 ACM SIGSAC Conference on Computer and Communications Security}} (Dallas, Texas, USA). \bibinfo{pages}{2231–2244}.
\newblock
\showISBNx{9781450349468}


\bibitem[Humayun et~al\mbox{.}(2021)]%
        {humayun2021internet}
\bibfield{author}{\bibinfo{person}{Mamoona Humayun}, \bibinfo{person}{NZ Jhanjhi}, \bibinfo{person}{Ahmed Alsayat}, {and} \bibinfo{person}{Vasaki Ponnusamy}.} \bibinfo{year}{2021}\natexlab{}.
\newblock \showarticletitle{Internet of things and ransomware: Evolution, mitigation and prevention}.
\newblock \bibinfo{journal}{\emph{Egyptian Informatics Journal}} \bibinfo{volume}{22}, \bibinfo{number}{1} (\bibinfo{year}{2021}), \bibinfo{pages}{105--117}.
\newblock


\bibitem[Hwang et~al\mbox{.}(2020)]%
        {Hwang2020two-stage}
\bibfield{author}{\bibinfo{person}{Jinsoo Hwang}, \bibinfo{person}{Jeankyung Kim}, \bibinfo{person}{Seunghwan Lee}, {and} \bibinfo{person}{Kichang Kim}.} \bibinfo{year}{2020}\natexlab{}.
\newblock \showarticletitle{Two-Stage Ransomware Detection Using Dynamic Analysis and Machine Learning Techniques}.
\newblock \bibinfo{journal}{\emph{Wireless Personal Communications}}  \bibinfo{volume}{112} (\bibinfo{date}{6} \bibinfo{year}{2020}), \bibinfo{pages}{2597--2609}.
\newblock
Issue 4.
\showISSN{1572834X}


\bibitem[Ibarra et~al\mbox{.}(2019)]%
        {ibarra2019ransomware}
\bibfield{author}{\bibinfo{person}{Jaime Ibarra}, \bibinfo{person}{Usman~Javed Butt}, \bibinfo{person}{Anh Do}, \bibinfo{person}{Hamid Jahankhani}, {and} \bibinfo{person}{Arshad Jamal}.} \bibinfo{year}{2019}\natexlab{}.
\newblock \showarticletitle{Ransomware impact to SCADA systems and its scope to critical infrastructure}. In \bibinfo{booktitle}{\emph{2019 IEEE 12th International Conference on Global Security, Safety and Sustainability (ICGS3)}}. IEEE, \bibinfo{pages}{1--12}.
\newblock


\bibitem[Intelligence(2022)]%
        {partyticket}
\bibfield{author}{\bibinfo{person}{CrowdStrike Intelligence}.} \bibinfo{year}{2022}\natexlab{}.
\newblock \bibinfo{title}{How to Decrypt the {P}arty{T}icket Ransomware Targeting Ukraine}.
\newblock \bibinfo{howpublished}{\url{https://www.crowdstrike.com/blog/how-to-decrypt-the-partyticket-ransomware-targeting-ukraine/}}.
\newblock


\bibitem[Jung and Won(2018)]%
        {jung2018ransomware}
\bibfield{author}{\bibinfo{person}{Sangmoon Jung} {and} \bibinfo{person}{Yoojae Won}.} \bibinfo{year}{2018}\natexlab{}.
\newblock \showarticletitle{Ransomware detection method based on context-aware entropy analysis}.
\newblock \bibinfo{journal}{\emph{Soft Computing}}  \bibinfo{volume}{22} (\bibinfo{year}{2018}), \bibinfo{pages}{6731--6740}.
\newblock


\bibitem[Karimi and Moattar(2017)]%
        {karimi2017android}
\bibfield{author}{\bibinfo{person}{Alireza Karimi} {and} \bibinfo{person}{Mohammad~Hosein Moattar}.} \bibinfo{year}{2017}\natexlab{}.
\newblock \showarticletitle{Android ransomware detection using reduced opcode sequence and image similarity}. In \bibinfo{booktitle}{\emph{2017 7th International Conference on Computer and Knowledge Engineering (ICCKE)}}. IEEE, \bibinfo{pages}{229--234}.
\newblock


\bibitem[Kaspersky({[n.\,d.]})]%
        {wannacry230}
\bibfield{author}{\bibinfo{person}{Kaspersky}.} \bibinfo{year}{[n.\,d.]}\natexlab{}.
\newblock \bibinfo{title}{Ransomware {W}anna{C}ry: All you need to know}.
\newblock \bibinfo{howpublished}{\url{https://usa.kaspersky.com/resource-center/threats/ransomware-wannacry}}.
\newblock


\bibitem[Keong~Ng et~al\mbox{.}(2020)]%
        {keong2020voterchoice}
\bibfield{author}{\bibinfo{person}{Chee Keong~Ng}, \bibinfo{person}{Sutharshan Rajasegarar}, \bibinfo{person}{Lei Pan}, \bibinfo{person}{Frank Jiang}, {and} \bibinfo{person}{Leo~Yu Zhang}.} \bibinfo{year}{2020}\natexlab{}.
\newblock \showarticletitle{VoterChoice: A ransomware detection honeypot with multiple voting framework}.
\newblock \bibinfo{journal}{\emph{Concurrency and Computation: Practice and Experience}} \bibinfo{volume}{32}, \bibinfo{number}{14} (\bibinfo{year}{2020}), \bibinfo{pages}{e5726}.
\newblock


\bibitem[Keshavarzi and Ghaffary(2020)]%
        {keshavarzi2020i2ce3}
\bibfield{author}{\bibinfo{person}{Masoudeh Keshavarzi} {and} \bibinfo{person}{Hamid~Reza Ghaffary}.} \bibinfo{year}{2020}\natexlab{}.
\newblock \showarticletitle{I2CE3: A dedicated and separated attack chain for ransomware offenses as the most infamous cyber extortion}.
\newblock \bibinfo{journal}{\emph{Computer Science Review}}  \bibinfo{volume}{36} (\bibinfo{year}{2020}), \bibinfo{pages}{100233}.
\newblock


\bibitem[Ketzaki et~al\mbox{.}(2020)]%
        {ketzaki2020behaviour}
\bibfield{author}{\bibinfo{person}{Eleni Ketzaki}, \bibinfo{person}{Petros Toupas}, \bibinfo{person}{Konstantinos~M Giannoutakis}, \bibinfo{person}{Anastasios Drosou}, {and} \bibinfo{person}{Dimitrios Tzovaras}.} \bibinfo{year}{2020}\natexlab{}.
\newblock \showarticletitle{A behaviour based ransomware detection using neural network models}. In \bibinfo{booktitle}{\emph{2020 10th International Conference on Advanced Computer Information Technologies (ACIT)}}. IEEE, \bibinfo{pages}{747--750}.
\newblock


\bibitem[Khan et~al\mbox{.}(2020)]%
        {khan2020digital}
\bibfield{author}{\bibinfo{person}{Firoz Khan}, \bibinfo{person}{Cornelius Ncube}, \bibinfo{person}{Lakshmana~Kumar Ramasamy}, \bibinfo{person}{Seifedine Kadry}, {and} \bibinfo{person}{Yunyoung Nam}.} \bibinfo{year}{2020}\natexlab{}.
\newblock \showarticletitle{A digital DNA sequencing engine for ransomware detection using machine learning}.
\newblock \bibinfo{journal}{\emph{IEEE Access}}  \bibinfo{volume}{8} (\bibinfo{year}{2020}), \bibinfo{pages}{119710--119719}.
\newblock


\bibitem[Kharaz et~al\mbox{.}(2016)]%
        {unveil}
\bibfield{author}{\bibinfo{person}{Amin Kharaz}, \bibinfo{person}{Sajjad Arshad}, \bibinfo{person}{Collin Mulliner}, \bibinfo{person}{William Robertson}, {and} \bibinfo{person}{Engin Kirda}.} \bibinfo{year}{2016}\natexlab{}.
\newblock \showarticletitle{{UNVEIL}: A {Large-Scale}, Automated Approach to Detecting Ransomware}. In \bibinfo{booktitle}{\emph{25th USENIX Security Symposium (USENIX Security 16)}}. \bibinfo{pages}{757--772}.
\newblock
\showISBNx{978-1-931971-32-4}


\bibitem[Kim et~al\mbox{.}(2018)]%
        {kim2018white}
\bibfield{author}{\bibinfo{person}{Dae-Youb Kim}, \bibinfo{person}{Geun-Yeong Choi}, {and} \bibinfo{person}{Ji-Hoon Lee}.} \bibinfo{year}{2018}\natexlab{}.
\newblock \showarticletitle{White list-based ransomware real-time detection and prevention for user device protection}. In \bibinfo{booktitle}{\emph{2018 IEEE International Conference on Consumer Electronics (ICCE)}}. IEEE, \bibinfo{pages}{1--5}.
\newblock


\bibitem[Kim et~al\mbox{.}(2022)]%
        {Kim2022}
\bibfield{author}{\bibinfo{person}{Giyoon Kim}, \bibinfo{person}{Soram Kim}, \bibinfo{person}{Soojin Kang}, {and} \bibinfo{person}{Jongsung Kim}.} \bibinfo{year}{2022}\natexlab{}.
\newblock \bibinfo{title}{A Method for Decrypting Data Infected with {H}ive Ransomware}.
\newblock \bibinfo{howpublished}{\url{https://arxiv.org/abs/2202.08477v1}}.
\newblock


\bibitem[Kiru and Jantan(2019)]%
        {kiru2019age}
\bibfield{author}{\bibinfo{person}{Muhammad~Ubale Kiru} {and} \bibinfo{person}{Aman~B Jantan}.} \bibinfo{year}{2019}\natexlab{}.
\newblock \showarticletitle{The age of ransomware: Understanding ransomware and its countermeasures.}
\newblock In \bibinfo{booktitle}{\emph{Artificial Intelligence and Security Challenges in Emerging Networks}}. \bibinfo{publisher}{IGI Global}, \bibinfo{pages}{1--37}.
\newblock


\bibitem[Kok et~al\mbox{.}(2019)]%
        {kok2019ransomware}
\bibfield{author}{\bibinfo{person}{S Kok}, \bibinfo{person}{Azween Abdullah}, \bibinfo{person}{N Jhanjhi}, {and} \bibinfo{person}{Mahadevan Supramaniam}.} \bibinfo{year}{2019}\natexlab{}.
\newblock \showarticletitle{Ransomware, threat and detection techniques: A review}.
\newblock \bibinfo{journal}{\emph{Int. J. Comput. Sci. Netw. Secur}} \bibinfo{volume}{19}, \bibinfo{number}{2} (\bibinfo{year}{2019}), \bibinfo{pages}{136}.
\newblock


\bibitem[Kok et~al\mbox{.}(2020)]%
        {kok2020evaluation}
\bibfield{author}{\bibinfo{person}{SH Kok}, \bibinfo{person}{A Azween}, {and} \bibinfo{person}{NZ Jhanjhi}.} \bibinfo{year}{2020}\natexlab{}.
\newblock \showarticletitle{Evaluation metric for crypto-ransomware detection using machine learning}.
\newblock \bibinfo{journal}{\emph{Journal of Information Security and Applications}}  \bibinfo{volume}{55} (\bibinfo{year}{2020}), \bibinfo{pages}{102646}.
\newblock


\bibitem[Kok et~al\mbox{.}(2022)]%
        {Kok2022}
\bibfield{author}{\bibinfo{person}{S.~H. Kok}, \bibinfo{person}{Azween Abdullah}, {and} \bibinfo{person}{N.~Z. Jhanjhi}.} \bibinfo{year}{2022}\natexlab{}.
\newblock \showarticletitle{Early detection of crypto-ransomware using pre-encryption detection algorithm}.
\newblock \bibinfo{journal}{\emph{Journal of King Saud University - Computer and Information Sciences}}  \bibinfo{volume}{34} (\bibinfo{date}{5} \bibinfo{year}{2022}), \bibinfo{pages}{1984--1999}.
\newblock
Issue 5.
\showISSN{1319-1578}


\bibitem[Kolodenker et~al\mbox{.}(2017)]%
        {paybreak}
\bibfield{author}{\bibinfo{person}{Eugene Kolodenker}, \bibinfo{person}{William Koch}, \bibinfo{person}{Gianluca Stringhini}, {and} \bibinfo{person}{Manuel Egele}.} \bibinfo{year}{2017}\natexlab{}.
\newblock \showarticletitle{Pay{B}reak: Defense Against Cryptographic Ransomware}. In \bibinfo{booktitle}{\emph{Proceedings of the 2017 ACM on Asia Conference on Computer and Communications Security}} (Abu Dhabi, United Arab Emirates) \emph{(\bibinfo{series}{ASIA CCS '17})}. \bibinfo{pages}{599–611}.
\newblock
\showISBNx{9781450349444}


\bibitem[Larsen et~al\mbox{.}(2021)]%
        {Larsen2021}
\bibfield{author}{\bibinfo{person}{Erik Larsen}, \bibinfo{person}{David Noever}, {and} \bibinfo{person}{Korey MacVittie}.} \bibinfo{year}{2021}\natexlab{}.
\newblock \bibinfo{title}{A Survey of Machine Learning Algorithms for Detecting Ransomware Encryption Activity}.
\newblock \bibinfo{howpublished}{\url{https://arxiv.org/abs/2110.07636v1}}.
\newblock


\bibitem[Lee et~al\mbox{.}(2017)]%
        {lee2017make}
\bibfield{author}{\bibinfo{person}{Jeonghwan Lee}, \bibinfo{person}{Jinwoo Lee}, {and} \bibinfo{person}{Jiman Hong}.} \bibinfo{year}{2017}\natexlab{}.
\newblock \showarticletitle{How to make efficient decoy files for ransomware detection?}. In \bibinfo{booktitle}{\emph{Proceedings of the International Conference on Research in Adaptive and Convergent Systems}}. \bibinfo{pages}{208--212}.
\newblock


\bibitem[Lee et~al\mbox{.}(2019)]%
        {lee2019machine}
\bibfield{author}{\bibinfo{person}{Kyungroul Lee}, \bibinfo{person}{Sun-Young Lee}, {and} \bibinfo{person}{Kangbin Yim}.} \bibinfo{year}{2019}\natexlab{}.
\newblock \showarticletitle{Machine learning based file entropy analysis for ransomware detection in backup systems}.
\newblock \bibinfo{journal}{\emph{IEEE Access}}  \bibinfo{volume}{7} (\bibinfo{year}{2019}), \bibinfo{pages}{110205--110215}.
\newblock


\bibitem[Lee et~al\mbox{.}(2018)]%
        {lee2018ransomware}
\bibfield{author}{\bibinfo{person}{Kyungroul Lee}, \bibinfo{person}{Kangbin Yim}, {and} \bibinfo{person}{Jung~Taek Seo}.} \bibinfo{year}{2018}\natexlab{}.
\newblock \showarticletitle{Ransomware prevention technique using key backup}.
\newblock \bibinfo{journal}{\emph{Concurrency and Computation: Practice and Experience}} \bibinfo{volume}{30}, \bibinfo{number}{3} (\bibinfo{year}{2018}), \bibinfo{pages}{e4337}.
\newblock


\bibitem[Lee et~al\mbox{.}(2020)]%
        {Lee2020Magniber}
\bibfield{author}{\bibinfo{person}{Sehoon Lee}, \bibinfo{person}{Myungseo Park}, {and} \bibinfo{person}{Jongsung Kim}.} \bibinfo{year}{2020}\natexlab{}.
\newblock \showarticletitle{Magniber v2 Ransomware Decryption: Exploiting the Vulnerability of a Self-Developed Pseudo Random Number Generator}.
\newblock \bibinfo{journal}{\emph{Electronics 2021, Vol. 10, Page 16}}  \bibinfo{volume}{10} (\bibinfo{date}{12} \bibinfo{year}{2020}), \bibinfo{pages}{16}.
\newblock
Issue 1.
\showISSN{2079-9292}


\bibitem[Lu et~al\mbox{.}(2020)]%
        {lu2020ransomware}
\bibfield{author}{\bibinfo{person}{Tianliang Lu}, \bibinfo{person}{Yanhui Du}, \bibinfo{person}{Jing Wu}, {and} \bibinfo{person}{Yuxuan Bao}.} \bibinfo{year}{2020}\natexlab{}.
\newblock \showarticletitle{Ransomware detection based on an improved double-layer negative selection algorithm}. In \bibinfo{booktitle}{\emph{Testbeds and Research Infrastructures for the Development of Networks and Communications: 14th EAI International Conference, TridentCom 2019, Changsha, China, December 7-8, 2019, Proceedings 14}}. Springer, \bibinfo{pages}{46--61}.
\newblock


\bibitem[Lu et~al\mbox{.}(2017)]%
        {lu2017ransomware}
\bibfield{author}{\bibinfo{person}{Tianliang Lu}, \bibinfo{person}{Lu Zhang}, \bibinfo{person}{Shunye Wang}, {and} \bibinfo{person}{Qi Gong}.} \bibinfo{year}{2017}\natexlab{}.
\newblock \showarticletitle{Ransomware detection based on V-detector negative selection algorithm}. In \bibinfo{booktitle}{\emph{2017 International conference on security, pattern analysis, and cybernetics (SPAC)}}. IEEE, \bibinfo{pages}{531--536}.
\newblock


\bibitem[Maffia et~al\mbox{.}(2021)]%
        {maffia2021longitudinal}
\bibfield{author}{\bibinfo{person}{Lorenzo Maffia}, \bibinfo{person}{Dario Nisi}, \bibinfo{person}{Platon Kotzias}, \bibinfo{person}{Giovanni Lagorio}, \bibinfo{person}{Simone Aonzo}, {and} \bibinfo{person}{Davide Balzarotti}.} \bibinfo{year}{2021}\natexlab{}.
\newblock \bibinfo{title}{Longitudinal Study of the Prevalence of Malware Evasive Techniques}.
\newblock
\newblock
\showeprint[arxiv]{2112.11289}~[cs.CR]


\bibitem[Maigida et~al\mbox{.}(2019)]%
        {maigida2019systematic}
\bibfield{author}{\bibinfo{person}{Abdullahi~Mohammed Maigida}, \bibinfo{person}{Shafi’i~Muhammad Abdulhamid}, \bibinfo{person}{Morufu Olalere}, \bibinfo{person}{John~K Alhassan}, \bibinfo{person}{Haruna Chiroma}, {and} \bibinfo{person}{Emmanuel~Gbenga Dada}.} \bibinfo{year}{2019}\natexlab{}.
\newblock \showarticletitle{Systematic literature review and metadata analysis of ransomware attacks and detection mechanisms}.
\newblock \bibinfo{journal}{\emph{Journal of Reliable Intelligent Environments}}  \bibinfo{volume}{5} (\bibinfo{year}{2019}), \bibinfo{pages}{67--89}.
\newblock


\bibitem[Maiorca et~al\mbox{.}(2017)]%
        {maiorca2017r}
\bibfield{author}{\bibinfo{person}{Davide Maiorca}, \bibinfo{person}{Francesco Mercaldo}, \bibinfo{person}{Giorgio Giacinto}, \bibinfo{person}{Corrado~Aaron Visaggio}, {and} \bibinfo{person}{Fabio Martinelli}.} \bibinfo{year}{2017}\natexlab{}.
\newblock \showarticletitle{R-PackDroid: API package-based characterization and detection of mobile ransomware}. In \bibinfo{booktitle}{\emph{Proceedings of the symposium on applied computing}}. \bibinfo{pages}{1718--1723}.
\newblock


\bibitem[Maniath et~al\mbox{.}(2017)]%
        {maniath2017deep}
\bibfield{author}{\bibinfo{person}{Sumith Maniath}, \bibinfo{person}{Aravind Ashok}, \bibinfo{person}{Prabaharan Poornachandran}, \bibinfo{person}{VG Sujadevi}, \bibinfo{person}{Prem~Sankar AU}, {and} \bibinfo{person}{Srinath Jan}.} \bibinfo{year}{2017}\natexlab{}.
\newblock \showarticletitle{Deep learning LSTM based ransomware detection}. In \bibinfo{booktitle}{\emph{2017 Recent Developments in Control, Automation \& Power Engineering (RDCAPE)}}. IEEE, \bibinfo{pages}{442--446}.
\newblock


\bibitem[Maniath et~al\mbox{.}(2019)]%
        {maniath2019survey}
\bibfield{author}{\bibinfo{person}{Sumith Maniath}, \bibinfo{person}{Prabaharan Poornachandran}, {and} \bibinfo{person}{VG Sujadevi}.} \bibinfo{year}{2019}\natexlab{}.
\newblock \showarticletitle{Survey on prevention, mitigation and containment of ransomware attacks}. In \bibinfo{booktitle}{\emph{Security in Computing and Communications: 6th International Symposium, SSCC 2018, Bangalore, India, September 19--22, 2018, Revised Selected Papers 6}}. Springer, \bibinfo{pages}{39--52}.
\newblock


\bibitem[May and Laron(2019)]%
        {may2019combating}
\bibfield{author}{\bibinfo{person}{Michael~J May} {and} \bibinfo{person}{Etamar Laron}.} \bibinfo{year}{2019}\natexlab{}.
\newblock \showarticletitle{Combating ransomware using content analysis and complex file events}. In \bibinfo{booktitle}{\emph{2019 10th IFIP International Conference on New Technologies, Mobility and Security (NTMS)}}. IEEE, \bibinfo{pages}{1--5}.
\newblock


\bibitem[McIntosh et~al\mbox{.}(2023a)]%
        {access}
\bibfield{author}{\bibinfo{person}{Timothy McIntosh}, \bibinfo{person}{A.S.M. Kayes}, \bibinfo{person}{Yi-Ping~Phoebe Chen}, \bibinfo{person}{Alex Ng}, {and} \bibinfo{person}{Paul Watters}.} \bibinfo{year}{2023}\natexlab{a}.
\newblock \showarticletitle{Applying staged event-driven access control to combat ransomware}.
\newblock \bibinfo{journal}{\emph{Computers \& Security}}  \bibinfo{volume}{128} (\bibinfo{year}{2023}).
\newblock
\showISSN{0167-4048}


\bibitem[McIntosh et~al\mbox{.}(2023b)]%
        {mcintosh2023applying}
\bibfield{author}{\bibinfo{person}{Timothy McIntosh}, \bibinfo{person}{ASM Kayes}, \bibinfo{person}{Yi-Ping~Phoebe Chen}, \bibinfo{person}{Alex Ng}, {and} \bibinfo{person}{Paul Watters}.} \bibinfo{year}{2023}\natexlab{b}.
\newblock \showarticletitle{Applying staged event-driven access control to combat ransomware}.
\newblock \bibinfo{journal}{\emph{Computers \& Security}}  \bibinfo{volume}{128} (\bibinfo{year}{2023}), \bibinfo{pages}{103160}.
\newblock


\bibitem[McIntosh et~al\mbox{.}(2021)]%
        {mcintosh2021survey}
\bibfield{author}{\bibinfo{person}{Timothy McIntosh}, \bibinfo{person}{A.~S.~M. Kayes}, \bibinfo{person}{Yi-Ping~Phoebe Chen}, \bibinfo{person}{Alex Ng}, {and} \bibinfo{person}{Paul Watters}.} \bibinfo{year}{2021}\natexlab{}.
\newblock \showarticletitle{Ransomware Mitigation in the Modern Era: A Comprehensive Review, Research Challenges, and Future Directions}.
\newblock \bibinfo{journal}{\emph{ACM Comput. Surv.}} \bibinfo{volume}{54}, \bibinfo{number}{9}, Article \bibinfo{articleno}{197} (\bibinfo{date}{oct} \bibinfo{year}{2021}), \bibinfo{numpages}{36}~pages.
\newblock
\showISSN{0360-0300}


\bibitem[Medhat et~al\mbox{.}(2018)]%
        {medhat2018new}
\bibfield{author}{\bibinfo{person}{May Medhat}, \bibinfo{person}{Samir Gaber}, {and} \bibinfo{person}{Nashwa Abdelbaki}.} \bibinfo{year}{2018}\natexlab{}.
\newblock \showarticletitle{A new static-based framework for ransomware detection}. In \bibinfo{booktitle}{\emph{2018 IEEE 16th Intl Conf on Dependable, Autonomic and Secure Computing, 16th Intl Conf on Pervasive Intelligence and Computing, 4th Intl Conf on Big Data Intelligence and Computing and Cyber Science and Technology Congress (DASC/PiCom/DataCom/CyberSciTech)}}. IEEE, \bibinfo{pages}{710--715}.
\newblock


\bibitem[Mehnaz et~al\mbox{.}(2018)]%
        {Mehnaz2018rwguard}
\bibfield{author}{\bibinfo{person}{Shagufta Mehnaz}, \bibinfo{person}{Anand Mudgerikar}, {and} \bibinfo{person}{Elisa Bertino}.} \bibinfo{year}{2018}\natexlab{}.
\newblock \showarticletitle{{RWGuard}: A Real-Time Detection System Against Cryptographic Ransomware}. In \bibinfo{booktitle}{\emph{Research in Attacks, Intrusions, and Defenses}}. \bibinfo{pages}{114--136}.
\newblock
\showISBNx{978-3-030-00470-5}


\bibitem[Modi et~al\mbox{.}(2020)]%
        {modi2020detecting}
\bibfield{author}{\bibinfo{person}{Jaimin Modi}, \bibinfo{person}{Issa Traore}, \bibinfo{person}{Asem Ghaleb}, \bibinfo{person}{Karim Ganame}, {and} \bibinfo{person}{Sherif Ahmed}.} \bibinfo{year}{2020}\natexlab{}.
\newblock \showarticletitle{Detecting ransomware in encrypted web traffic}. In \bibinfo{booktitle}{\emph{Foundations and Practice of Security: 12th International Symposium, FPS 2019, Toulouse, France, November 5--7, 2019, Revised Selected Papers 12}}. Springer, \bibinfo{pages}{345--353}.
\newblock


\bibitem[Mohan and Kumar(2017)]%
        {mohan2017efficacy}
\bibfield{author}{\bibinfo{person}{Jithin~Chandra Mohan} {and} \bibinfo{person}{Renuka Kumar}.} \bibinfo{year}{2017}\natexlab{}.
\newblock \showarticletitle{On the efficacy of android ransomware detection techniques: A survey}.
\newblock \bibinfo{journal}{\emph{International Journal of Pure and Applied Mathematics}} \bibinfo{volume}{115}, \bibinfo{number}{8} (\bibinfo{year}{2017}), \bibinfo{pages}{115--120}.
\newblock


\bibitem[Moore(2016)]%
        {honeypot2016}
\bibfield{author}{\bibinfo{person}{Chris Moore}.} \bibinfo{year}{2016}\natexlab{}.
\newblock \showarticletitle{Detecting Ransomware with Honeypot Techniques}. In \bibinfo{booktitle}{\emph{2016 Cybersecurity and Cyberforensics Conference (CCC)}}. \bibinfo{pages}{77--81}.
\newblock
\urldef\tempurl%
\url{https://doi.org/10.1109/CCC.2016.14}
\showDOI{\tempurl}


\bibitem[Morato et~al\mbox{.}(2018)]%
        {morato2018ransomware}
\bibfield{author}{\bibinfo{person}{Daniel Morato}, \bibinfo{person}{Eduardo Berrueta}, \bibinfo{person}{Eduardo Maga{\~n}a}, {and} \bibinfo{person}{Mikel Izal}.} \bibinfo{year}{2018}\natexlab{}.
\newblock \showarticletitle{Ransomware early detection by the analysis of file sharing traffic}.
\newblock \bibinfo{journal}{\emph{Journal of Network and Computer Applications}}  \bibinfo{volume}{124} (\bibinfo{year}{2018}), \bibinfo{pages}{14--32}.
\newblock


\bibitem[Moussaileb et~al\mbox{.}(2018)]%
        {moussaileb2018ransomware}
\bibfield{author}{\bibinfo{person}{Routa Moussaileb}, \bibinfo{person}{Benjamin Bouget}, \bibinfo{person}{Aur{\'e}lien Palisse}, \bibinfo{person}{H{\'e}l{\`e}ne Le~Bouder}, \bibinfo{person}{Nora Cuppens}, {and} \bibinfo{person}{Jean-Louis Lanet}.} \bibinfo{year}{2018}\natexlab{}.
\newblock \showarticletitle{Ransomware's early mitigation mechanisms}. In \bibinfo{booktitle}{\emph{Proceedings of the 13th international conference on availability, reliability and security}}. \bibinfo{pages}{1--10}.
\newblock


\bibitem[Moussaileb et~al\mbox{.}(2021)]%
        {moussaileb2021survey}
\bibfield{author}{\bibinfo{person}{Routa Moussaileb}, \bibinfo{person}{Nora Cuppens}, \bibinfo{person}{Jean-Louis Lanet}, {and} \bibinfo{person}{H{\'e}l{\`e}ne~Le Bouder}.} \bibinfo{year}{2021}\natexlab{}.
\newblock \showarticletitle{A survey on windows-based ransomware taxonomy and detection mechanisms}.
\newblock \bibinfo{journal}{\emph{ACM Computing Surveys (CSUR)}} \bibinfo{volume}{54}, \bibinfo{number}{6} (\bibinfo{year}{2021}), \bibinfo{pages}{1--36}.
\newblock


\bibitem[Naseer et~al\mbox{.}(2020)]%
        {naseer2020windows}
\bibfield{author}{\bibinfo{person}{Ayesha Naseer}, \bibinfo{person}{Riffat Mir}, \bibinfo{person}{Azmat Mir}, {and} \bibinfo{person}{Muhammad Aleem}.} \bibinfo{year}{2020}\natexlab{}.
\newblock \showarticletitle{Windows-based Ransomware: A Survey.}
\newblock \bibinfo{journal}{\emph{Journal of Information Assurance \& Security}} \bibinfo{volume}{15}, \bibinfo{number}{3} (\bibinfo{year}{2020}).
\newblock


\bibitem[Noever and Noever(2021)]%
        {posse}
\bibfield{author}{\bibinfo{person}{David Noever} {and} \bibinfo{person}{Samantha~Miller Noever}.} \bibinfo{year}{2021}\natexlab{}.
\newblock \bibinfo{title}{{POSSE}: Patterns of Systems During Software Encryption}.
\newblock \bibinfo{howpublished}{\url{https://arxiv.org/abs/2109.12162}}.
\newblock


\bibitem[Olaimat et~al\mbox{.}(2021)]%
        {Olaimat2021}
\bibfield{author}{\bibinfo{person}{Mohammad~N. Olaimat}, \bibinfo{person}{Mohd~Aizaini Maarof}, {and} \bibinfo{person}{Bander Ali~S. Al-Rimy}.} \bibinfo{year}{2021}\natexlab{}.
\newblock \showarticletitle{Ransomware Anti-Analysis and Evasion Techniques: A Survey and Research Directions}.
\newblock \bibinfo{journal}{\emph{2021 3rd International Cyber Resilience Conference, CRC 2021}} (\bibinfo{date}{1} \bibinfo{year}{2021}).
\newblock
\showISBNx{9781665418447}


\bibitem[Oz et~al\mbox{.}(2023)]%
        {oz2023rob}
\bibfield{author}{\bibinfo{person}{Harun Oz}, \bibinfo{person}{Ahmet Aris}, \bibinfo{person}{Abbas Acar}, \bibinfo{person}{G{\"u}liz~Seray Tuncay}, \bibinfo{person}{Leonardo Babun}, {and} \bibinfo{person}{Selcuk Uluagac}.} \bibinfo{year}{2023}\natexlab{}.
\newblock \showarticletitle{{R{\o}B}: Ransomware over Modern Web Browsers}. In \bibinfo{booktitle}{\emph{32nd USENIX Security Symposium (USENIX Security 23)}}. \bibinfo{publisher}{USENIX Association}, \bibinfo{address}{Anaheim, CA}, \bibinfo{pages}{7073--7090}.
\newblock
\showISBNx{978-1-939133-37-3}


\bibitem[Oz et~al\mbox{.}(2022)]%
        {oz2022survey}
\bibfield{author}{\bibinfo{person}{Harun Oz}, \bibinfo{person}{Ahmet Aris}, \bibinfo{person}{Albert Levi}, {and} \bibinfo{person}{A~Selcuk Uluagac}.} \bibinfo{year}{2022}\natexlab{}.
\newblock \showarticletitle{A survey on ransomware: Evolution, taxonomy, and defense solutions}.
\newblock \bibinfo{journal}{\emph{ACM Computing Surveys (CSUR)}} \bibinfo{volume}{54}, \bibinfo{number}{11s} (\bibinfo{year}{2022}), \bibinfo{pages}{1--37}.
\newblock


\bibitem[Palisse et~al\mbox{.}(2017a)]%
        {palisse2017data}
\bibfield{author}{\bibinfo{person}{Aur{\'e}lien Palisse}, \bibinfo{person}{Antoine Durand}, \bibinfo{person}{H{\'e}l{\`e}ne Le~Bouder}, \bibinfo{person}{Colas Le~Guernic}, {and} \bibinfo{person}{Jean-Louis Lanet}.} \bibinfo{year}{2017}\natexlab{a}.
\newblock \showarticletitle{Data aware defense (DaD): towards a generic and practical ransomware countermeasure}. In \bibinfo{booktitle}{\emph{Secure IT Systems: 22nd Nordic Conference, NordSec 2017, Tartu, Estonia, November 8--10, 2017, Proceedings 22}}. Springer, \bibinfo{pages}{192--208}.
\newblock


\bibitem[Palisse et~al\mbox{.}(2017b)]%
        {palisse2017ransomware}
\bibfield{author}{\bibinfo{person}{Aur{\'e}lien Palisse}, \bibinfo{person}{H{\'e}l{\`e}ne Le~Bouder}, \bibinfo{person}{Jean-Louis Lanet}, \bibinfo{person}{Colas Le~Guernic}, {and} \bibinfo{person}{Axel Legay}.} \bibinfo{year}{2017}\natexlab{b}.
\newblock \showarticletitle{Ransomware and the legacy crypto API}. In \bibinfo{booktitle}{\emph{Risks and Security of Internet and Systems: 11th International Conference, CRiSIS 2016, Roscoff, France, September 5-7, 2016, Revised Selected Papers 11}}. Springer, \bibinfo{pages}{11--28}.
\newblock


\bibitem[Popoola et~al\mbox{.}(2017)]%
        {popoola2017ransomware}
\bibfield{author}{\bibinfo{person}{Segun~I Popoola}, \bibinfo{person}{Ujioghosa~B Iyekekpolo}, \bibinfo{person}{Samuel~O Ojewande}, \bibinfo{person}{Faith~O Sweetwilliams}, \bibinfo{person}{Samuel~N John}, {and} \bibinfo{person}{Aderemi~A Atayero}.} \bibinfo{year}{2017}\natexlab{}.
\newblock \showarticletitle{Ransomware: Current trend, challenges, and research directions}. In \bibinfo{booktitle}{\emph{Proceedings of the World Congress on Engineering and Computer Science}}, Vol.~\bibinfo{volume}{1}. \bibinfo{pages}{169--174}.
\newblock


\bibitem[Quinkert et~al\mbox{.}(2018)]%
        {raptor}
\bibfield{author}{\bibinfo{person}{Florian Quinkert}, \bibinfo{person}{Thorsten Holz}, \bibinfo{person}{KSM~Tozammel Hossain}, \bibinfo{person}{Emilio Ferrara}, {and} \bibinfo{person}{Kristina Lerman}.} \bibinfo{year}{2018}\natexlab{}.
\newblock \bibinfo{title}{{RAPTOR}: Ransomware Attack PredicTOR}.
\newblock \bibinfo{howpublished}{\url{https://arxiv.org/abs/1803.01598}}.
\newblock
\urldef\tempurl%
\url{https://doi.org/10.48550/ARXIV.1803.01598}
\showDOI{\tempurl}


\bibitem[Ramesh and Menen(2020)]%
        {ramesh2020automated}
\bibfield{author}{\bibinfo{person}{Gowtham Ramesh} {and} \bibinfo{person}{Anjali Menen}.} \bibinfo{year}{2020}\natexlab{}.
\newblock \showarticletitle{Automated dynamic approach for detecting ransomware using finite-state machine}.
\newblock \bibinfo{journal}{\emph{Decision Support Systems}}  \bibinfo{volume}{138} (\bibinfo{year}{2020}), \bibinfo{pages}{113400}.
\newblock


\bibitem[Razaulla et~al\mbox{.}(2023)]%
        {survey23}
\bibfield{author}{\bibinfo{person}{Salwa Razaulla}, \bibinfo{person}{Claude Fachkha}, \bibinfo{person}{Christine Markarian}, \bibinfo{person}{Amjad Gawanmeh}, \bibinfo{person}{Wathiq Mansoor}, \bibinfo{person}{Benjamin C.~M. Fung}, {and} \bibinfo{person}{Chadi Assi}.} \bibinfo{year}{2023}\natexlab{}.
\newblock \showarticletitle{The Age of Ransomware: A Survey on the Evolution, Taxonomy, and Research Directions}.
\newblock \bibinfo{journal}{\emph{IEEE Access}}  \bibinfo{volume}{11} (\bibinfo{year}{2023}), \bibinfo{pages}{40698--40723}.
\newblock
\urldef\tempurl%
\url{https://doi.org/10.1109/ACCESS.2023.3268535}
\showDOI{\tempurl}


\bibitem[Reddy et~al\mbox{.}(2021)]%
        {reddy2021machine}
\bibfield{author}{\bibinfo{person}{Bheemidi~Vikram Reddy}, \bibinfo{person}{Gutha~Jaya Krishna}, \bibinfo{person}{Vadlamani Ravi}, {and} \bibinfo{person}{Dipankar Dasgupta}.} \bibinfo{year}{2021}\natexlab{}.
\newblock \showarticletitle{Machine learning and feature selection based ransomware detection using hexacodes}. In \bibinfo{booktitle}{\emph{Evolution in Computational Intelligence: Frontiers in Intelligent Computing: Theory and Applications (FICTA 2020), Volume 1}}. Springer, \bibinfo{pages}{583--597}.
\newblock


\bibitem[Rehman et~al\mbox{.}(2019)]%
        {rehman2019security}
\bibfield{author}{\bibinfo{person}{Habib~ur Rehman}, \bibinfo{person}{Eiad Yafi}, \bibinfo{person}{Mohammed Nazir}, {and} \bibinfo{person}{Khurram Mustafa}.} \bibinfo{year}{2019}\natexlab{}.
\newblock \showarticletitle{Security assurance against cybercrime ransomware}. In \bibinfo{booktitle}{\emph{Intelligent Computing \& Optimization 1}}. Springer, \bibinfo{pages}{21--34}.
\newblock


\bibitem[Reidys et~al\mbox{.}({[n.\,d.]})]%
        {RSSD}
\bibfield{author}{\bibinfo{person}{Benjamin Reidys}, \bibinfo{person}{Peng Liu}, {and} \bibinfo{person}{Jian Huang}.} \bibinfo{year}{[n.\,d.]}\natexlab{}.
\newblock \showarticletitle{{RSSD}: Defend against Ransomware with Hardware-Isolated Network-Storage Codesign and Post-Attack Analysis}. In \bibinfo{booktitle}{\emph{Proceedings of the 27th ACM International Conference on Architectural Support for Programming Languages and Operating Systems}} (Lausanne, Switzerland) \emph{(\bibinfo{series}{ASPLOS '22})}. \bibinfo{address}{New York, NY, USA}, \bibinfo{pages}{726–739}.
\newblock
\showISBNx{9781450392051}


\bibitem[Roy and Chen(2021)]%
        {roy2021deepran}
\bibfield{author}{\bibinfo{person}{Krishna~Chandra Roy} {and} \bibinfo{person}{Qian Chen}.} \bibinfo{year}{2021}\natexlab{}.
\newblock \showarticletitle{{DeepRan}: Attention-based {BiLSTM} and {CRF} for ransomware early detection and classification}.
\newblock \bibinfo{journal}{\emph{Information Systems Frontiers}}  \bibinfo{volume}{23} (\bibinfo{year}{2021}), \bibinfo{pages}{299--315}.
\newblock


\bibitem[Scaife et~al\mbox{.}(2016)]%
        {cryptolock}
\bibfield{author}{\bibinfo{person}{Nolen Scaife}, \bibinfo{person}{Henry Carter}, \bibinfo{person}{Patrick Traynor}, {and} \bibinfo{person}{Kevin R.~B. Butler}.} \bibinfo{year}{2016}\natexlab{}.
\newblock \showarticletitle{Crypto{L}ock (and Drop It): Stopping Ransomware Attacks on User Data}. In \bibinfo{booktitle}{\emph{2016 IEEE 36th International Conference on Distributed Computing Systems (ICDCS)}}. \bibinfo{pages}{303--312}.
\newblock


\bibitem[Scalas et~al\mbox{.}(2019)]%
        {scalas2019effectiveness}
\bibfield{author}{\bibinfo{person}{Michele Scalas}, \bibinfo{person}{Davide Maiorca}, \bibinfo{person}{Francesco Mercaldo}, \bibinfo{person}{Corrado~Aaron Visaggio}, \bibinfo{person}{Fabio Martinelli}, {and} \bibinfo{person}{Giorgio Giacinto}.} \bibinfo{year}{2019}\natexlab{}.
\newblock \showarticletitle{On the effectiveness of system API-related information for Android ransomware detection}.
\newblock \bibinfo{journal}{\emph{Computers \& Security}}  \bibinfo{volume}{86} (\bibinfo{year}{2019}), \bibinfo{pages}{168--182}.
\newblock


\bibitem[{S}cram{B}ox(2016)]%
        {AES-brute-force}
\bibfield{author}{\bibinfo{person}{{S}cram{B}ox}.} \bibinfo{year}{2016}\natexlab{}.
\newblock \bibinfo{title}{How long would it take to brute force AES-256?}
\newblock \bibinfo{howpublished}{\url{https://scrambox.com/article/brute-force-aes/}}.
\newblock


\bibitem[Sendner et~al\mbox{.}(2022)]%
        {database-ransomware}
\bibfield{author}{\bibinfo{person}{Christoph Sendner}, \bibinfo{person}{Lukas Iffländer}, \bibinfo{person}{Sebastian Schindler}, \bibinfo{person}{Michael Jobst}, \bibinfo{person}{Alexandra Dmitrienko}, {and} \bibinfo{person}{Samuel Kounev}.} \bibinfo{year}{2022}\natexlab{}.
\newblock \showarticletitle{Ransomware Detection in Databases through Dynamic Analysis of Query Sequences}. In \bibinfo{booktitle}{\emph{2022 IEEE Conference on Communications and Network Security (CNS)}}. \bibinfo{pages}{326--334}.
\newblock
\urldef\tempurl%
\url{https://doi.org/10.1109/CNS56114.2022.9947244}
\showDOI{\tempurl}


\bibitem[Sgandurra et~al\mbox{.}(2016)]%
        {EldeRan}
\bibfield{author}{\bibinfo{person}{Daniele Sgandurra}, \bibinfo{person}{Luis Muñoz-González}, \bibinfo{person}{Rabih Mohsen}, {and} \bibinfo{person}{Emil~C. Lupu}.} \bibinfo{year}{2016}\natexlab{}.
\newblock \bibinfo{title}{Automated Dynamic Analysis of Ransomware: Benefits, Limitations and use for Detection}.
\newblock \bibinfo{howpublished}{\url{https://arxiv.org/abs/1609.03020}}.
\newblock
\urldef\tempurl%
\url{https://doi.org/10.48550/ARXIV.1609.03020}
\showDOI{\tempurl}


\bibitem[Sharmeen et~al\mbox{.}(2020)]%
        {sharmeen2020avoiding}
\bibfield{author}{\bibinfo{person}{Shaila Sharmeen}, \bibinfo{person}{Yahye~Abukar Ahmed}, \bibinfo{person}{Shamsul Huda}, \bibinfo{person}{Bari~{\c{S}} Ko{\c{c}}er}, {and} \bibinfo{person}{Mohammad~Mehedi Hassan}.} \bibinfo{year}{2020}\natexlab{}.
\newblock \showarticletitle{Avoiding future digital extortion through robust protection against ransomware threats using deep learning based adaptive approaches}.
\newblock \bibinfo{journal}{\emph{IEEE Access}}  \bibinfo{volume}{8} (\bibinfo{year}{2020}), \bibinfo{pages}{24522--24534}.
\newblock


\bibitem[Shaukat and Ribeiro(2018)]%
        {shaukat2018ransomwall}
\bibfield{author}{\bibinfo{person}{Saiyed~Kashif Shaukat} {and} \bibinfo{person}{Vinay~J Ribeiro}.} \bibinfo{year}{2018}\natexlab{}.
\newblock \showarticletitle{RansomWall: A layered defense system against cryptographic ransomware attacks using machine learning}. In \bibinfo{booktitle}{\emph{2018 10th international conference on communication systems \& networks (COMSNETS)}}. IEEE, \bibinfo{pages}{356--363}.
\newblock


\bibitem[Shi et~al\mbox{.}(2017)]%
        {Shi2017}
\bibfield{author}{\bibinfo{person}{Hao Shi}, \bibinfo{person}{Jelena Mirkovic}, {and} \bibinfo{person}{Abdulla Alwabel}.} \bibinfo{year}{2017}\natexlab{}.
\newblock \showarticletitle{Handling Anti-Virtual Machine Techniques in Malicious Software}.
\newblock \bibinfo{journal}{\emph{ACM Transactions on Privacy and Security (TOPS)}}  \bibinfo{volume}{21} (\bibinfo{date}{12} \bibinfo{year}{2017}), \bibinfo{pages}{31}.
\newblock
Issue 1.
\showISSN{24712574}


\bibitem[Shinde et~al\mbox{.}(2016)]%
        {shinde2016ransomware}
\bibfield{author}{\bibinfo{person}{Rhythima Shinde}, \bibinfo{person}{Pieter Van~der Veeken}, \bibinfo{person}{Stijn Van~Schooten}, {and} \bibinfo{person}{Jan van~den Berg}.} \bibinfo{year}{2016}\natexlab{}.
\newblock \showarticletitle{Ransomware: Studying transfer and mitigation}. In \bibinfo{booktitle}{\emph{2016 International Conference on Computing, Analytics and Security Trends (CAST)}}. IEEE, \bibinfo{pages}{90--95}.
\newblock


\bibitem[Silva et~al\mbox{.}(2019)]%
        {silva2019survey}
\bibfield{author}{\bibinfo{person}{Juan A~Herrera Silva}, \bibinfo{person}{Lorena Isabel~Barona L{\'o}pez}, \bibinfo{person}{{\'A}ngel Leonardo~Valdivieso Caraguay}, {and} \bibinfo{person}{Myriam Hern{\'a}ndez-{\'A}lvarez}.} \bibinfo{year}{2019}\natexlab{}.
\newblock \showarticletitle{A survey on situational awareness of ransomware attacks—detection and prevention parameters}.
\newblock \bibinfo{journal}{\emph{Remote Sensing}} \bibinfo{volume}{11}, \bibinfo{number}{10} (\bibinfo{year}{2019}).
\newblock


\bibitem[Singh et~al\mbox{.}(2023)]%
        {SINGH2023108601}
\bibfield{author}{\bibinfo{person}{Jaskaran Singh}, \bibinfo{person}{Keshav Sharma}, \bibinfo{person}{Mohammad Wazid}, {and} \bibinfo{person}{Ashok~Kumar Das}.} \bibinfo{year}{2023}\natexlab{}.
\newblock \showarticletitle{SINN-RD: Spline interpolation-envisioned neural network-based ransomware detection scheme}.
\newblock \bibinfo{journal}{\emph{Computers and Electrical Engineering}}  \bibinfo{volume}{106} (\bibinfo{year}{2023}), \bibinfo{pages}{108601}.
\newblock
\showISSN{0045-7906}


\bibitem[{S}onic{W}all(2021)]%
        {9attempts}
\bibfield{author}{\bibinfo{person}{{S}onic{W}all}.} \bibinfo{year}{2021}\natexlab{}.
\newblock \bibinfo{title}{‘The Year of Ransomware’ Continues with Unprecedented Late-Summer Surge}.
\newblock \bibinfo{howpublished}{\url{https://www.sonicwall.com/news/sonicwall-the-year-of-ransomware-continues-with-unprecedented-late-summer-surge/}}.
\newblock


\bibitem[Sophos({[n.\,d.]})]%
        {sophos2021}
\bibfield{author}{\bibinfo{person}{Sophos}.} \bibinfo{year}{[n.\,d.]}\natexlab{}.
\newblock \bibinfo{title}{The State of Ransomware 2021}.
\newblock \bibinfo{howpublished}{\url{https://secure2.sophos.com/en-us/medialibrary/pdfs/whitepaper/sophos-state-of-ransomware-2021-wp.pdf}}.
\newblock


\bibitem[Sophos(2018)]%
        {77protection}
\bibfield{author}{\bibinfo{person}{Sophos}.} \bibinfo{year}{2018}\natexlab{}.
\newblock \bibinfo{title}{Businesses Impacted by Repeated Ransomware Attacks and Failing to Close the Gap on Exploits, According to Sophos Global Survey}.
\newblock \bibinfo{howpublished}{\url{https://www.sophos.com/en-us/press-office/press-releases/2018/01/businesses-impacted-by-repeated-ransomware-attacks-according-to-sophos-global-survey}}.
\newblock


\bibitem[Su et~al\mbox{.}(2018)]%
        {su2018detecting}
\bibfield{author}{\bibinfo{person}{Dan Su}, \bibinfo{person}{Jiqiang Liu}, \bibinfo{person}{Xiaoyang Wang}, {and} \bibinfo{person}{Wei Wang}.} \bibinfo{year}{2018}\natexlab{}.
\newblock \showarticletitle{Detecting Android locker-ransomware on chinese social networks}.
\newblock \bibinfo{journal}{\emph{IEEE Access}}  \bibinfo{volume}{7} (\bibinfo{year}{2018}), \bibinfo{pages}{20381--20393}.
\newblock


\bibitem[Sultan et~al\mbox{.}(2018)]%
        {Sultan2018}
\bibfield{author}{\bibinfo{person}{Hirra Sultan}, \bibinfo{person}{Aqeel Khalique}, \bibinfo{person}{Safdar Tanweer}, {and} \bibinfo{person}{Shah~Imran Alam}.} \bibinfo{year}{2018}\natexlab{}.
\newblock \showarticletitle{A Survey on Ransomware: Evolution, Growth, AND Impact}.
\newblock \bibinfo{journal}{\emph{International Journal of Advanced Research in Computer Science}}  \bibinfo{volume}{9} (\bibinfo{date}{4} \bibinfo{year}{2018}), \bibinfo{pages}{802--810}.
\newblock
Issue 2.
\showISSN{0976-5697}


\bibitem[Takeuchi et~al\mbox{.}(2018)]%
        {detection-svm}
\bibfield{author}{\bibinfo{person}{Yuki Takeuchi}, \bibinfo{person}{Kazuya Sakai}, {and} \bibinfo{person}{Satoshi Fukumoto}.} \bibinfo{year}{2018}\natexlab{}.
\newblock \showarticletitle{Detecting Ransomware Using Support Vector Machines}. In \bibinfo{booktitle}{\emph{Workshop Proceedings of the 47th International Conference on Parallel Processing}} (Eugene, OR, USA) \emph{(\bibinfo{series}{ICPP Workshops '18})}. Article \bibinfo{articleno}{1}, \bibinfo{numpages}{6}~pages.
\newblock
\showISBNx{9781450365239}


\bibitem[Tang et~al\mbox{.}(2020)]%
        {ransomspector}
\bibfield{author}{\bibinfo{person}{Fei Tang}, \bibinfo{person}{Boyang Ma}, \bibinfo{person}{Jinku Li}, \bibinfo{person}{Fengwei Zhang}, \bibinfo{person}{Jipeng Su}, {and} \bibinfo{person}{Jianfeng Ma}.} \bibinfo{year}{2020}\natexlab{}.
\newblock \showarticletitle{Ransom{S}pector: An introspection-based approach to detect crypto ransomware}.
\newblock \bibinfo{journal}{\emph{Computers \& Security}}  \bibinfo{volume}{97} (\bibinfo{year}{2020}), \bibinfo{pages}{101997}.
\newblock
\showISSN{0167-4048}


\bibitem[Thummapudi et~al\mbox{.}(2023)]%
        {hardware23}
\bibfield{author}{\bibinfo{person}{Kumar Thummapudi}, \bibinfo{person}{Palden Lama}, {and} \bibinfo{person}{Rajendra~V. Boppana}.} \bibinfo{year}{2023}\natexlab{}.
\newblock \showarticletitle{Detection of Ransomware Attacks Using Processor and Disk Usage Data}.
\newblock \bibinfo{journal}{\emph{IEEE Access}}  \bibinfo{volume}{11} (\bibinfo{year}{2023}), \bibinfo{pages}{51395--51407}.
\newblock
\urldef\tempurl%
\url{https://doi.org/10.1109/ACCESS.2023.3279819}
\showDOI{\tempurl}


\bibitem[Times(2021)]%
        {pipeline}
\bibfield{author}{\bibinfo{person}{The New~York Times}.} \bibinfo{year}{2021}\natexlab{}.
\newblock \bibinfo{title}{Colonial Pipeline Paid Roughly \$5 Million in Bitcoin to Hackers}.
\newblock \bibinfo{howpublished}{\url{https://www.nytimes.com/2021/05/13/technology/colonial-pipeline-ransom.html}}.
\newblock


\bibitem[Verma et~al\mbox{.}(2018)]%
        {verma2018analysing}
\bibfield{author}{\bibinfo{person}{Mayank Verma}, \bibinfo{person}{Ponnurangam Kumarguru}, \bibinfo{person}{Shuva~Brata Deb}, {and} \bibinfo{person}{Anuradha Gupta}.} \bibinfo{year}{2018}\natexlab{}.
\newblock \showarticletitle{Analysing indicator of compromises for ransomware: Leveraging IOCs with machine learning techniques}. In \bibinfo{booktitle}{\emph{2018 IEEE International Conference on Intelligence and Security Informatics (ISI)}}. IEEE, \bibinfo{pages}{154--159}.
\newblock


\bibitem[Vinayakumar et~al\mbox{.}(2017)]%
        {vinayakumar2017evaluating}
\bibfield{author}{\bibinfo{person}{R Vinayakumar}, \bibinfo{person}{KP Soman}, \bibinfo{person}{KK~Senthil Velan}, {and} \bibinfo{person}{Shaunak Ganorkar}.} \bibinfo{year}{2017}\natexlab{}.
\newblock \showarticletitle{Evaluating shallow and deep networks for ransomware detection and classification}. In \bibinfo{booktitle}{\emph{2017 international conference on advances in computing, communications and informatics (ICACCI)}}. IEEE, \bibinfo{pages}{259--265}.
\newblock


\bibitem[Wang et~al\mbox{.}(2019)]%
        {MimosaFTL}
\bibfield{author}{\bibinfo{person}{Peiying Wang}, \bibinfo{person}{Shijie Jia}, \bibinfo{person}{Bo Chen}, \bibinfo{person}{Luning Xia}, {and} \bibinfo{person}{Peng Liu}.} \bibinfo{year}{2019}\natexlab{}.
\newblock \showarticletitle{MimosaFTL: Adding Secure and Practical Ransomware Defense Strategy to Flash Translation Layer}. In \bibinfo{booktitle}{\emph{Proceedings of the Ninth ACM Conference on Data and Application Security and Privacy}} (Richardson, Texas, USA) \emph{(\bibinfo{series}{CODASPY '19})}. \bibinfo{address}{New York, NY, USA}, \bibinfo{pages}{327–338}.
\newblock
\showISBNx{9781450360999}


\bibitem[Wang et~al\mbox{.}(2018)]%
        {wang2018automatically}
\bibfield{author}{\bibinfo{person}{ZiHan Wang}, \bibinfo{person}{ChaoGe Liu}, \bibinfo{person}{Jing Qiu}, \bibinfo{person}{ZhiHong Tian}, \bibinfo{person}{Xiang Cui}, {and} \bibinfo{person}{Shen Su}.} \bibinfo{year}{2018}\natexlab{}.
\newblock \showarticletitle{Automatically traceback RDP-based targeted ransomware attacks}.
\newblock \bibinfo{journal}{\emph{Wireless Communications and Mobile Computing}}  \bibinfo{volume}{2018} (\bibinfo{year}{2018}), \bibinfo{pages}{1--13}.
\newblock


\bibitem[Wani and Revathi(2020)]%
        {wani2020ransomware}
\bibfield{author}{\bibinfo{person}{Azka Wani} {and} \bibinfo{person}{S Revathi}.} \bibinfo{year}{2020}\natexlab{}.
\newblock \showarticletitle{Ransomware protection in loT using software defined networking}.
\newblock \bibinfo{journal}{\emph{Int. J. Electr. Comput. Eng}} \bibinfo{volume}{10}, \bibinfo{number}{3} (\bibinfo{year}{2020}), \bibinfo{pages}{3166--3175}.
\newblock


\bibitem[Yamany et~al\mbox{.}(2022)]%
        {yamany2022new}
\bibfield{author}{\bibinfo{person}{Bahaa Yamany}, \bibinfo{person}{Mahmoud~Said Elsayed}, \bibinfo{person}{Anca~D Jurcut}, \bibinfo{person}{Nashwa Abdelbaki}, {and} \bibinfo{person}{Marianne~A Azer}.} \bibinfo{year}{2022}\natexlab{}.
\newblock \showarticletitle{A New Scheme for Ransomware Classification and Clustering Using Static Features}.
\newblock \bibinfo{journal}{\emph{Electronics}} \bibinfo{volume}{11}, \bibinfo{number}{20} (\bibinfo{year}{2022}), \bibinfo{pages}{3307}.
\newblock


\bibitem[Yong~Wong et~al\mbox{.}(2021)]%
        {inside-look}
\bibfield{author}{\bibinfo{person}{Miuyin Yong~Wong}, \bibinfo{person}{Matthew Landen}, \bibinfo{person}{Manos Antonakakis}, \bibinfo{person}{Douglas~M. Blough}, \bibinfo{person}{Elissa~M. Redmiles}, {and} \bibinfo{person}{Mustaque Ahamad}.} \bibinfo{year}{2021}\natexlab{}.
\newblock \showarticletitle{An Inside Look into the Practice of Malware Analysis}. In \bibinfo{booktitle}{\emph{Proceedings of the 2021 ACM SIGSAC Conference on Computer and Communications Security}} (Virtual Event, Republic of Korea). \bibinfo{pages}{3053–3069}.
\newblock
\showISBNx{9781450384544}


\bibitem[Yuste and Pastrana(2021)]%
        {Yuste2021Avaddon}
\bibfield{author}{\bibinfo{person}{Javier Yuste} {and} \bibinfo{person}{Sergio Pastrana}.} \bibinfo{year}{2021}\natexlab{}.
\newblock \showarticletitle{Avaddon ransomware: An in-depth analysis and decryption of infected systems}.
\newblock \bibinfo{journal}{\emph{Computers \& Security}}  \bibinfo{volume}{109} (\bibinfo{year}{2021}).
\newblock
\showISSN{0167-4048}


\bibitem[Zhang et~al\mbox{.}(2020)]%
        {opcode2020}
\bibfield{author}{\bibinfo{person}{Bin Zhang}, \bibinfo{person}{Wentao Xiao}, \bibinfo{person}{Xi Xiao}, \bibinfo{person}{Arun~Kumar Sangaiah}, \bibinfo{person}{Weizhe Zhang}, {and} \bibinfo{person}{Jiajia Zhang}.} \bibinfo{year}{2020}\natexlab{}.
\newblock \showarticletitle{Ransomware classification using patch-based CNN and self-attention network on embedded N-grams of opcodes}.
\newblock \bibinfo{journal}{\emph{Future Generation Computer Systems}}  \bibinfo{volume}{110} (\bibinfo{year}{2020}), \bibinfo{pages}{708--720}.
\newblock


\bibitem[Zhang et~al\mbox{.}(2019)]%
        {opcode2019}
\bibfield{author}{\bibinfo{person}{Hanqi Zhang}, \bibinfo{person}{Xi Xiao}, \bibinfo{person}{Francesco Mercaldo}, \bibinfo{person}{Shiguang Ni}, \bibinfo{person}{Fabio Martinelli}, {and} \bibinfo{person}{Arun~Kumar Sangaiah}.} \bibinfo{year}{2019}\natexlab{}.
\newblock \showarticletitle{Classification of ransomware families with machine learning based on N-gram of opcodes}.
\newblock \bibinfo{journal}{\emph{Future Generation Computer Systems}}  \bibinfo{volume}{90} (\bibinfo{year}{2019}), \bibinfo{pages}{211--221}.
\newblock


\bibitem[Zhou et~al\mbox{.}(2023)]%
        {zhou2023limits}
\bibfield{author}{\bibinfo{person}{Chijin Zhou}, \bibinfo{person}{Lihua Guo}, \bibinfo{person}{Yiwei Hou}, \bibinfo{person}{Zhenya Ma}, \bibinfo{person}{Quan Zhang}, \bibinfo{person}{Mingzhe Wang}, \bibinfo{person}{Zhe Liu}, {and} \bibinfo{person}{Yu Jiang}.} \bibinfo{year}{2023}\natexlab{}.
\newblock \showarticletitle{Limits of I/O Based Ransomware Detection: An Imitation Based Attack}. In \bibinfo{booktitle}{\emph{2023 IEEE Symposium on Security and Privacy (SP)}}. IEEE Computer Society, \bibinfo{pages}{2584--2601}.
\newblock


\bibitem[Zhou et~al\mbox{.}(2020)]%
        {zhou2020evaluation}
\bibfield{author}{\bibinfo{person}{Jiaxing Zhou}, \bibinfo{person}{Miyuki Hirose}, \bibinfo{person}{Yoshio Kakizaki}, {and} \bibinfo{person}{Atsuo Inomata}.} \bibinfo{year}{2020}\natexlab{}.
\newblock \showarticletitle{Evaluation to Classify Ransomware Variants based on Correlations between APIs.}. In \bibinfo{booktitle}{\emph{ICISSP}}. \bibinfo{pages}{465--472}.
\newblock


\bibitem[Zimba and Chishimba(2019)]%
        {zimba2019understanding}
\bibfield{author}{\bibinfo{person}{Aaron Zimba} {and} \bibinfo{person}{Mumbi Chishimba}.} \bibinfo{year}{2019}\natexlab{}.
\newblock \showarticletitle{Understanding the evolution of ransomware: paradigm shifts in attack structures}.
\newblock \bibinfo{journal}{\emph{International Journal of computer network and information security}} \bibinfo{volume}{11}, \bibinfo{number}{1} (\bibinfo{year}{2019}), \bibinfo{pages}{26}.
\newblock


\bibitem[Zuhair and Selamat(2019)]%
        {zuhair2019rands}
\bibfield{author}{\bibinfo{person}{Hiba Zuhair} {and} \bibinfo{person}{Ali Selamat}.} \bibinfo{year}{2019}\natexlab{}.
\newblock \showarticletitle{RANDS: A machine learning-based anti-ransomware tool for windows platforms}.
\newblock In \bibinfo{booktitle}{\emph{Advancing Technology Industrialization Through Intelligent Software Methodologies, Tools and Techniques}}. \bibinfo{publisher}{IOS Press}, \bibinfo{pages}{573--587}.
\newblock


\bibitem[Zuhair et~al\mbox{.}(2020)]%
        {zuhair2020multi}
\bibfield{author}{\bibinfo{person}{Hiba Zuhair}, \bibinfo{person}{Ali Selamat}, {and} \bibinfo{person}{Ondrej Krejcar}.} \bibinfo{year}{2020}\natexlab{}.
\newblock \showarticletitle{A multi-tier streaming analytics model of 0-day ransomware detection using machine learning}.
\newblock \bibinfo{journal}{\emph{Applied Sciences}} \bibinfo{volume}{10}, \bibinfo{number}{9} (\bibinfo{year}{2020}), \bibinfo{pages}{3210}.
\newblock


\end{thebibliography}

\section*{Appendix}

\begin{table}
\centering

\caption{Evaluation results of commercial malware scanners. The plain file column shows the results of scanning plain .exe ransomware samples. Simple password and complex password columns show the scanning results for simple password- and complex password-compressed files (in .zip format), respectively. The number before the slash is the number of successful detections. The number after is the number of total scanners available for this sample. This number varies slightly for each sample. Last row shows the average number of detection on 54 samples. SHA256 is the first 4 digits of the sample’s SHA256 hash. The samples are uploaded for scanning in May 2022 and December 2022.}

\scriptsize\begin{tabular}{
>{\columncolor[HTML]{EFEFEF}}c c
>{\columncolor[HTML]{EFEFEF}}c c
>{\columncolor[HTML]{EFEFEF}}c }
\hline
\cellcolor[HTML]{C0C0C0}\textbf{\begin{tabular}[c]{@{}c@{}}Ransomware\\ Family\end{tabular}} & \cellcolor[HTML]{C0C0C0}\textbf{SHA256} & \cellcolor[HTML]{C0C0C0}\textbf{\begin{tabular}[c]{@{}c@{}}Plain\\ File\end{tabular}} & \cellcolor[HTML]{C0C0C0}\textbf{\begin{tabular}[c]{@{}c@{}}Simple\\ Password\end{tabular}} & \cellcolor[HTML]{C0C0C0}\textbf{\begin{tabular}[c]{@{}c@{}}Complex\\ Password\end{tabular}} \\ \hline
\cellcolor[HTML]{EFEFEF}                                                                     & 000e                                    & 56/71                                                                                 & 0/62                                                                                       & 0/62                                                                                        \\ \hhline{|>{\arrayrulecolor{tgray}}->{\arrayrulecolor{tblack}}|-|-|-|-|} 
\cellcolor[HTML]{EFEFEF}                                                                     & 0047                                    & 62/71                                                                                 & 0/64                                                                                       & 0/61                                                                                        \\ \hhline{|>{\arrayrulecolor{tgray}}->{\arrayrulecolor{tblack}}|-|-|-|-|}
\multirow{-3}{*}{\cellcolor[HTML]{EFEFEF}7ev3n}                                              & 0084                                    & 57/70                                                                                 & 0/62                                                                                       & 0/62                                                                                        \\ \hline
Alcatraz                                                                                     & 9185                                    & 52/70                                                                                 & 1/61                                                                                       & 0/61                                                                                        \\ \hline
\cellcolor[HTML]{EFEFEF}                                                                     & 7188                                    & 48/64                                                                                 & 2/59                                                                                       & 1/59                                                                                        \\ \hhline{|>{\arrayrulecolor{tgray}}->{\arrayrulecolor{tblack}}|-|-|-|-|}
\multirow{-2}{*}{\cellcolor[HTML]{EFEFEF}AvosLocker}                                         & f810                                    & 51/66                                                                                 & 1/57                                                                                       & 0/58                                                                                        \\ \hline
Babuk                                                                                        & eb18                                    & 54/67                                                                                 & 1/58                                                                                       & 0/59                                                                                        \\ \hline
\cellcolor[HTML]{EFEFEF}                                                                     & 9a55                                    & 52/70                                                                                 & 1/62                                                                                       & 1/62                                                                                        \\ \hhline{|>{\arrayrulecolor{tgray}}->{\arrayrulecolor{tblack}}|-|-|-|-|}
\multirow{-2}{*}{\cellcolor[HTML]{EFEFEF}BlackBasta}                                         & 7883                                    & 60/70                                                                                 & 1/62                                                                                       & 1/62                                                                                        \\ \hline
Cerber                                                                                       & 0cd2                                    & 49/69                                                                                 & 1/57                                                                                       & 0/58                                                                                        \\ \hline
Dharma                                                                                       & dc5b                                    & 52/68                                                                                 & 0/62                                                                                       & 0/62                                                                                        \\ \hline
DoejoCrypt                                                                                   & e044                                    & 56/72                                                                                 & 0/62                                                                                       & 0/64                                                                                        \\ \hline
HelloKitty                                                                                   & fa72                                    & 59/69                                                                                 & 1/57                                                                                       & 0/60                                                                                        \\ \hline
Hive                                                                                         & 47db                                    & 49/69                                                                                 & 1/58                                                                                       & 0/59                                                                                        \\ \hline
\cellcolor[HTML]{EFEFEF}                                                                     & 9c74                                    & 48/67                                                                                 & 2/58                                                                                       & 1/61                                                                                        \\ \hhline{|>{\arrayrulecolor{tgray}}->{\arrayrulecolor{tblack}}|-|-|-|-|} 
\cellcolor[HTML]{EFEFEF}                                                                     & 3ae9                                    & 59/68                                                                                 & 1/62                                                                                       & 1/62                                                                                        \\ \hhline{|>{\arrayrulecolor{tgray}}->{\arrayrulecolor{tblack}}|-|-|-|-|} 
\multirow{-3}{*}{\cellcolor[HTML]{EFEFEF}Jigsaw}                                             & df04                                    & 56/72                                                                                 & 1/62                                                                                       & 1/62                                                                                        \\ \hline
Karma                                                                                        & 6c98                                    & 55/68                                                                                 & 2/58                                                                                       & 1/58                                                                                        \\ \hline
\cellcolor[HTML]{EFEFEF}                                                                     & a2ad                                    & 55/70                                                                                 & 1/58                                                                                       & 0/61                                                                                        \\ \hhline{|>{\arrayrulecolor{tgray}}->{\arrayrulecolor{tblack}}|-|-|-|-|} 
\multirow{-2}{*}{\cellcolor[HTML]{EFEFEF}LockBit}                                            & dec4                                    & 59/69                                                                                 & 1/58                                                                                       & 0/57                                                                                        \\ \hline
LockFile                                                                                     & 2a23                                    & 53/68                                                                                 & 1/58                                                                                       & 0/57                                                                                        \\ \hline
\cellcolor[HTML]{EFEFEF}                                                                     & 1264                                    & 49/68                                                                                 & 2/57                                                                                       & 1/58                                                                                        \\ \hhline{|>{\arrayrulecolor{tgray}}->{\arrayrulecolor{tblack}}|-|-|-|-|} 
\cellcolor[HTML]{EFEFEF}                                                                     & a0cc                                    & 53/69                                                                                 & 2/59                                                                                       & 1/61                                                                                        \\ \hhline{|>{\arrayrulecolor{tgray}}->{\arrayrulecolor{tblack}}|-|-|-|-|} 
\multirow{-3}{*}{\cellcolor[HTML]{EFEFEF}Lorenz}                                             & edc2                                    & 55/69                                                                                 & 1/57                                                                                       & 0/60                                                                                        \\ \hline
MarraCrypt                                                                                   & be88                                    & 60/71                                                                                 & 0/62                                                                                       & 0/62                                                                                        \\ \hline
\cellcolor[HTML]{EFEFEF}                                                                     & 0abb                                    & 49/67                                                                                 & 1/59                                                                                       & 0/59                                                                                        \\ \hhline{|>{\arrayrulecolor{tgray}}->{\arrayrulecolor{tblack}}|-|-|-|-|} 
\multirow{-2}{*}{\cellcolor[HTML]{EFEFEF}MedusaLocker}                                       & f5fb                                    & 51/68                                                                                 & 2/59                                                                                       & 1/59                                                                                        \\ \hline
\cellcolor[HTML]{EFEFEF}                                                                     & 4dc8                                    & 61/69                                                                                 & 1/58                                                                                       & 0/59                                                                                        \\ \hhline{|>{\arrayrulecolor{tgray}}->{\arrayrulecolor{tblack}}|-|-|-|-|} 
\cellcolor[HTML]{EFEFEF}                                                                     & 44f1                                    & 57/69                                                                                 & 1/58                                                                                       & 0/61                                                                                        \\ \hhline{|>{\arrayrulecolor{tgray}}->{\arrayrulecolor{tblack}}|-|-|-|-|} 
\multirow{-3}{*}{\cellcolor[HTML]{EFEFEF}Mespinoza}                                          & af99                                    & 58/69                                                                                 & 1/57                                                                                       & 0/58                                                                                        \\ \hline
MountLocker                                                                                  & 5eae                                    & 56/69                                                                                 & 1/59                                                                                       & 0/59                                                                                        \\ \hline
\cellcolor[HTML]{EFEFEF}                                                                     & 9dde                                    & 52/68                                                                                 & 1/58                                                                                       & 1/59                                                                                        \\ \hhline{|>{\arrayrulecolor{tgray}}->{\arrayrulecolor{tblack}}|-|-|-|-|} 
\cellcolor[HTML]{EFEFEF}                                                                     & 265d                                    & 58/67                                                                                 & 2/58                                                                                       & 1/58                                                                                        \\ \hhline{|>{\arrayrulecolor{tgray}}->{\arrayrulecolor{tblack}}|-|-|-|-|} 
\multirow{-3}{*}{\cellcolor[HTML]{EFEFEF}Phobos}                                             & 8710                                    & 62/70                                                                                 & 2/59                                                                                       & 1/61                                                                                        \\ \hline
Ragnarok                                                                                     & db8b                                    & 52/69                                                                                 & 1/61                                                                                       & 0/59                                                                                        \\ \hline
\cellcolor[HTML]{EFEFEF}                                                                     & 9eb7                                    & 61/68                                                                                 & 1/60                                                                                       & 0/59                                                                                        \\ \hhline{|>{\arrayrulecolor{tgray}}->{\arrayrulecolor{tblack}}|-|-|-|-|} 
\multirow{-2}{*}{\cellcolor[HTML]{EFEFEF}Ryuk}                                               & 40b8                                    & 58/68                                                                                 & 1/58                                                                                       & 0/60                                                                                        \\ \hline
Sage                                                                                         & ac27                                    & 65/71                                                                                 & 0/62                                                                                       & 1/62                                                                                        \\ \hline
SatanCryptor                                                                                 & dd28                                    & 56/71                                                                                 & 0/64                                                                                       & 0/62                                                                                        \\ \hline
Snatch                                                                                       & edad                                    & 53/69                                                                                 & 1/57                                                                                       & 0/61                                                                                        \\ \hline
\cellcolor[HTML]{EFEFEF}                                                                     & 9b11                                    & 61/68                                                                                 & 1/58                                                                                       & 0/58                                                                                        \\ \hhline{|>{\arrayrulecolor{tgray}}->{\arrayrulecolor{tblack}}|-|-|-|-|} 
\multirow{-2}{*}{\cellcolor[HTML]{EFEFEF}Sodinokibi}                                         & fd16                                    & 60/70                                                                                 & 1/57                                                                                       & 0/61                                                                                        \\ \hline
Sugar                                                                                        & 1d4f                                    & 52/67                                                                                 & 1/57                                                                                       & 0/60                                                                                        \\ \hline
SunCrypt                                                                                     & 759f                                    & 58/72                                                                                 & 1/62                                                                                       & 1/62                                                                                        \\ \hline
TellYouThePass                                                                               & 7697                                    & 47/68                                                                                 & 2/61                                                                                       & 1/59                                                                                        \\ \hline
TeslaCrypt                                                                                   & 4de6                                    & 61/71                                                                                 & 0/61                                                                                       & 0/62                                                                                        \\ \hline
\cellcolor[HTML]{EFEFEF}                                                                     & d609                                    & 61/72                                                                                 & 0/62                                                                                       & 0/62                                                                                        \\ \hhline{|>{\arrayrulecolor{tgray}}->{\arrayrulecolor{tblack}}|-|-|-|-|}
\cellcolor[HTML]{EFEFEF}                                                                     & ee03                                    & 61/72                                                                                 & 0/62                                                                                       & 0/64                                                                                        \\ \hhline{|>{\arrayrulecolor{tgray}}->{\arrayrulecolor{tblack}}|-|-|-|-|} 
\multirow{-3}{*}{\cellcolor[HTML]{EFEFEF}Venus}                                              & 59b0                                    & 58/72                                                                                 & 0/61                                                                                       & 0/64                                                                                        \\ \hline
\cellcolor[HTML]{EFEFEF}                                                                     & 7a92                                    & 60/67                                                                                 & 1/59                                                                                       & 0/50                                                                                        \\ \hhline{|>{\arrayrulecolor{tgray}}->{\arrayrulecolor{tblack}}|-|-|-|-|} 
\multirow{-2}{*}{\cellcolor[HTML]{EFEFEF}VirLock}                                            & f4b1                                    & 63/71                                                                                 & 1/59                                                                                       & 0/61                                                                                        \\ \hline
VoidCrypt                                                                                    & 4b78                                    & 53/71                                                                                 & 0/62                                                                                       & 0/60                                                                                        \\ \hline
WannaCry                                                                                     & ed01                                    & 60/67                                                                                 & 3/59                                                                                       & 2/59                                                                                        \\ \hline
Xorist                                                                                       & fb54                                    & 62/71                                                                                 & 1/59                                                                                       & 0/59                                                                                        \\ \hline
\textbf{Average}                                                                                       &                                     & \textbf{56}                                                                                & \textbf{0.96}                                                                                      & \textbf{0.33}                                                                                        \\ \hline
\end{tabular}

\label{tab:scanner}
\end{table}

\begin{table}
\centering
\caption{Top 29 important APIs for identifying ransomware execution from feature importance analysis based on a random forest model.}
\small 

\begin{tabular}{
>{\columncolor[HTML]{EFEFEF}}c c
>{\columncolor[HTML]{EFEFEF}}c c}
\hline
\multicolumn{2}{c}{\cellcolor[HTML]{C0C0C0}\textbf{API}} & \multicolumn{2}{c}{\cellcolor[HTML]{C0C0C0}\textbf{API}}                  \\ \hline
1              & RegEnumKeyExW                           & 16                                           & NtCreateKey                \\ \hline
2              & CreateDirectoryW                        & 17                                           & LoadResource               \\ \hline
3              & DrawTextExW                             & 18                                           & GetDiskFreeSpaceExW        \\ \hline
4              & CoInitializeEx                          & 19                                           & EnumWindows                \\ \hline
5              & NtDeleteKey                             & 20                                           & RegOpenKeyExW              \\ \hline
6              & SHGetFolderPathW                        & 21                                           & NtQueryKey                 \\ \hline
7              & GetFileInformationByHandleEx            & 22                                           & NtQueryValueKey            \\ \hline
8              & GetForegroundWindow                     & 23                                           & NtSetValueKey              \\ \hline
9              & NtQueryAttributesFile                   & 24                                           & CreateActCtxW              \\ \hline
10             & DeviceIoControl                         & 25                                           & GetSystemTimeAsFileTime    \\ \hline
11             & SearchPathW                             & 26                                           & GetSystemWindowsDirectoryW \\ \hline
12             & SetFileTime                             & 27                                           & SetErrorMode               \\ \hline
13             & SendNotifyMessageW                      & 28                                           & GetFileVersionInfoSizeW    \\ \hline
14             & GetSystemMetrics                        & 29                                           & NtOpenMutant               \\ \hline
15             & GetKeyState                             & \multicolumn{1}{l}{\cellcolor[HTML]{EFEFEF}} & \multicolumn{1}{l}{}       \\ \hline
\end{tabular}
\label{tab:top29}
\end{table}

\begin{table}
\centering
\caption{ Version information of antivirus software tested.}

\small \begin{tabular}{
>{\columncolor[HTML]{EFEFEF}}l l
>{\columncolor[HTML]{EFEFEF}}l }
\hline
\cellcolor[HTML]{C0C0C0}\textbf{} & \cellcolor[HTML]{C0C0C0}\textbf{Version/Build} & \cellcolor[HTML]{C0C0C0}\textbf{License} \\ \hline
Antivirus A                       & N/A                                            & N/A                                      \\ \hline
Antivirus B                       & N/A                                            & N/A                                      \\ \hline
Bitdefender                       & 26.0.18.75                                     & 30-day trail                             \\ \hline
Malwarebytes                      & 4.5.10.200                                     & 14-day trail                             \\ \hline
Kaspersky                         & 21.3.10.391 (h)                                & 30-day trail                             \\ \hline
McAfee                            & 16.0 R31                                       & 30-day trail                             \\ \hline
Norton                            & 22.22.4.11                                     & 30-day trail                             \\ \hline
360                               & 10.8.0.1465                                    & free                                     \\ \hline
\end{tabular}

\label{tab:av-version}
\end{table}

\begin{table*}[h]
\centering

\caption{The full SHA-256 hashes of the ransomware samples we measure. The exact samples can be found using the hashes. Year is the (possible) compilation time from the executable file. 1969 and 2010 might be intentional for anti-analysis reasons. Entropy is the file entropy calculated by Detect it Easy (DiE). A sample is packed if the entropy is over 6.5. Samples marked with yes in the used for evaluation column are used for testing antivirus software.}

\scriptsize\begin{tabular}{
>{\columncolor[HTML]{EFEFEF}}c c
>{\columncolor[HTML]{EFEFEF}}c c
>{\columncolor[HTML]{EFEFEF}}c c}
\hline
\cellcolor[HTML]{C0C0C0}\textbf{\begin{tabular}[c]{@{}c@{}}Ransomware\\ Family\end{tabular}} & \cellcolor[HTML]{C0C0C0}\textbf{\begin{tabular}[c]{@{}c@{}}Compiled\\ Year\end{tabular}} & \cellcolor[HTML]{C0C0C0}\textbf{SHA256}                          & \cellcolor[HTML]{C0C0C0}\textbf{Entropy} & \cellcolor[HTML]{C0C0C0}\textbf{\begin{tabular}[c]{@{}c@{}}if packed\end{tabular}} & \cellcolor[HTML]{C0C0C0}\textbf{\begin{tabular}[c]{@{}c@{}}Used for\\ evaluation\end{tabular}} \\ \hline
\cellcolor[HTML]{EFEFEF}                                                                     & 2016                                                                                     & 000ec059ab4eaefd2591449c6581b34748d3f90ef1688b9ec6daf5ab58d5da73 & 6.40 (80\%)                              &                                                                                      &                                                                                                \\ \hhline{|>{\arrayrulecolor{tgray}}->{\arrayrulecolor{tblack}}|-|-|-|-|-|} 
\cellcolor[HTML]{EFEFEF}                                                                     & 2016                                                                                     & 0047aed5ba539ab2e56e78d47b0ae8673d4f221bf5106987f66437e6eb0978ba & 6.38 (79\%)                              &                                                                                      &                                                                                                \\ \hhline{|>{\arrayrulecolor{tgray}}->{\arrayrulecolor{tblack}}|-|-|-|-|-|} 
\multirow{-3}{*}{\cellcolor[HTML]{EFEFEF}7ev3n}                                              & 2016                                                                                     & 0084af770e99180fcdc6778c513d36384cf4b3ff24d0f8bc62ecaa76651be616 & 6.40 (80\%)                              &                                                                                      &                                                                                                \\ \hline
Alcatraz                                                                                     & 2016                                                                                     & 918504ede26bb9a3aa315319da4d3549d64531afba593bfad71a653292899fec & 6.48 (81\%)                              &                                                                                      &                                                                                                \\ \hline
\cellcolor[HTML]{EFEFEF}                                                                     & 2021                                                                                     & 718810b8eeb682fc70df602d952c0c83e028c5a5bfa44c506756980caf2edebb & 6.63 (82\%)                              & yes                                                                                  & yes                                                                                            \\ \hhline{|>{\arrayrulecolor{tgray}}->{\arrayrulecolor{tblack}}|-|-|-|-|-|} 
\multirow{-2}{*}{\cellcolor[HTML]{EFEFEF}AvosLocker}                                         & 2021                                                                                     & f810deb1ba171cea5b595c6d3f816127fb182833f7a08a98de93226d4f6a336f & 6.63 (82\%)                              & yes                                                                                  &                                                                                                \\ \hline
Babuk                                                                                        & 2021                                                                                     & eb180fcc43380b15013d9fe42e658fc6f6c32cf23426ef10b89bc6548d40523b & 5.73 (71\%)                              &                                                                                      & yes                                                                                            \\ \hline
\cellcolor[HTML]{EFEFEF}                                                                     & 2022                                                                                     & 9a55f55886285eef7ffabdd55c0232d1458175b1d868c03d3e304ce7d98980bc & 6.62 (82\%)                              & yes                                                                                  &                                                                                                \\ \hhline{|>{\arrayrulecolor{tgray}}->{\arrayrulecolor{tblack}}|-|-|-|-|-|} 
\multirow{-2}{*}{\cellcolor[HTML]{EFEFEF}BlackBasta}                                         & 2022                                                                                     & 7883f01096db9bcf090c2317749b6873036c27ba92451b212b8645770e1f0b8a & 6.62 (82\%)                              &                                                                                      &                                                                                                \\ \hline
Cerber                                                                                       & 2015                                                                                     & 0cd28b912cf4d9898a6f03c4edfd73d1d90faf971ad84b28c6c254408ad7630f & 7.86 (98\%)                              & yes                                                                                  & yes                                                                                            \\ \hline
Dharma                                                                                       & 2017                                                                                     & dc5ba84e57cf8d8dfcb8fb2de6f842786428fc46c34d8a3e02c8119bbd9f7584 & 7.23 (90\%)                              & yes                                                                                  &                                                                                                \\ \hline
DoejoCrypt                                                                                   & 2021                                                                                     & e044d9f2d0f1260c3f4a543a1e67f33fcac265be114a1b135fd575b860d2b8c6 & 6.99 (87\%)                              & yes                                                                                  &                                                                                                \\ \hline
HelloKitty                                                                                   & 2020                                                                                     & fa722d0667418d68c4935e1461010a8f730f02fa1f595ee68bd0768fd5d1f8bb & 5.98 (26\%)                              &                                                                                      & yes                                                                                            \\ \hline
Hive                                                                                         & 1969                                                                                     & 47dbb2594cd5eb7015ef08b7fb803cd5adc1a1fbe4849dc847c0940f1ccace35 & 6.06 (75\%)                              &                                                                                      & yes                                                                                            \\ \hline
\cellcolor[HTML]{EFEFEF}                                                                     & 2020                                                                                     & 9c748a69c48b79e6422b3bea1766e415de5532cb7ba2b9673d5a51163e6c1df2 & 7.98 (99\%)                              & yes                                                                                  & yes                                                                                            \\ \hhline{|>{\arrayrulecolor{tgray}}->{\arrayrulecolor{tblack}}|-|-|-|-|-|} 
\cellcolor[HTML]{EFEFEF}                                                                     & 2016                                                                                     & 3ae96f73d805e1d3995253db4d910300d8442ea603737a1428b613061e7f61e7 & 7.68 (95\%)                              & yes                                                                                  &                                                                                                \\ \hhline{|>{\arrayrulecolor{tgray}}->{\arrayrulecolor{tblack}}|-|-|-|-|-|}  
\multirow{-3}{*}{\cellcolor[HTML]{EFEFEF}Jigsaw}                                             & 2020                                                                                     & df049efbfa7ac0b76c8daff5d792c550c7a7a24f6e9e887d01a01013c9caa763 & 7.61 (95\%)                              & yes                                                                                  &                                                                                                \\ \hline
Karma                                                                                        & 2021                                                                                     & 6c98d424ab1b9bfba683eda340fef6540ffe4ec4634f4b95cf9c70fe4ab2de90 & 5.87 (73\%)                              &                                                                                      & yes                                                                                            \\ \hline
\cellcolor[HTML]{EFEFEF}                                                                     & 2021                                                                                     & a2ad5cc5045a1645f07da7eab14ba13eb69ab7286204f61ba6a4226bfade7f17 & 6.68 (83\%)                              & yes                                                                                  & yes                                                                                            \\ \hhline{|>{\arrayrulecolor{tgray}}->{\arrayrulecolor{tblack}}|-|-|-|-|-|}  
\multirow{-2}{*}{\cellcolor[HTML]{EFEFEF}LockBit}                                            & 2021                                                                                     & dec4ca3a0863919f85c2a1a4a7e607e68063a9be1719ccb395353fe4a2d087e5 & 6.68 (83\%)                              & yes                                                                                  &                                                                                                \\ \hline
LockFile                                                                                     & 2021                                                                                     & 2a23fac4cfa697cc738d633ec00f3fbe93ba22d2498f14dea08983026fdf128a & 7.92 (98\%)                              & yes                                                                                  & yes                                                                                            \\ \hline
\cellcolor[HTML]{EFEFEF}                                                                     & 2021                                                                                     & 1264b40feaa824d5ba31cef3c8a4ede230c61ef71c8a7994875deefe32bd8b3d & 6.26 (78\%)                              &                                                                                      & yes                                                                                            \\ \hhline{|>{\arrayrulecolor{tgray}}->{\arrayrulecolor{tblack}}|-|-|-|-|-|} 
\cellcolor[HTML]{EFEFEF}                                                                     & 2021                                                                                     & a0ccb9019b90716c8ee1bc0829e0e04cf7166be2f25987abbc8987e65cef2e6f & 6.31 (78\%)                              &                                                                                      &                                                                                                \\ \hhline{|>{\arrayrulecolor{tgray}}->{\arrayrulecolor{tblack}}|-|-|-|-|-|} 
\multirow{-3}{*}{\cellcolor[HTML]{EFEFEF}Lorenz}                                             & 2021                                                                                     & edc2070fd8116f1df5c8d419189331ec606d10062818c5f3de865cd0f7d6db84 & 6.27 (78\%)                              &                                                                                      &                                                                                                \\ \hline
MarraCrypt                                                                                   & 2020                                                                                     & be88512c9250a558a3524e1c3bbd0299517cb0d6c3fb749c22df32033bf081e8 & 7.40 (92\%)                              & yes                                                                                  &                                                                                                \\ \hline
\cellcolor[HTML]{EFEFEF}                                                                     & 2021                                                                                     & f5fb7fa5231c18f0951c755c4cb0ec07b0889b5e320f42213cbf6bbbe499ad31 & 5.57 (69\%)                              &                                                                                      &                                                                                                \\ \hhline{|>{\arrayrulecolor{tgray}}->{\arrayrulecolor{tblack}}|-|-|-|-|-|} 
\multirow{-2}{*}{\cellcolor[HTML]{EFEFEF}MedusaLocker}                                       & 2021                                                                                     & 0abb4a302819cdca6c9f56893ca2b52856b55a0aa68a3cb8bdcd55dcc1fad9ad & 5.57 (69\%)                              &                                                                                      &                                                                                                \\ \hline
\cellcolor[HTML]{EFEFEF}                                                                     & 2020                                                                                     & 4dc802894c45ec4d119d002a7569be6c99a9bba732d0057364da9350f9d3659b & 6.65 (83\%)                              & yes                                                                                  & yes                                                                                            \\ \hhline{|>{\arrayrulecolor{tgray}}->{\arrayrulecolor{tblack}}|-|-|-|-|-|}  
\cellcolor[HTML]{EFEFEF}                                                                     & 2021                                                                                     & 44f1def68aef34687bfacf3668e56873f9d603fc6741d5da1209cc55bdc6f1f9 & 6.65 (83\%)                              & yes                                                                                  &                                                                                                \\ \hhline{|>{\arrayrulecolor{tgray}}->{\arrayrulecolor{tblack}}|-|-|-|-|-|} 
\multirow{-3}{*}{\cellcolor[HTML]{EFEFEF}Mespinoza}                                          & 2020                                                                                     & af99b482eb0b3ff976fa719bf0079da15f62a6c203911655ed93e52ae05c4ac8 & 6.65 (83\%)                              & yes                                                                                  &                                                                                                \\ \hline
MountLocker                                                                                  & 2020                                                                                     & 5eae13527d4e39059025c3e56dad966cf67476fe7830090e40c14d0a4046adf0 & 4.00 (50\%)                              &                                                                                      & yes                                                                                            \\ \hline
\cellcolor[HTML]{EFEFEF}                                                                     & 2020                                                                                     & 9dde984b21a00bc3307c28bd81f229500b795ce4e908b6f8cb5fbd338b22b8e1 & 3.38 (42\%)                              &                                                                                      & yes                                                                                            \\ \hhline{|>{\arrayrulecolor{tgray}}->{\arrayrulecolor{tblack}}|-|-|-|-|-|}  
\cellcolor[HTML]{EFEFEF}                                                                     & 2020                                                                                     & 265d1ae339e9397976d9328b2c84aca61a7cb6c0bca9f2f8dc213678e2b2ad86 & 6.97 (87\%)                              & yes                                                                                  &                                                                                                \\ \hhline{|>{\arrayrulecolor{tgray}}->{\arrayrulecolor{tblack}}|-|-|-|-|-|} 
\multirow{-3}{*}{\cellcolor[HTML]{EFEFEF}Phobos}                                             & 2020                                                                                     & 8710ad8fb2938326655335455987aa17961b2496a345a7ed9f4bbfcb278212bc & 6.70 (83\%)                              & yes                                                                                  &                                                                                                \\ \hline
Ragnarok                                                                                     & 2020                                                                                     & db8b499d613b604a439bca37c3be2f578bdfcde1b2271eccbcf22db85996e785 & 6.73 (84\%)                              & yes                                                                                  &                                                                                                \\ \hline
\cellcolor[HTML]{EFEFEF}                                                                     & 2021                                                                                     & 9eb7abf2228ad28d8b7f571e0495d4a35da40607f04355307077975e271553b8 & 6.42 (80\%)                              & yes                                                                                  & yes                                                                                            \\ \hhline{|>{\arrayrulecolor{tgray}}->{\arrayrulecolor{tblack}}|-|-|-|-|-|}  
\multirow{-2}{*}{\cellcolor[HTML]{EFEFEF}Ryuk}                                               & 2020                                                                                     & 40b865d1c3ab1b8544bcf57c88edd30679870d40b27d62feb237a19f0c5f9cd1 & 5.11 (63\%)                              &                                                                                      &                                                                                                \\ \hline
Sage                                                                                         & 2017                                                                                     & ac2736be4501b8c6823ebcf7241ceda38c3071418fb43c08b30f54f1a45d07e0 & 6.54 (81\%)                              & yes                                                                                  &                                                                                                \\ \hline
SatanCryptor                                                                                 & 1969                                                                                     & dd286a4d79d0f4c2b906073c7f46680252ca09c1c39b0dc12c92097c56662876 & 7.91 (98\%)                              & yes                                                                                  &                                                                                                \\ \hline
Snatch                                                                                       & 1969                                                                                     & edade6616334f3d313ac3ea7c3e432d8d9461cddad8e2ec3a94ffdc6e336a94e & 7.89 (98\%)                              & yes                                                                                  & yes                                                                                            \\ \hline
\cellcolor[HTML]{EFEFEF}                                                                     & 2021                                                                                     & 9b11711efed24b3c6723521a7d7eb4a52e4914db7420e278aa36e727459d59dd & 6.14 (76\%)                              &                                                                                      & yes                                                                                            \\ \hhline{|>{\arrayrulecolor{tgray}}->{\arrayrulecolor{tblack}}|-|-|-|-|-|}  
\multirow{-2}{*}{\cellcolor[HTML]{EFEFEF}Sodinokibi}                                         & 2021                                                                                     & fd164c4c121371f94cfd3a034ad8cf8edc7c0f7141a8f4c9da1683d41b212a87 & 6.76 (84\%)                              & yes                                                                                  &                                                                                                \\ \hline
Sugar                                                                                        & 2021                                                                                     & 1d4f0f02e613ccbbc47e32967371aa00f8d3dfcf388c39f0c55a911b8256f654 & 7.89 (98\%)                              & yes                                                                                  & yes                                                                                            \\ \hline
SunCrypt                                                                                     & 1969                                                                                     & 759f2b24be12e208903b00f9719db71a332ddf8252986c26afbcda9f32623bc4 & 6.88 (86\%)                              & yes                                                                                  &                                                                                                \\ \hline
TellYouThePass                                                                               & 1969                                                                                     & 76960749ed11d97582923e31d59115910ae74d8753c8e92f918f604ca8a0d26d & 6.04 (75\%)                              &                                                                                      & yes                                                                                            \\ \hline
TeslaCrypt                                                                                   & 2005                                                                                     & 4de6675c089aad8a52993b1a21afd06dc7086f4ea948755c09a7a8471e4fddbd & 7.57 (94\%)                              & yes                                                                                  &                                                                                                \\ \hline
\cellcolor[HTML]{EFEFEF}                                                                     & 2022                                                                                     & d6098f0d579273528b28b0b49c8b72b6f9908aef9e1ba0ec5da0874fa8c92266 & 7.04 (88\%)                              & yes                                                                                  &                                                                                                \\ \hhline{|>{\arrayrulecolor{tgray}}->{\arrayrulecolor{tblack}}|-|-|-|-|-|} 
\cellcolor[HTML]{EFEFEF}                                                                     & 2022                                                                                     & ee036f333a0c4a24d9aa09848e635639e481695a9209474900eb71c9e453256b & 7.04 (88\%)                              & yes                                                                                  &                                                                                                \\ \hhline{|>{\arrayrulecolor{tgray}}->{\arrayrulecolor{tblack}}|-|-|-|-|-|}  
\multirow{-3}{*}{\cellcolor[HTML]{EFEFEF}Venus}                                              & 2022                                                                                     & 59b05789e5ac3d47c0a3d0f3e4ccacb2667cb7367e42adb9a3cbb108a538fc77 & 7.04 (88\%)                              & yes                                                                                  &                                                                                                \\ \hline
\cellcolor[HTML]{EFEFEF}                                                                     & 2015                                                                                     & 7a92e23a6842cb51c9959892b83aa3be633d56ff50994e251b4fe82be1f2354c & 7.96 (99\%)                              & yes                                                                                  & yes                                                                                            \\ \hhline{|>{\arrayrulecolor{tgray}}->{\arrayrulecolor{tblack}}|-|-|-|-|-|}  
\multirow{-2}{*}{\cellcolor[HTML]{EFEFEF}VirLock}                                            & 2015                                                                                     & f4b11885a3056fc56efdedbc0dd71fae152368e4c2e96a3481c6dff21e9d75aa & 7.94 (99\%)                              & yes                                                                                  &                                                                                                \\ \hline
VoidCrypt                                                                                    & 2021                                                                                     & 4b78968928cfa5437ffdd56a39a5ea8c10a7b6dc5d3f342d003260088876b3cf & 7.96 (99\%)                              & yes                                                                                  &                                                                                                \\ \hline
WannaCry                                                                                     & 2010                                                                                     & ed01ebfbc9eb5bbea545af4d01bf5f1071661840480439c6e5babe8e080e41aa & 8.00 (99\%)                              & yes                                                                                  & yes                                                                                            \\ \hline
Xorist                                                                                       & 2012                                                                                     & fb54a1b85ab37cdee346e06cf716cbe0b071f4833020823595f3b69614c5446e & 7.20 (90\%)                              & yes                                                                                  &                                                                                                \\ \hline
\end{tabular}

\label{tab:sha256}
\end{table*}

\begin{table*}[]
\centering
\caption{The list of benign samples used. SHA\-256 is the hash of the sample we run. The \# alter column shows the number of malware scanners that mark the samples as malicious (false positives). The number after the slash is the number of scanners available for scanning this sample. The samples are uploaded for scanning in July 2022 and December 2022.}

\scriptsize \begin{tabular}{
>{\columncolor[HTML]{EFEFEF}}l l
>{\columncolor[HTML]{EFEFEF}}l c}
\hline
\cellcolor[HTML]{C0C0C0}\textbf{Software} & \cellcolor[HTML]{C0C0C0}\textbf{File Name}            & \cellcolor[HTML]{C0C0C0}\textbf{SHA-256}                         & \cellcolor[HTML]{C0C0C0}\textbf{\begin{tabular}[c]{@{}c@{}}\# Alters\\ from\\ Malware\\Scanners\end{tabular}} \\ \hline
7 Zip                                     & 7z2200-x64.exe                                        & 0b01c258a2e9857de86bd845deef59953cff283e6ed030dba3da529262484b00 & 1/68                                                                                                       \\ \hline
Atom                                      & atomsetup.exe                                         & ca69560bbc0f868301b1797580ce0d5dfe9a7822b0917897c2f3542393dde358 & 0/63                                                                                                       \\ \hline
Chorme                                    & ChromeSetup.exe                                       & 72222838e052e5151ecda0427eb0502b7a9395403b8be89f9a177aa8e9b43a5d & 0/69                                                                                                       \\ \hline
Discord                                   & DiscordSetup.exe                                      & ee9f94706055735af63117f1e6c80c0a50c72444d6a44f751ae2e33934910b58 & 0/66                                                                                                       \\ \hline
Ditto                                     & dittosetup\_64bit\_3\_21\_134\_0.exe                  & db4d049b9dde36b45659d97d88cbe35f2fbb3f31b8fd8ebbe682f1b700aabe7e & 0/59                                                                                                       \\ \hline
Dropbox                                   & dropbox\_99.4.501\_offline\_installer.exe             & f8b83cc9b7172002f2767c53696ec8e1a84af21d4d19bd6d9151c03d4e2521ea & 0/66                                                                                                       \\ \hline
Firefox                                   & Firefox Installer.exe                                 & 9b14ca825c3ce54440a32217e976fce33e4d2ab9492deb558943406023ef8c68 & 1/69                                                                                                       \\ \hline
Git                                       & git-2.37.1-32-bit.exe                                 & 714069fe4291c4ca7a51f7e7e81b0c94038590294f3b9e0981456a664c92966b & 0/67                                                                                                       \\ \hline
LibreOffice                               & LibreOffice\_7.3.4\_Win\_x64.msi                      & 509c70c1c8136805480146b55e4bad5dc73b11ee47b4682b43cf07670109e176 & 0/40                                                                                                       \\ \hline
NotePad++                                 & npp.8.4.3.installer.exe                               & 367893ed67fb585446bef612f8774e5f35eff9c2f89e9e89c006dce8f61d8128 & 0/69                                                                                                       \\ \hline
Python 3.8                                & python-3.8.5.exe                                      & f5fe57aeaa90ff4c5afed629b51880b53e4cabd0ebcadb33f56ca56fa1654de8 & 0/69                                                                                                       \\ \hline
Adobe PDF Reader DC                       & readerdc64\_en\_l\_cra\_install.exe                   & f492b470a1b60a5075cd4ebd5b52fa1274f4292a2c8dbd571af208c5b8690b7c & 0/64                                                                                                       \\ \hline
Spotify                                   & SpotifySetup.exe                                      & a8e15459a613063f3fc47ca1d77239615834831a2df3fdf4c7a270ff70a298d8 & 0/69                                                                                                       \\ \hline
Steam                                     & SteamSetup.exe                                        & 874788b45dfc043289ba05387e83f27b4a046004a88a4c5ee7c073187ff65b9d & 0/69                                                                                                       \\ \hline
TeamViewer                                & teamviewer\_setup.exe                                 & e463f3a11c4eafc698906876d610702fc9227a65183a30104579d8912ecdefe4 & 1/68                                                                                                       \\ \hline
Tree Size                                 & TreeSizeFreeSetup.exe                                 & 4de19445df877ef4df981fbead9440cf4a8832a284ea0e753ff1e7dd41dc10fa & 0/69                                                                                                       \\ \hline
WhatsApp                                  & whatsappsetup.exe                                     & 8aefba89a391331d8d3ad08f988c2a5ba0d69d04069f03a3121a01573da7be6c & 0/66                                                                                                       \\ \hline
WinRAR                                    & winrar-x32-611.exe                                    & 59276c49519ebd5194b95622c1c81d4b2c45d14eb6b07ea6d9f2b37c9c7bbf93 & 1/68                                                                                                       \\ \hline
WinZIP                                    & winzip26-home.exe                                     & a9ed6c5db282c4d42f4fd232627dab25b2f777b561a0065998a82b9e668d9f70 & 1/69                                                                                                       \\ \hline
Zoom meeting                              & zoominstallerfull.exe                                 & cdb3a3b20d65db7e51e345aa32075bc37b99dc8de86c5df950409bd56168da53 & 5/66                                                                                                       \\ \hline
NetBean IDE                               & apache-netbeans-13-bin-windows-x64.exe                & a06ea580a2bfe50bdc8c9791fed5c6032ce8330b16e0c6c5dbf6c9e1c931dc9e & 0/61                                                                                                       \\ \hline
Bitdefender                               & bitdefender\_avplus\_v10.exe                          & b973f4fe1f3bb9de06abd2f615d2f47c3c52810ee09d17255a6ba3c0a65eb801 & 0/65                                                                                                       \\ \hline
Dev-C++                                   & dev-cpp\_5.11\_tdm-gcc\_4.9.2\_setup.exe              & faad96bbcc51f115c9edd691785d1309e7663b67dcfcf7c11515c3d28c9c0f1f & 1/67                                                                                                       \\ \hline
Evernote                                  & Evernote-10.48.4-win-ddl-ga-3760-5f4dcc5719-setup.exe & a81ec8d119abaaef31cb46125f50c0089f054d085ca3b1f6927b48f3be40e9de & 0/66                                                                                                       \\ \hline
Google Drive                              & GoogleDriveSetup.exe                                  & 2cb39f4b8e640944c83e7eee34f0f886b58df23fa4141c225714cd6646b96575 & 0/67                                                                                                       \\ \hline
Grammarly                                 & grammarlysetup.exe                                    & 368e252f2e066bda82b8524c4ac939e5728154a334b991153a4fcbc3a2320f14 & 0/69                                                                                                       \\ \hline
iCloud                                    & iCloudSetup.exe                                       & 4cfd20d13cdce2b5c435f2ddaf4ee4c81d976461846bf3b954e8af6cbcdeb9f7 & 0/67                                                                                                       \\ \hline
KeePass password safe                     & keepass-2.52-setup.exe                                & da403bc2e91132d1c1e0c49f585441e4cd430c8195ca8af38adc2ea300de52cb & 0/71                                                                                                       \\ \hline
MalwareBytes                              & mbsetup.exe                                           & 057ac0f95e80abc5c73d9aefbc4e5e1bb778c2c154bf65c35435a34cdaf3da94 & 0/72                                                                                                       \\ \hline
Microsoft Teams                           & microsoft-teams-1-5-00-28567.exe                      & dcca2a974c673e21f3b5b11cee955fb20b14903c3218cef3b9d2b061cc8a0c30 & 0/61                                                                                                       \\ \hline
OneDrive                                  & OneDriveSetup.exe                                     & 83d2429a8568ee4ea0ed002c0897560c6b0a3e0b2a66f72a4149a521d461c6e7 & 0/71                                                                                                       \\ \hline
Paint.Net                                 & paint.net.4.0.21.install.exe                          & 088a02864e8daf807584fdd14ba3ed191979db0af301a318e7c1e8fc4c03dcbd & 3/68                                                                                                       \\ \hline
QuickTime                                 & quicktimeinstaller.exe                                & 56eff77b029b5f56c47d11fe58878627065dbeacbc3108d50d98a83420152c2b & 0/66                                                                                                       \\ \hline
Skype                                     & Skype-8.90.0.405.exe                                  & d073e31487c5584f12b263d0372288c049a1d316a75151801bb7e6ebb39766b1 & 0/69                                                                                                       \\ \hline
Slack                                     & SlackSetup.exe                                        & 0682a25eae6bfe3bc42e949aa4af0274c983690b726922217befb02d8e2f5306 & 0/69                                                                                                       \\ \hline
Screen Split                              & SS\_Setup\_6-57.exe                                   & 78a9ba1748686eecc181e97151c813acc92edcd35c05a673f24823ae0bb2a8ec & 0/66                                                                                                       \\ \hline
Mozilla Thunderbrid                       & Thunderbird Setup 91.3.1.exe                          & 39a502318a8b10bc25d9547b9c48fb646f7083d7161a4da7cee48ceabe77c65e & 0/53                                                                                                       \\ \hline
WordWeb                                   & wordweb10.exe                                         & 27142582b89e0fa2ca6a9d5036eec3bde140109aa9632f9b5eac30933a082080 & 1/70                                                                                                       \\ \hline
\end{tabular}

\label{tab:benign}
\end{table*}




\end{document}